\documentclass[article,reprint,onecolumn,11pt,superscriptaddress,preprintnumbers,floatfix,nofootinbib,longbibliography]{revtex4-2}

\usepackage{epsfig,amsmath,amssymb,color,comment,physics}
\usepackage{algorithm,algpseudocode}
\usepackage[makeroom]{cancel}
\usepackage{mathrsfs}
\usepackage{float}
\usepackage[normalem]{ulem}
\usepackage[countmax]{subfloat}
\usepackage[normalem]{ulem}
\usepackage[english]{babel}
\usepackage{dsfont}
\usepackage{braket}
\usepackage[bookmarks=true,colorlinks,linkcolor=Blue,urlcolor=NavyBlue,citecolor=RoyalBlue]{hyperref}
\usepackage[dvipsnames]{xcolor}
\usepackage{graphicx}
\usepackage[caption=false]{subfig} 
\usepackage{cleveref}
\usepackage{tikz}
\usepackage{standalone}
\usepackage{overpic}

\graphicspath{{./Figs/}}
\usepackage{amsmath, amssymb,latexsym,amsfonts,bm,hyperref}
\usepackage{orcidlink}
\usetikzlibrary{patterns, positioning, calc, decorations.pathreplacing}

\usepackage{orcidlink}
\usepackage{amsmath}
\usepackage{booktabs}
\usepackage{graphicx, color}
\usepackage{amssymb}
\usepackage{epsfig}
\usepackage{soul}
\usepackage{makecell, multirow}
\usepackage{bm}
\usepackage{booktabs}
\usepackage{blkarray}
\usepackage{bm, amsmath, amsfonts, amssymb,xfrac}
\usepackage{multirow, tabularx, dcolumn}
\usepackage{mathtools, leftidx, braket, slashed, cancel, bigdelim}
\usepackage{soul}
\usepackage{pgf}
\definecolor{link_blue}{RGB}{51,102,204}

\floatname{algorithm}{Stage}

\usepackage{orcidlink}

\begin{document}
\title{Gauge-Covariant Wavelets for Quark-Propagator
Sparsening and Distillation in Lattice QCD}

\author{Debsubhra Chakraborty${}^{\orcidlink{0000-0001-5815-4182}}$}
\email{debsubhra.chakraborty@tifr.res.in}
\affiliation{Department of Theoretical Physics, Tata Institute of Fundamental Research, \\ Homi Bhabha Road, Mumbai 400005, India}

\author{Nilmani Mathur${}^{\orcidlink{0000-0003-2422-7317}}$}
\email{nilmani@theory.tifr.res.in}
\affiliation{Department of Theoretical Physics, Tata Institute of Fundamental Research, \\
Homi Bhabha Road, Mumbai 400005, India}

\author{Mrinmoy Basak${}^{\orcidlink{0000-0001-8965-0824}}$}
\affiliation{Department of Theoretical Physics, Tata Institute of Fundamental Research, \\ Homi Bhabha Road, Mumbai 400005, India}

\preprint{TIFR/TH/26-28}

\begin{abstract}
We develop a novel gauge-covariant wavelet framework for computing hadronic correlation functions in lattice QCD.
Two complementary approaches connect spatial localization and
resolution-scale separation to propagator sparsening and
distillation. A gauge-covariant fast wavelet transform, based
on compactly supported biorthogonal filters, coarse-grains
quark propagators and reduces early-time excited-state
contamination relative to direct decimation at equal spatial
reduction. A second construction applies a resolvent filter
of the spatial covariant Laplacian to localized sources,
generating a distillation subspace without computing Laplacian
eigenpairs. We detail its gauge covariance, 
smearing profiles, and implementation in exact and stochastic
distillation. Spatial localization enables unbiased
Hansen--Hurwitz sampling of hadronic contractions, reducing
their computational cost substantially. We also examine momentum aliasing
induced by fixed coarse grids and the use of random grid
shifts to remove unwanted contributions in expectation.
Numerical benchmarks of the proposed methods are performed with single hadrons, the $I=2$
two-pion system, the deuteron channel, and boosted nucleons.
The extracted energy levels agree very well with full or conventional
distillation calculations within statistical uncertainties.
We quantify the costs of on-demand basis construction and contractions,
demonstrate contraction speedups at comparable statistical
precision. These studies establish a practical foundation of wavelet-based propagator
compression and operator construction for multi-operator
calculations of hadronic and nuclear systems, particularly at larger lattice volumes.
\end{abstract}
\maketitle

\section{Introduction}
\label{sec:introduction}

Hadronic correlation functions are central to lattice-QCD
determinations of spectra, scattering amplitudes, and matrix
elements.  In lattice QCD, their computation requires interpolating operators with adequate overlap
onto the states of interest and efficient algorithms for combining quark propagators into the required Wick
contractions. The computational demands increase with the
size of the operator basis used in generalized-eigenvalue analyses \cite{MICHAEL198558,luscherwolff1990}
and are particularly challenging for multihadron and nuclear
systems and hadrons at nonzero momentum.
 Although advances in algorithms
and high-performance computing have established lattice QCD
as a precision tool for hadron physics
\cite{gattringerlang2009,kronfeld2022}, the cost of constructing
and evaluating a large number of correlation functions, particularly in large-volume lattices,  continues to limit the  
scope of such calculations.

An important strategy is to represent quark fields in a
reduced space that retains adequate overlap with the relevant
hadronic states. Distillation implements this idea by projecting
quark fields onto the subspace spanned by the lowest eigenmodes
of the gauge-covariant spatial Laplacian
\cite{hadronspectrum2009}. The resulting factorization separates
quark propagation from operator construction, allowing projected
propagators, or perambulators, to be reused across many
interpolating operators. This makes distillation particularly
effective for constructing correlation matrices for
generalized-eigenvalue problem (GEVP) 
\cite{MICHAEL198558,luscherwolff1990} analyses. Its stochastic extension,
stochastic LapH, combines noise restricted to the distillation
subspace with dilution projectors to reduce the number of
Dirac inversions required to estimate quark propagation
\cite{PhysRevD.83.114505}. Further developments include optimized
distillation profiles \cite{Knechtli:2022bji, Hollwieser:2022doy},
momentum-adapted constructions \cite{
PhysRevD.93.094515,PhysRevD.103.034502},
and localized bases for stochastic evaluation of hadronic
contractions \cite{mv6l-m8rq}.

Several computational challenges, however, remain, particularly
for multihadron calculations on large spatial volumes. At fixed physical smearing, the number of retained
Laplacian modes grows approximately linearly with the physical
spatial volume. The eigensolve, eigenvector storage, and repeated input/output
required for reuse across multiple projects therefore become
increasingly costly. Stochastic estimation reduces the inversion cost
but does not remove the need to construct the Laplacian
subspace. Moreover, evaluating hadronic contractions within
that subspace can itself be costly, particularly for baryonic, multihadron 
systems with a large correlation matrix GEVP setup within large volumes. Localized conventional
distillation addresses part of this problem by changing the
basis within the retained Laplacian eigenspace
\cite{mv6l-m8rq}. It therefore preserves the original
distillation projector, but still requires the underlying
eigenvectors.

Quark-propagator sparsening offers a complementary approach
\cite{Detmold_2021,xfmp-c2d4, khhk-qw9n}. Instead of constructing a spectral
subspace, it reduces the spatial support on which contractions
are evaluated. Restricting a sink sum to a regular coarse
sublattice of spacing $N$ in lattice units reduces the number
of contributing sites by a factor of $N^3$. For contractions
whose cost is proportional to this number, the spatial
reduction translates directly into computational savings.
However, a direct decimation modifies momentum projection
and can enhance excited-state contributions. Smearing before
decimation can suppress these effects
\cite{xfmp-c2d4}, while uniform random shifts of
the coarse-grid origin provide an unbiased estimator of the
corresponding full spatial sum \cite{khhk-qw9n}.
Consequently, reducing the number of contraction sites must
be considered together with its effects on spectral overlap,
momentum contamination, and statistical precision.

Multiresolution analysis (MRA) provides a framework for
organizing these spatial and spectral reductions
\cite{daubechies1988,mallat1989}. In an MRA, nested
approximation spaces represent a function at successive
resolutions, while complementary detail spaces encode the
information added upon refinement. In that, scaling functions span
the approximation spaces, and wavelets resolve the details.
For suitable function classes, localization in position
and scale supports efficient representations and controlled
approximation, as exploited in signal compression, denoising,
and numerical differential equations \cite{mallat2009}.
These features suggest two applications to computing hadronic correlation functions. The first one is filtering before spatial coarsening, and second is 
constructing localized approximation spaces for distillation.

The physical motivation for such representations follows
from the multiscale structure of quantum field theory.
In QCD, short-distance fluctuations coexist with confinement,
hadronic correlation lengths, and longer nuclear scales.
Localized descriptions of quantum field theories have a long history, including
Wilson's wave-packet constructions \cite{wilson1965}.
Wavelet methods have subsequently been investigated in
Hamiltonian formulations and studies of entanglement
\cite{bulutpolyzou2013,PhysRevLett.116.140403,george2022,basak2026a,basak2026b}, regularization of gauge theories~\cite{10.1143/PTP.94.1135}, 
in proposals for continuum field-theory simulations
\cite{neuberger2018}, and in connections to multiscale
entanglement renormalization and quantum circuits
\cite{kaplantilloy2026}. Multiscale representations, in some aspects, also
enter lattice-QCD solvers through domain decomposition
and adaptive multigrid \cite{luscher2007,babich2010}.
The objective here is to use spatial localization and
resolution-dependent filtering directly in the construction
and contraction of hadronic operators, which have been largely unexplored.

In this work we introduce
a novel gauge-covariant multiresolution framework and present
representative applications to lattice-QCD correlation
functions. We develop its 
two complementary
realizations in detail and examine their numerical
implementation, stochastic estimators, and coarse-grid
effects. The first realization is propagator sparsening
through a gauge-covariant fast wavelet transform. We use
low-pass filters from compactly supported
Cohen--Daubechies--Feauveau biorthogonal wavelets
\cite{cdf1992}, inserting parallel transporters to combine
propagators at different spatial sites covariantly.
Successive filtering and coarsening then provide a
resolution-dependent representation of the propagators.
We compare this construction with direct decimation
and examine how filtering and coarse-grid shifts affect
the resulting correlation functions.

The second realization, which we call \emph{wavelet
distillation}, constructs a localized approximation space
by applying a resolvent-type filter of the gauge-covariant
spatial Laplacian to sources on a coarse sublattice.
The filtered sources are generally nonorthogonal;
their Gram matrix defines an orthogonal projector onto
their span, or equivalently an orthonormal basis can be
obtained through L\"owdin orthogonalization \cite{Lowdin1950,Lowdin1970}.
Unlike a basis rotation within conventional distillation \cite{mv6l-m8rq},
this procedure constructs an approximation
space directly, without first computing Laplacian
eigenpairs. The resolvent filter can be applied through
sparse linear solves which are computationally much cheaper. Agreement with conventional
distillation is therefore a numerical test of the
spectral information accessible to the two subspaces,
rather than a consequence of their algebraic equivalence.

We formulate both exact and stochastic distillation in
this basis. In the stochastic construction, diluted noise
vectors estimate quark propagation within the approximation
space, while a separate Hansen--Hurwitz (HH) estimator \cite{HansenHurwitz} samples
the indices entering hadronic contractions. These two
estimators address different costs: Dirac inversions and
contraction work, respectively. Spatial localization
concentrates the weight of hadronic elementals and
provides an importance-sampling distribution for the
latter. We describe the conditions required for unbiased
estimation and compare the resulting precision and
computational cost with exact contraction of the
stochastically estimated propagators.

A further issue is the translation structure introduced
by the coarse grid. With its origin held fixed, the
construction can couple a nominally momentum-projected
operator to aliased momenta. This is particularly relevant
for boosted states, where an alias may have lower energy
than the target state, and hence, through mixing, can lower the energy of the lowest state. We analyze these contributions
and the use of independent random source and sink grid shifts
to eliminate them in expectation. Comparisons with and
without shifts distinguish changes in state overlap
from the recovery of the intended momentum projection.

Our numerical studies include single-hadron correlation
functions, the $I=2$ two-pion spectrum from a
multi-operator analysis within GEVP, the deuteron channel, and
boosted nucleons. We compare the extracted energies
and statistical uncertainties with full or conventional
distillation calculations and assess the computational
costs of basis construction and contractions. We also
discuss on-demand basis generation and its implications
for storage and input/output. These benchmarks establish
the performance of the methods for the systems studied
and identify the considerations relevant to extending
them to larger volumes and operator bases.

The paper is organized as follows. Sec.~\ref{rev:dist_spars}
briefly reviews distillation and propagator sparsening in lattice
QCD. In Sec.~\ref{sec:wavelets-mra}, we introduce  the multiresolution analysis and wavelets for lattice QCD calculations. 
Sec.~\ref{sec:wavelet:Sparsening} develops quark-field sparsening
using the fast wavelet transform (FWT), with numerical benchmarks
presented in Subsec.~\ref{sec:wavelet:results}.
Sec.~\ref{sec:wavelet-distillation} introduces wavelet distillation
and benchmarks its performance for several hadronic correlators,
including boosted nucleon correlators. In Sec.~\ref{sec:wavelet-stochastic},
we develop stochastic wavelet distillation and Hansen--Hurwitz
sampling, present numerical benchmarks,
and analyze statistical precision and computational costs.
We summarize by discussing the results and prospects for
future applications.

\section{Distillation and Sparsening}\label{rev:dist_spars}
In this section, we briefly review the construction of the \emph{distillation}~\cite{hadronspectrum2009} and \emph{quark field sparsening}~\cite{Detmold_2021, PhysRevD.103.014514, xfmp-c2d4, khhk-qw9n} methodologies. The separation of scales,  the insensitivity of low-lying hadronic properties to the high-frequency modes of the quark fields is the central idea underlying both methodologies. The distillation method, introduced in Ref.~\cite{hadronspectrum2009}, provides a framework for removing the high-frequency components of the quark fields and for estimating quark propagators (quark lines) within the approximate subspace spanned by the low-frequency components, which is sufficient for studying low-lying hadronic properties. In this methodology, one defines a projector operator $V(t)=\sum_{I=1}^{n_v}\xi_I(t)\xi_I^{\dagger}(t)$, constructed from the lowest $n_v$ eigenvectors, $\xi_I(t)$, of the 3D spatial, gauge-covariant Laplacian,
\begin{equation}
  \nabla^2(t)=\sum_{\mu=\pm1}^{\pm3}U_{\mathbf{x},\mu}(t)\delta_{\mathbf{x}+\hat{\mu},\mathbf{y}}-6\delta_{\mathbf{x},\mathbf{y}},
\end{equation}
defined on the lattice. The projected, smoothed quark fields are then defined as $\tilde{q}(t) = V(t)q(t)$. Here, $U_{\mathbf{x},\mu}(t)$ denotes the standard parallel transporter of unit lattice length in the direction $\mu$, defined at the lattice site $(\mathbf{x},t)$,
\begin{equation}
  -\nabla^2(t)\, \xi_I(t) = \lambda_I(t)\, \xi_I(t), \qquad I \in \{1,\dots,n_v\},
  \label{eq:laplacian-eigenproblem}
\end{equation}
where $\xi_I(t)$ and $\lambda_I(t)$ denote, respectively, the eigenvectors and eigenvalues of the 3D Laplacian at timeslice $t$ on the lattice. The all-to-all quark propagator, restricted to the distillation subspace spanned by the $n_v$ eigenvectors, $\xi_I(t)$, and known as the quark perambulator, is given by
\begin{equation}
  \tau_{IJ}(t_f,t_i) = \xi_I^\dagger(t_f)\, D^{-1}\, \xi_J(t_i),
  \qquad I,J \in \{1,2,\dots,n_v\}.
  \label{eq:distillation-propagator}
\end{equation}
Here, $D$ is the four dimensional Dirac operator defined on the lattice. This construction removes the high-frequency modes of the quark fields that are irrelevant for low-lying hadron physics. 
One of the main advantages of this framework is that only $4n_v$ Dirac-operator inversions per timeslice are required to compute the hadronic correlation functions within the distillation subspace, irrespective of the number or specific form of the interpolating operators used. These developments have made possible the extraction of excited-state spectra for single-hadron systems, as well as the investigation of multi-hadron systems, including their scattering properties using a large basis of interpolating operators within GEVP analyses. As a result, the past decade has seen remarkable advances in hadron spectroscopy and scattering studies \cite{RevModPhys.90.025001,Green:2026jgv}.

These advantages however introduce two important scaling bottlenecks. Maintaining a fixed ultraviolet resolution requires an approximately constant density of Laplacian eigenmodes, causing $n_v$ to grow linearly with the spatial volume; the eigensolve and associated eigenvector storage and I/O therefore become increasingly expensive on large lattices. Moreover, although distillation reduces the number of Dirac inversions, the Wick contractions needed to assemble correlation functions remain computationally demanding~\cite{khhk-qw9n,mv6l-m8rq}. This cost is particularly severe for multibaryon systems~\cite{Detmold:2019ghl,PhysRevD.87.114512,PhysRevD.110.114505} and can also be substantial for multimeson operator bases containing local tetraquark interpolators~\cite{khhk-qw9n}.

\noindent\emph{Quark-field sparsening} was introduced in Refs.~\cite{Detmold_2021, PhysRevD.103.014514}, to address this problem by evaluating correlation
functions on a coarse spatial sublattice
\cite{Detmold_2021,PhysRevD.103.014514}. Defining
\begin{equation}
\mathcal{G}(\mathbf{x},t)
=
\langle\Omega|
\mathcal{O}(\mathbf{x},t)
\mathcal{O}^{\dagger}(\mathbf{0},0)
|\Omega\rangle ,
\end{equation}
the full and sparse correlators for an interpolating field, $\mathcal{O}$, are
\begin{align}
C_{\mathrm{full}}(\mathbf{p},t)
&=
\sum_{\mathbf{x}\in\Lambda_3}
e^{-i\mathbf{p}\cdot\mathbf{x}}\,
\mathcal{G}(\mathbf{x},t),
\label{eq:full-correlator}\\
C_{\mathrm{sparse}}(\mathbf{p},t)
&=
\sum_{\mathbf{x}\in\widetilde{\Lambda}_3(N)}
e^{-i\mathbf{p}\cdot\mathbf{x}}\,
\mathcal{G}(\mathbf{x},t),
\label{eq:sparse-correlator}
\end{align}
where
\begin{align}
\Lambda_3
&=
\left\{
\mathbf{n}\in\mathbb{Z}^3
\,\middle|\,
0\leq n_i<L
\right\},
\nonumber\\
\widetilde{\Lambda}_3(N)
&=
\left\{
\mathbf{n}\in\Lambda_3
\,\middle|\,
n_i=0\ \mathrm{mod}\ N
\right\},
\label{eq:coarse-sublattice}
\end{align}
with \(L\) divisible by \(N\). Since
\(
|\widetilde{\Lambda}_3(N)|=|\Lambda_3|/N^3
\),
sparsening nominally reduces the spatial contraction cost by a factor of
\(N^3\), while retaining the low-energy spectral information
\cite{Detmold_2021,PhysRevD.103.014514}. Direct decimation can, however,
enhance excited-state contamination at early Euclidean times. To address this issue, Ref.~\cite{xfmp-c2d4} explores how smearing the quark propagators before using them to compute the sparse correlation function can reduce the enhanced excited-state artifacts at early Euclidean times. In Ref.~\cite{khhk-qw9n}, the coarse-grid origin is sampled uniformly
on each gauge configuration, to obtain an unbiased estimator of
the full correlation function upon averaging over the shifts.

Distillation and sparsening thus exploit the same physical separation:
low-energy hadronic observables are comparatively insensitive to
short-distance quark-field fluctuations. 
Multiresolution analysis using wavelets provides
a systematic realization of this separation by decomposing a field into
nested approximation and detail spaces
\cite{daubechies1988,mallat1989,Mallat1989_1,Daubechies1992_1}. The associated wavelets are localized in both position and frequency domains.  In this work we introduce a gauge-covariant multiresolution framework by constructing an approximation subspace, in the spirit of \emph{distillation}, where we replace the eigenvector basis, $\{\xi_I\}$, of Eq.~\eqref{eq:laplacian-eigenproblem} with a basis of localized \emph{scaling functions} from a suitably chosen wavelet family. Since a wavelet basis is, by construction, localized in coordinate space, it directly gives rise to a sparsity that can be exploited to speed up Wick contractions, while still furnishing a low-rank, gauge-covariant projector that can serve as a genuine distillation-type basis for variational (GEVP) analyses. Additionally, we construct a fast wavelet transform based on compactly supported Cohen--Daubechies--Feauveau (CDF) biorthogonal wavelets \cite{cdf1992} and use it to coarse-grain quark propagators for computing sparse correlation functions, in place of the direct decimation shown in Eq.~\eqref{eq:sparse-correlator}, while retaining a controlled multiresolution decomposition.

\section{Wavelets and Multiresolution Analysis}
\label{sec:wavelets-mra}
The mathematical framework underlying both wavelet-based constructions used in this work is \emph{multiresolution analysis} (MRA)~\cite{daubechies1988, mallat1989, Mallat1989_1, Daubechies1992_1}. An MRA of $L^2(\mathbb{R}^d)$, the space of square-integrable functions defined on $\mathbb{R}^d$, organizes a signal into a ladder of nested approximation subspaces, $V_s$, together with complementary detail spaces, $W_s$, defined through $V_{s-1} = V_s \oplus W_s$. The approximation subspaces, $V_s$, satisfy the following properties:
\begin{equation}
  \cdots \subset V_2 \subset V_1 \subset V_0 \subset L^2(\mathbb{R}^d), \qquad \bigcap_s V_s = \{0\}, \qquad \overline{\bigcup_s V_s} = L^2(\mathbb{R}^d).
  \label{eq:mra-ladder}
\end{equation}
 $V_s$ captures the content of the signal at resolution $s$ (coarser scales corresponding to larger $s$), while $W_s$ carries the detail lost in passing from resolution $s-1$ to resolution $s$. In the classical construction, each $V_s$ is generated by integer translates of a single \emph{scaling function} $\varphi$, dilated by $2^{-s}$, with $\varphi$ satisfying a two-scale refinement relation in one-dimension,
\begin{align}
  &\varphi(x) = \sqrt{2}\sum_r h(r)\, \varphi(2x-r),
  \label{eq:two-scale}\\
  &\psi(x) = \sqrt{2}\sum_r g(r)\, \varphi(2x-r),\label{eq:wavelet1D}
\end{align}
for a finite low-pass filter $h(r)$, which depends on the specific wavelet family. An analogous relation in Eq.~\eqref{eq:wavelet1D} with a high-pass filter $g(r)$ defines the \emph{wavelet} function $\psi$ spanning $W_s$. Compactly supported orthonormal~\cite{daubechies1988} and biorthogonal~\cite{cdf1992} solutions of Eq.~\eqref{eq:two-scale} give rise to the \emph{fast wavelet transform} (FWT): a filter-bank algorithm in which coefficients at resolution $s$ are obtained from those at $s-1$ by convolution with $h$ followed by decimation, with reconstruction proceeding through the reverse operation.

This dilation-based route is natural on a fixed, translationally invariant lattice, which we adopt for \emph{quark field sparsening}. Since the dilation operator is not inherently gauge covariant, we insert Wilson lines to make sure the scaling Eq.~\eqref{eq:two-scale} transforms covariantly as discussed in detail in Sec.~\ref{sec:wavelet:Sparsening}. We specifically use a gauge-covariant FWT constructed from compactly supported CDF biorthogonal wavelets~\cite{cdf1992}, though the inclusion of Wilson lines means that perfect reconstruction of the quark propagator from its wavelet decomposition components is no longer exact. Note, this imperfection will not affect its application to sparsening that we discuss in Sec.~\ref{sec:wavelet:Sparsening}.

Alternatively, one can avoid inserting Wilson lines altogether by defining scaling functions that transform covariantly under gauge transformations by construction. In this route to multiresolution analysis, scale is defined \emph{spectrally}, through a function of a positive semi-definite, gauge-covariant operator $L$~\cite{CoifmanMaggioni2006, Hammond2011}, rather than \emph{geometrically} through spatial dilation. Given the spectral decomposition $L = \sum_\ell \lambda_\ell\, \chi_\ell \chi_\ell^\dagger$, a kernel $g:\mathbb{R}^+ \to \mathbb{R}^+$, and a scale parameter $t$, the associated wavelet operator is $T_g^t = g(tL)$. Localized \emph{spectral wavelets} are then obtained by acting with this operator on a delta source placed at a chosen vertex $n$,
\begin{equation}
  \psi_{t,n} = g(tL)\,\delta_n = \sum_\ell g(t\lambda_\ell)\, \chi_\ell^*(n)\, \chi_\ell.
  \label{eq:spectral-wavelet}
\end{equation}
A band-pass kernel, satisfying $g(0)=0$ and $g(\lambda)\to0$ as $\lambda\to\infty$, generates the localized functions that span the detail space, while a complementary scaling-function kernel $h$ with $h(0)>0$ supplies the low-frequency content, with $\phi_n = h(L)\delta_n$ spanning the approximation space. The spatial localization of $\psi_{t,n}$ and $\phi_n$ around $n$ follows from the smoothness of the kernels and the sparsity of $L$~\cite{Hammond2011}. Originally introduced for diffusion operators on graphs and data sets~\cite{CoifmanMaggioni2006} and formalized for general spectral kernels in Ref.~\cite{Hammond2011}, this construction requires no explicit diagonalization beyond evaluating a matrix function of $L$ on delta sources. It generalizes directly to any operator for which $g(L)$ can be efficiently computed, including the gauge-covariant spatial Laplacian $\nabla^2(t)$ that we employ for \emph{wavelet distillation} in Subsec.~\ref{sec:wavelet-construction}. In that application, however, we restrict our attention to the single-scale approximation subspace, which suffices to isolate the low-momentum modes relevant for distillation, rather than constructing a full hierarchy of nested approximation and detail spaces.

The two constructions therefore realize the same underlying idea of decomposing a field into approximation spaces via basis functions localized in both position and scale through complementary structural routes: (i) a dilation-based filter bank for the CDF wavelets used in \emph{quark field sparsening}, and (ii) a spectral, operator-based construction for the localized Laplacian-resolvent basis used in \emph{wavelet distillation}. We develop each realization, and its covariantization to a non-Abelian gauge background, in the two sections that follow.

\section{Quark-field sparsening with the fast wavelet transform}
\label{sec:wavelet:Sparsening}

The cost of hadronic contractions is proportional to the number of spatial sites over which the sink is summed. Quark-field sparsening~\cite{Detmold_2021,PhysRevD.103.014514} reduces this number by evaluating the sink only on a regular sublattice of spacing $N$, lowering the contraction cost by a factor $N^3$. A direct decimation of the sink, however, has two artifacts. It enhances excited-state contamination at early Euclidean times~\cite{xfmp-c2d4}, and a sum over a coarse grid projects not only onto the desired momentum but also onto momenta that differ from it by reciprocal vectors of the coarse grid. Smearing the propagators before decimating reduces the first effect~\cite{xfmp-c2d4}, and displacing the coarse grid randomly on every configuration removes the second in the ensemble average~\cite{PhysRevD.103.014514,khhk-qw9n}.

The fast wavelet transform (FWT) instead combines filtering and decimation in a controlled way. Each step applies a compact, symmetric low-pass filter and then halves the resolution, so that the propagator is smoothed on precisely the scale removed by the subsequent decimation. Iterating the step yields a hierarchy of coarse-grained propagators on grids of spacing $2^sa$, $s=1,2,\dots$, which reduces the cost of the sink contractions by $8^s$. Here, $a$ is the lattice spacing. We apply the transform once per propagator and reuse for all hadronic channels.

The section is organized as follows. Subsec.~\ref{sec:wavelet:fwt} defines the FWT on the lattice and the filters we use, and Sec.~\ref{sec:wavelet:gcfwt} discusses its gauge covariance. Subsec.~\ref{sec:wavelet:aliasing} discusses the momentum aliasing induced by the coarse grid and its removal by random grid shifts. Numerical setup is described in Subsec.~\ref{sec:fwt-setup}. Finally, in Sec.~\ref{sec:wavelet:results}, we compare results
from direct decimation and FWT-based sparsening with those from full correlators for the pion, $\rho$ meson, nucleon, and
two-nucleon system in the deuteron channel.

\subsection{Fast wavelet transform on the lattice}
\label{sec:wavelet:fwt}

Consider a field $f(\mathbf n)$ on a periodic spatial lattice with $L/a$ sites per direction, and associate it with resolution level $s=0$, $f^{(0)}\equiv f$. The level-$s$ approximation $f^{(s)}$ lives on the coarse lattice,
\begin{equation}
  \Lambda_s=\big\{\mathbf n\;\big|\;n_i\in\{0,1,\dots,L/(2^sa)-1\},\ i=1,2,3\big\},
  \label{eq:coarse_lattice}
\end{equation}
with spacing $2^sa$, where the site $\mathbf n\in\Lambda_s$ corresponds to the site $2^s\mathbf n$ of the original lattice. Successive levels are related by the scaling relation of the FWT,
\begin{equation}
  f^{(s)}(\mathbf n)=\sum_{\mathbf r}h(\mathbf r)\,f^{(s-1)}(2\mathbf n+\mathbf r),\qquad \mathbf n\in\Lambda_s,
  \label{eq:FWT}
\end{equation}
with arguments of $f^{(s-1)}$ taken modulo $L/(2^{s-1}a)$, and $h(\mathbf r)$ the low-pass filter of the chosen wavelet family.

We use Cohen-Daubechies-Feauveau (CDF) biorthogonal wavelets~\cite{cdf1992}, whose compact support keeps the transform local and computationally inexpensive, and whose symmetry is required to preserve parity. Relaxing orthogonality to biorthogonality is essential here, since  
no real-valued compactly supported orthonormal wavelet, apart from the discontinuous Haar wavelet, can be symmetric, as shown  by Daubechies ~\cite{Daubechies1992_1, daubechies1988}. The three-dimensional filter is the tensor product $h(\mathbf r)=k(r_1)\,k(r_2)\,k(r_3)$ of the symmetric one-dimensional filter $k(n)=k(-n)$ listed in Table~\ref{tab:FilterCoefficients}. We use two members of the family: Bior~4,4 (also known as CDF~9/7) and the wider Bior~5,5.

\begin{table}[htbp]
\centering
\renewcommand{\arraystretch}{1.1}
\begin{tabular}{l c c}
\hline\hline\\[-2.5ex]
$K,N,\tilde N$ & $n$ & $k(n)$\\[0.5ex]
\hline\\[-2ex]
$K=4$          & $0$       & $\phantom{-}0.602949018$\\
$N=4$          & $\pm1$    & $\phantom{-}0.266864118$\\
$\tilde N=4$   & $\pm2$    & $-0.078223266$\\
               & $\pm3$    & $-0.016864118$\\
               & $\pm4$    & $\phantom{-}0.026748757$\\[1ex]
\hline\\[-2ex]
$K=5$          & $0$       & $\phantom{-}0.636046869922$\\
$N=5$          & $\pm1$    & $\phantom{-}0.337150822538$\\
$\tilde N=5$   & $\pm2$    & $-0.066117805605$\\
               & $\pm3$    & $-0.096666153049$\\
               & $\pm4$    & $-0.001905629356$\\
               & $\pm5$    & $\phantom{-}0.009515330511$\\[0.5ex]
\hline\hline
\end{tabular}
\caption{One-dimensional low-pass filter coefficients $k(n)$ of the CDF wavelets Bior~4,4 ($K=4$) and Bior~5,5 ($K=5$), normalized to $\sum_nk(n)=1$ \cite{cdf1992}.  $N$ and $\tilde N$ are the numbers of vanishing moments of the primal and dual wavelets. The normalization only rescales the coarse-grained propagators and drops out of all energies.}
\label{tab:FilterCoefficients}
\end{table}

\subsection{Gauge-covariant fast wavelet transform}
\label{sec:wavelet:gcfwt}

Quark fields at different sites transform independently under local gauge transformations. Hence, Eq.~\eqref{eq:FWT} cannot be applied to them directly. We therefore parallel-transport each field entering the sum to the common site $2\mathbf n$ before combining them. For the quark propagator $Q^{(0)}(\mathbf x,t_f;\mathbf x_i,t_i)$ from a source at $(\mathbf x_i,t_i)$, the gauge-covariant FWT is defined by,
\begin{equation}
  Q^{(s)}(\mathbf n,t_f;\mathbf x_i,t_i)=\sum_{\mathbf r}h(\mathbf r)\,\mathcal W^{(s-1)}(2\mathbf n,2\mathbf n+\mathbf r;t_f)\,Q^{(s-1)}(2\mathbf n+\mathbf r,t_f;\mathbf x_i,t_i),\qquad \mathbf n\in\Lambda_s .
  \label{eq:GaugeCovariantFWT1}
\end{equation}
The transform acts only on the spatial sink coordinate, separately on every sink time slice. The transform can as well be applied at the spatial source coordinates for a symmetric treatment. 

The transporter $\mathcal W^{(s-1)}(\mathbf m,\mathbf m+\mathbf r;t)$ connects two sites of the level-$(s-1)$ lattice. Since many shortest paths connect them, we average over the $3!=6$ paths built from straight segments along the three axes, taken in all orders,
\begin{equation}
  \mathcal W^{(s-1)}(\mathbf m,\mathbf m+\mathbf r)=\frac16\sum_{\sigma\in S_3}\prod_{i=1}^{3}\mathcal L_{\sigma_i}\big(\mathbf m_{i-1},r_{\sigma_i}\big),
  \qquad \mathbf m_0=\mathbf m,\quad \mathbf m_i=\mathbf m_{i-1}+r_{\sigma_i}\hat\sigma_i,
  \label{eq:PathAverage}
\end{equation}
with $\sigma_i\equiv\sigma(i)$ and the product ordered from left to right. Here $\mathcal L_j(\mathbf m,\ell)=\bar U^{(s-1)}_j(\mathbf m)\,\bar U^{(s-1)}_j(\mathbf m+\hat j)\cdots\bar U^{(s-1)}_j(\mathbf m+(\ell-1)\hat j)$ is the straight line of length $\ell$ in direction $j$, with Hermitian conjugate links for $\ell<0$ and $\mathcal L_j=1$ for $\ell=0$. To suppress ultraviolet gauge fluctuations, the lines are built from links $\bar U^{(s-1)}$ obtained from the level-$(s-1)$ links $U^{(s-1)}$ by one step of HYP smearing~\cite{Hasenfratz:2001hp} with $(\alpha_1,\alpha_2,\alpha_3)=(0.75,0.60,0.30)$. We also find that the APE smearing~\cite{APESMEARING} gives results of comparable quality.

The recursion requires gauge links on each coarse lattice. We obtain them by blocking: the coarse link in direction $j$ is the product of the two smeared fine links it spans, projected onto SU(3), as, 
\begin{equation}
  U^{(s)}_j(\mathbf n)=\mathcal P_{\mathrm{SU}(3)}\Big[\bar U^{(s-1)}_j(2\mathbf n)\,\bar U^{(s-1)}_j(2\mathbf n+\hat j)\Big],\qquad U^{(0)}_j\equiv U_j,
  \label{eq:CoarseLinks}
\end{equation}
and the level-$s$ links are smeared again before they are used in the next step.

Under a gauge transformation on the sink time slice, $Q^{(s-1)}(\mathbf x)\to\Omega^{(s-1)}(\mathbf x)\,Q^{(s-1)}(\mathbf x)$ and $\mathcal W^{(s-1)}(\mathbf m,\mathbf m+\mathbf r)\to\Omega^{(s-1)}(\mathbf m)\,\mathcal W^{(s-1)}(\mathbf m,\mathbf m+\mathbf r)\,\Omega^{(s-1)\dagger}(\mathbf m+\mathbf r)$. Hence $Q^{(s)}(\mathbf n)$ transforms covariantly with $\Omega^{(s)}(\mathbf n)\equiv\Omega^{(s-1)}(2\mathbf n)$, and the coarse links of Eq.~\eqref{eq:CoarseLinks} transform in the same way. Covariance thus holds at every level, and all hadronic correlation functions constructed from the coarse propagators are gauge invariant.

We compute the correlation functions at level $s$ by performing the usual sink contractions with $Q^{(s)}$ and summing the sink over $\Lambda_s$. Since this involves $(L/2^sa)^3$ instead of $(L/a)^3$ sites, the contraction cost is reduced by $8^s$. 
The transform cost scales linearly in the lattice volume and the number
of filter coefficients, and is applied once per propagator,
independently of the number of hadronic channels subsequently
contracted.

\subsection{Aliasing and random sublattice shifts}
\label{sec:wavelet:aliasing}

A sum over a grid of spacing $2^sa$, $a$ being the original lattice spacing,  projects onto zero momentum only up to the reciprocal vectors of that grid: it also admits the momenta $\mathbf P=2\pi\mathbf k/(2^sa)$ with $\mathbf k\in\mathbb Z^3\setminus\{\mathbf 0\}$. States with these momenta are heavier than the zero-momentum ground state and they contaminate the correlator as additional excited states. The smallest aliased momentum, $2\pi/(2^sa)$, decreases with the level. The contamination is therefore less at $s=1,2$ but can extend to large Euclidean times at $s=3$. Decimation, which keeps the propagator only on the coarse sites, couples to these states with full weight, whereas the low-pass filter of the FWT suppresses the high-momentum components of the sink that feed them.

The aliased contributions carry phases $e^{i\mathbf P\cdot\bar{\mathbf c}}$ that depend on the offset $\bar{\mathbf c}$ of the coarse grid relative to the source. If $\bar{\mathbf c}$ is drawn uniformly from $\{0,\dots,2^s-1\}^3$ on each configuration, independently of the gauge field, these phases average to zero, and the offset-averaged correlator becomes
\begin{equation}
  \big\langle C^{(s)}_{\bar{\mathbf c}}(\mathbf{p}, t)\big\rangle_{\bar{\mathbf c}}=8^{-s}\,\mathcal{C}^{(s)}(\mathbf p,t),
  \label{eq:unbiased}
\end{equation}
where $\mathcal{C}^{(s)}(\mathbf p,t)$ is the momentum-projected correlator obtained by evaluating the level-$s$ sink at every site of the fine lattice. The aliased terms are thereby converted from a systematic bias into statistical fluctuations with zero mean. For direct decimation, $\mathcal{C}^{(s)}$ serves as an unbiased estimator of the full fine-grid correlator, as is well known from grid sparsening and position-space sampling~\cite{PhysRevD.103.014514,khhk-qw9n}. For the FWT, by contrast, $\mathcal{C}^{(s)}$ corresponds to a point-to-smeared correlator with a wavelet-filtered sink. It shares the same energy spectrum as the full correlator but features modified overlap factors; in particular, because the source and sink are asymmetric, its effective mass is not guaranteed to approach the plateau from above. While this can be remedied by applying the same filter at the source, we restrict ourselves to the point-source setup in the present work.

In the recursive transform, uniformly distributed offsets are generated level by level. At each level we draw an independent $\mathbf c^{(s)}\in\{0,1\}^3$ and replace $2\mathbf n\to2\mathbf n+\mathbf c^{(s)}$ in Eqs.~\eqref{eq:GaugeCovariantFWT1} and \eqref{eq:CoarseLinks}. In practice, the level-$(s-1)$ propagator and gauge links are both translated by $\mathbf c^{(s)}$ before the transform, which keeps the transporters consistent with the propagator and preserves gauge covariance. The accumulated offset,
\begin{equation}
  \bar{\mathbf c}^{(s)}=\sum_{l=1}^{s}2^{l-1}\,\mathbf c^{(l)}\in\{0,\dots,2^s-1\}^3,
  \label{eq:composite_shift}
\end{equation}
is then uniformly distributed, as required by Eq.~\eqref{eq:unbiased}. We use the same offsets for the FWT and for decimation, so that both schemes are compared on identical grids. As we show later this random shift helps in attaining the correct energy levels while keeping the benefits of FWT.

\subsection{Numerical setup}
\label{sec:fwt-setup}

Calculations are performed on an $N_f = 2+1+1$ MILC gauge ensemble~\cite{PhysRevD.87.054505} generated with Highly Improved Staggered Quark (HISQ) sea dynamics~\cite{PhysRevD.75.054502}, featuring a lattice volume of $L^3\times T = 24^3\times 64$, a lattice spacing of $a \approx 0.1207$~fm, and a sea pion mass of $305$~MeV. In the valence sector, we employ a Wilson-clover quark action tuned to a pion mass of $689$~MeV, computing quark propagators from a single point source at the origin on each of 100 gauge configurations. Note the FWT does not depend on either sea or valence quark mass. The propagators are coarse-grained at the sink to levels $s=1,2,3$, i.e.\ to $12^3$, $6^3$ and $3^3$ sites, reducing the number of sink sites from 13\,824 to 1728, 216 and 27. We compare three schemes: the gauge-covariant FWT with the Bior~4,4 and Bior~5,5 filters, and direct decimation. Each is evaluated on a fixed grid ($\bar{\mathbf c}=\mathbf 0$) and also with one set of random offsets per configuration. The level-0 correlator, labelled ``full'', is common to all schemes and serves as the reference.

We compute zero-momentum correlators of the pion, the $\rho$ meson (averaged over polarizations), the positive-parity nucleon, and the neutron--proton system in the deuteron channel ($NP$), with the following local operators,
\begin{eqnarray}
  \mathcal{O}_{\pi}(\mathbf{x}, t) &=& \overline{d}(\mathbf{x},t)\gamma_5u(\mathbf{x}, t),\quad \mathcal{O}^j_{\rho}(\mathbf{x}, t) = \overline{d}(\mathbf{x},t)\gamma_ju(\mathbf{x}, t),\\
  \mathcal N_0^{\alpha}(\mathbf x,t)&=&\epsilon_{abc}\,u_a^{\alpha}(\mathbf x,t)\,u_b^{\beta}(\mathbf x,t)\,(C\gamma_5)_{\beta\delta}\,d_c^{\delta}(\mathbf x,t),\label{eq:protonOps}\\
  \mathcal N_1^{\alpha}(\mathbf x,t)&=&\epsilon_{abc}\,d_a^{\alpha}(\mathbf x,t)\,d_b^{\beta}(\mathbf x,t)\,(C\gamma_5)_{\beta\delta}\,u_c^{\delta}(\mathbf x,t),\label{eq:neutronOps}\\
  \mathcal O_D^{j}(\mathbf x,t)&=&\tfrac{1}{\sqrt2}\epsilon_{IJ}\mathcal N_I^{\alpha}(\mathbf x,t)\,(C\gamma_jP_+)_{\alpha\beta}\,\mathcal N_J^{\beta}(\mathbf x,t).\label{eq:DeuteronOps1}
\end{eqnarray}
Meson correlators are folded as $C_{\mathrm fold}(t)=[C(t)+C(T-t)]/2$, and their effective masses are defined as $am_{\rm eff}(t)=\operatorname{arccosh}\{[C(t-1)+C(t+1)]/[2C(t)]\}$. For the baryonic channels we use the forward half of the correlator and $am_{\rm eff}(t)=\ln[C(t)/C(t+1)]$. For each hadron we extract their lowest energy level from correlated single-state fits, $A\cosh[E(T/2-t)]$ for mesons and $Ae^{-Et}$ for baryons, with the covariance matrix estimated once on the full ensemble and kept fixed for the central fit and all $100$ bootstrap samples. The same windows are used for all schemes and grid choices. Since all schemes and levels are analyzed on the same bootstrap samples, the difference $\Delta E=E-E_{\rm full}$ is determined with correlated errors much smaller than those of the individual energies. 

\subsection{Results from FWT}
\label{sec:wavelet:results}
We tabulate all the fit results in Appendix~\ref{app:fwt-fits}. 
Figures~\ref{fig:fwt_meff_pion}--\ref{fig:fwt_meff_np} illustrate and compare the effective masses of the three schemes (Decimation, FWTs with Bior 4,4, and Bior 5,5) with those of the full correlator at levels $s=1,2,3$, on a fixed grid (upper rows) and with random offsets (lower rows).

\begin{figure}[htbp]
\centering
\includegraphics[width=0.97\textwidth]{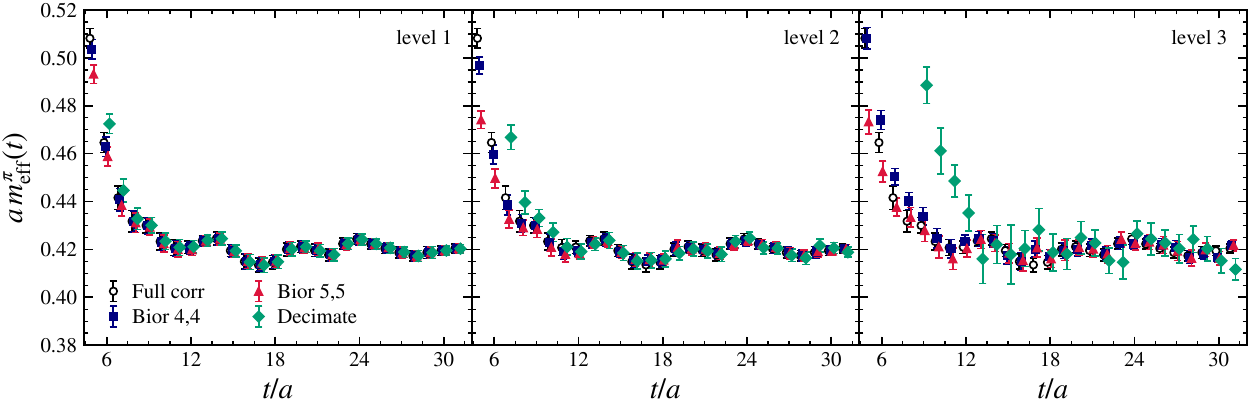}\\[1ex]
\includegraphics[width=0.97\textwidth]{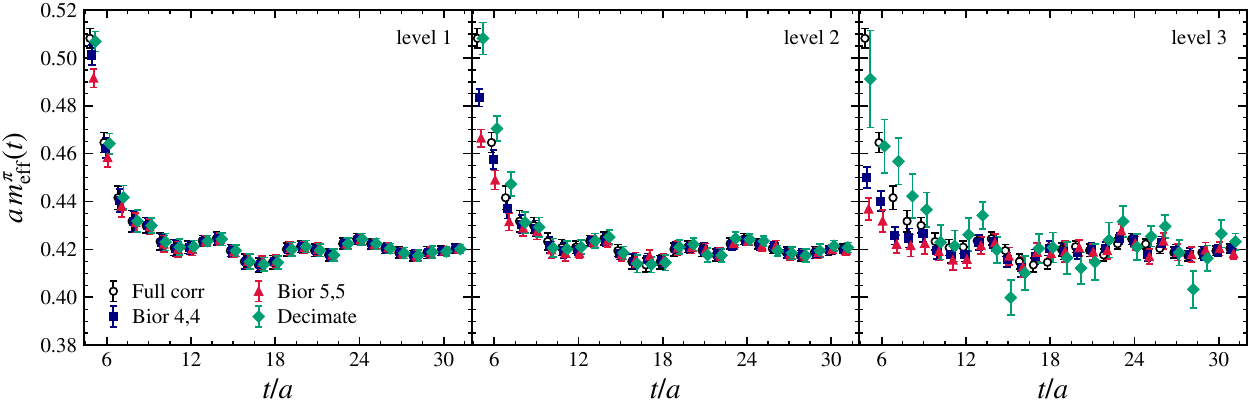}
\caption{Pion effective masses at levels $s=1,2,3$ (left to right) on a fixed coarse grid (top) and with random sublattice shifts (bottom). Open circles: full correlator ($s=0$), identical in all panels. Squares, triangles and diamonds: FWT with Bior~4,4, FWT with Bior~5,5, and decimation, respectively. For presentation clarity we displace points horizontally.}
\label{fig:fwt_meff_pion}
\end{figure}

\begin{figure}[htbp]
\centering
\includegraphics[width=0.96\textwidth]{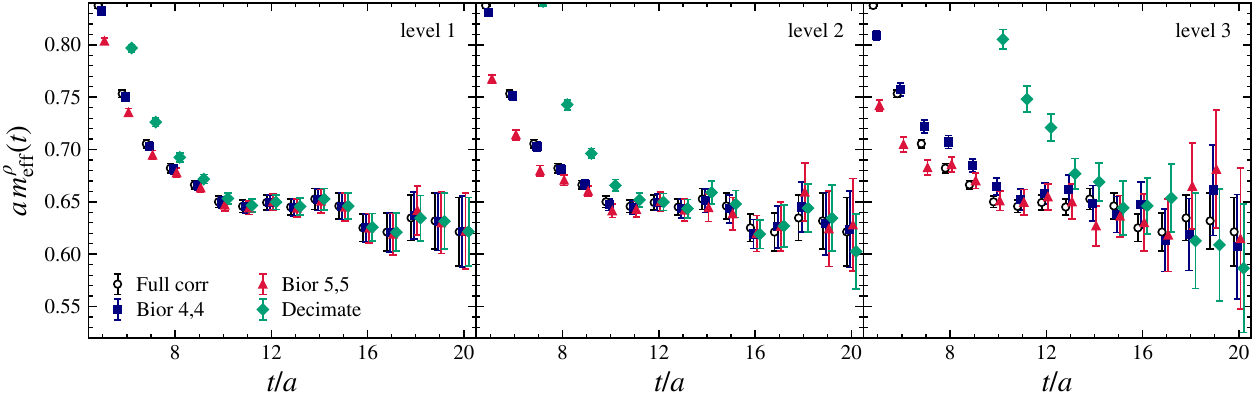}\\[1ex]
\includegraphics[width=0.96\textwidth]{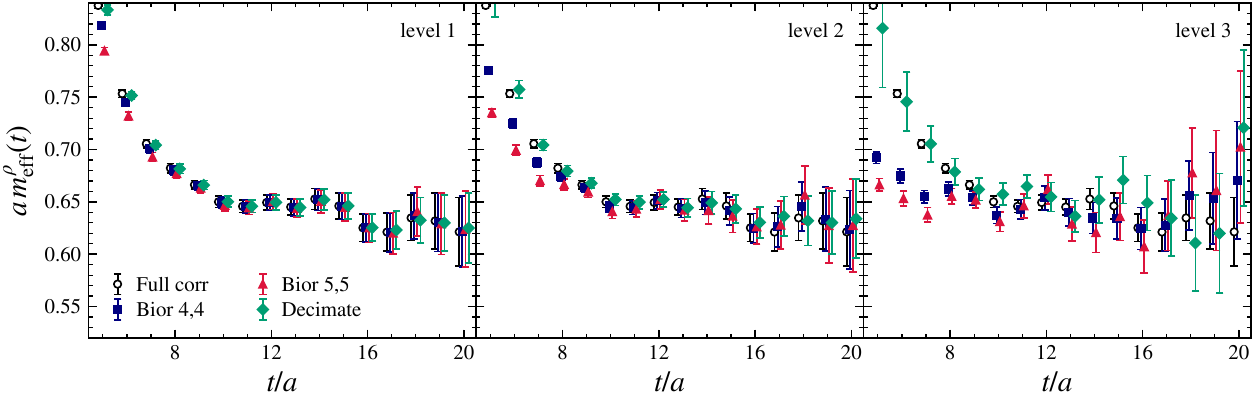}
\caption{As Fig.~\ref{fig:fwt_meff_pion}, for the $\rho$ meson averaged over polarizations.}
\label{fig:fwt_meff_rho}
\end{figure}

\begin{figure}[htbp]
\centering
\includegraphics[width=0.95\textwidth]{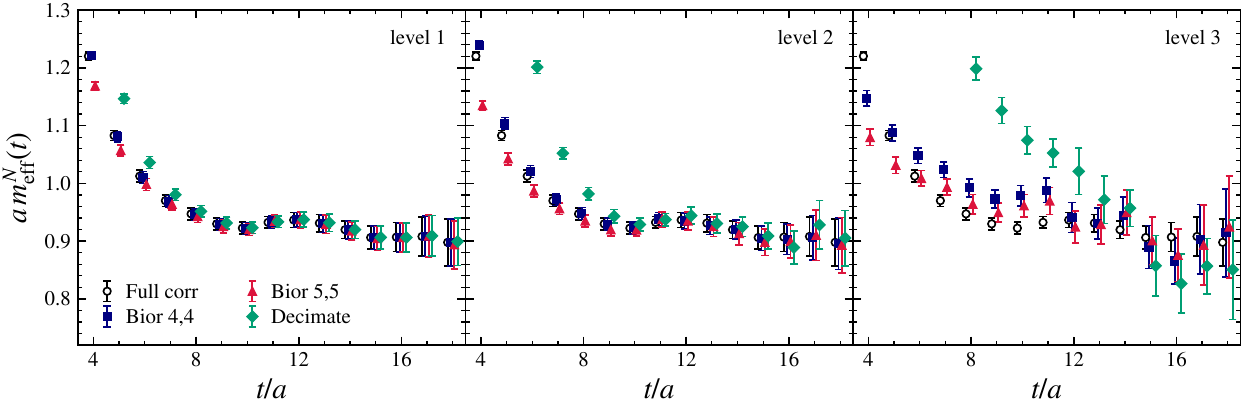}\\[1ex]
\includegraphics[width=0.95\textwidth]{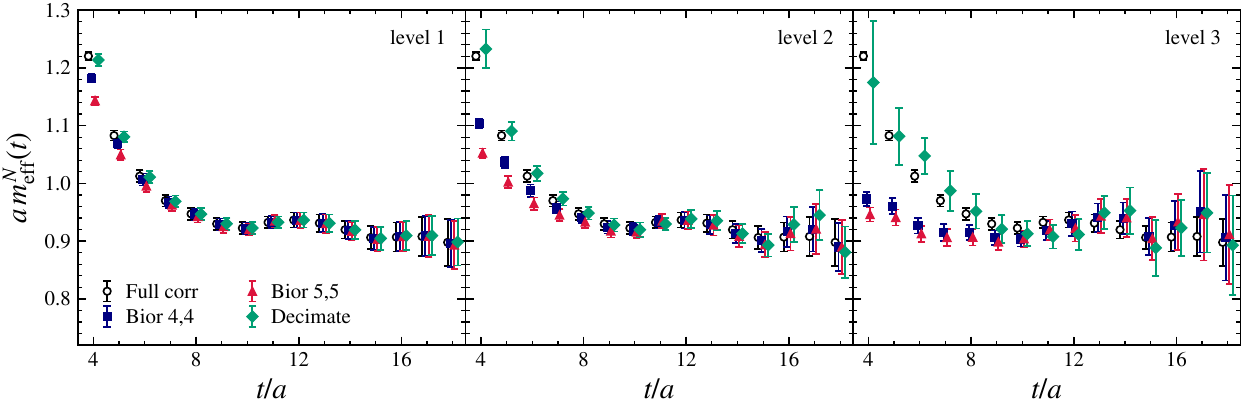}
\caption{As Fig.~\ref{fig:fwt_meff_pion}, for the nucleon.}
\label{fig:fwt_meff_nucleon}
\end{figure}

\begin{figure}[htbp]
\centering
\includegraphics[width=0.95\textwidth]{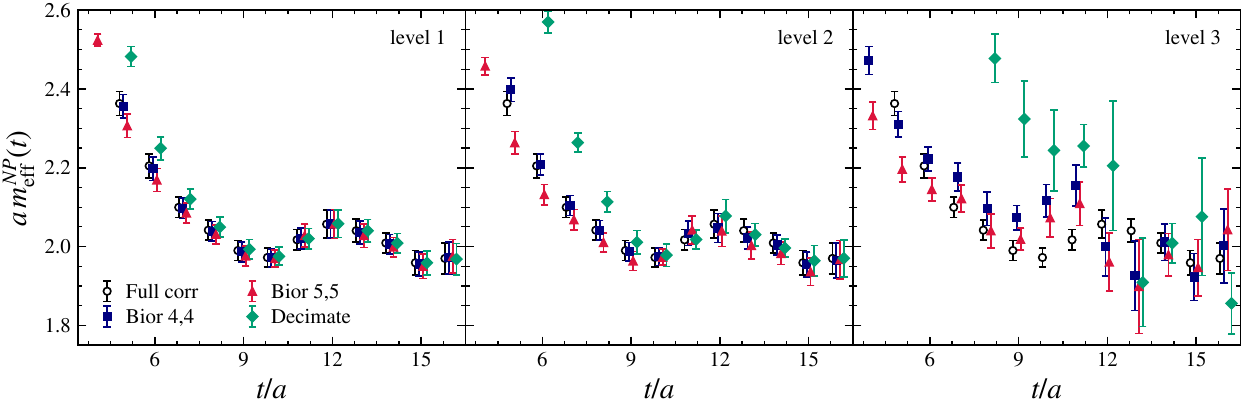}\\[1ex]
\includegraphics[width=0.95\textwidth]{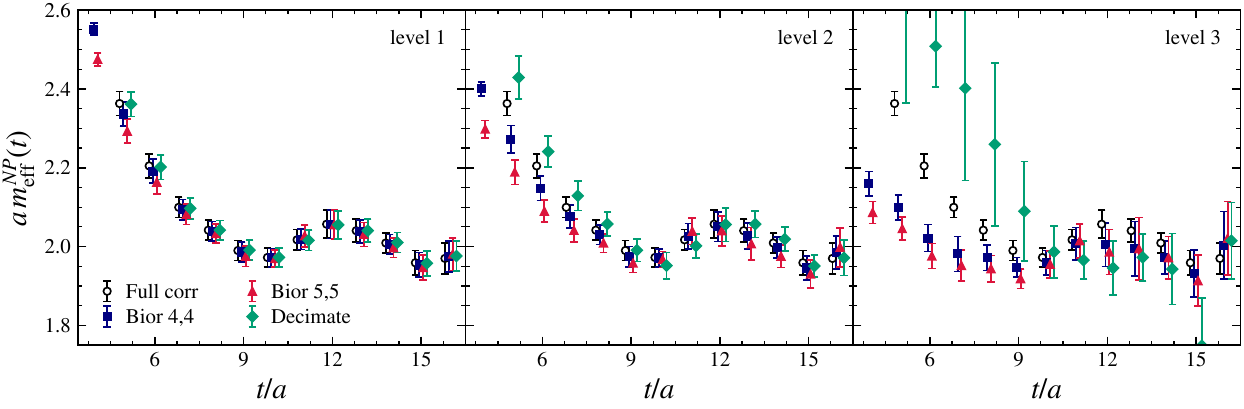}
\caption{As Fig.~\ref{fig:fwt_meff_pion}, for the neutron--proton system in the deuteron channel.}
\label{fig:fwt_meff_np}
\end{figure}

We summarize the main findings below.

\paragraph{Fixed coarse grid:}
At $s=1$ all schemes agree with the full correlator beyond the first few time slices. At coarser levels, decimation develops an enhancement of the effective mass that grows with $s$ as expected from the aliased states discussed in Subsec.~\ref{sec:wavelet:aliasing}. The FWT correlators are much less affected. At $s=3$ they show only a mild enhancement at intermediate times, more pronounced for Bior~4,4 than for the wider Bior~5,5, and their energies agree with the full ones within uncertainties at all levels.

\paragraph{Random sublattice shifts:}
With random offsets, the enhancement of the decimated effective masses disappears on average, as expected from Eq.~\eqref{eq:unbiased}. The aliased contributions now appear as larger fluctuations, most visibly at early times at $s=2,3$. The FWT effective masses at $s=2$ and $3$ now approach the plateau earlier than those of the full correlator, more so for Bior~5,5. For all channels, the FWT energies agree with the full ones at all levels.

\paragraph{Statistical precision and cost:}
The loss of statistical precision is modest. At $s=1$ it is negligible. At $s=2$, with 64 times fewer sink sites, the FWT errors grow by up to about $10\%$ for all channels. At $s=3$, with only 27 sink sites, the pion errors grow by at most about $20\%$, while the baryonic errors roughly double. At this level the FWT is more precise than decimation in every channel and for both grid choices. For contraction-dominated calculations, level $s=2$ therefore offers a favorable balance: a 64-fold reduction of the sink contraction cost, single-hadron energies that agree with the full ones at the sub-percent level.
We use the transform in this way for the hexaquark contractions  that we will be discussing later in Subsec.~\ref{sec:wavelet-fv-spectra}.

\section{Wavelet distillation}
\label{sec:wavelet-distillation}

In the distillation method~\cite{hadronspectrum2009} one replaces the quark field on each time slice by its projection $\tilde q(t)=V(t)\,q(t)$ onto the span of the $n_v$ lowest eigenvectors of the gauge-covariant spatial Laplacian. The projector fixes the quark-field smearing,
while the factorization of propagation and operator construction
allows the resulting perambulators to be reused across large
operator bases.  

At fixed smearing resolution, however, $n_v$ grows approximately linearly with the spatial volume, increasing the costs of basis construction, eigenvector
storage and input/output, and Wick contractions. Moreover,
the spatially extended Laplacian eigenvectors generally
produce dense hadron elementals, limiting opportunities
to exploit locality in the contractions.

In this section we construct an alternative distillation space spanned by
localized, gauge-covariant scaling functions.
These are
obtained by applying a resolvent filter of the spatial
Laplacian to color sources on a coarse sublattice, implementing
the spectral construction described in
Sec.~\ref{sec:wavelets-mra}. We refer to this approach as
\emph{wavelet distillation}.  Each basis function is generated
through a sequence of a few well-conditioned sparse linear solves, without computing
Laplacian eigenpairs. For the parameters studied here,
the reduced basis-construction cost permits regeneration
on demand, avoiding persistent basis storage and the
associated disk input/output. Spatial localization also
concentrates the weight of the hadron elementals in a small
subset of entries, enabling the stochastic contraction
methods developed in Sec.~\ref{sec:wavelet-stochastic}.
The resulting subspace is constructed directly and is
generally distinct from the truncated Laplacian eigenspace.
Agreement with conventional distillation therefore tests
whether the two spaces retain the spectral information
relevant to the hadronic states studied; it does not follow
from a change of basis within the same subspace.

The section is organized as follows. In Subsec.~\ref{sec:wavelet-construction} we construct the basis, establish its gauge covariance, and show how random shifts of the coarse grid restore translation invariance in the ensemble average. Subsec.~\ref{sec:wavelet-benchmark} describes the numerical setup and compares the basis functions and the resulting smearing profile with those of conventional distillation. In Subsec.~\ref{sec:wavelet-fv-spectra} we compare finite-volume energies at zero total momentum for the nucleon, the $I=2$ two-pion system, and the deuteron channel with both dibaryon and hexaquark operators. Finally, in Subsec.~\ref{sec:wavelet-highmom}, we extend this formalism to compute correlation functions at finite momenta.

\subsection{Construction of the basis}
\label{sec:wavelet-construction}
For a positive semi-definite operator such as $L(t)\equiv-\nabla^2(t)$, a function $f(L(t))$ can be applied to a vector without diagonalization whenever $f$ has a convenient closed form~\cite{higham2008functions}. Choosing $f$ to be a power of the resolvent,
\begin{equation}
{1\over {(\mathrm I+\tau L(t))}}
= {1\over {(\mathrm I-\tau\nabla^2(t))}},    
\end{equation}
we define the basis functions
\begin{equation}
  \xi_{n,a}(\mathbf x,t) = \big(\mathrm I-\tau\,\nabla^2(t)\big)^{-m}\,\delta_{\mathbf x,\,\mathbf x_n+\mathbf s}\;e_a,
  \quad
  \mathbf x_n\in\widetilde\Lambda_3(N)\equiv\{\mathbf x\in\Lambda_3\;|\;x_i \bmod N=0\},
  \quad \mathbf s\in\mathbb Z_N^3 .
  \label{eq:wavelet-basis-def}
\end{equation}
A color-vector delta source $\delta_{\mathbf x,\mathbf x_n+\mathbf s}\,e_a$, with $e_a$ a unit vector in color space, is placed on every site of a coarse sublattice of spacing $N$, displaced by the offset $\mathbf s$ (taken modulo the lattice extent). The unshifted grid corresponds to $\mathbf s=\mathbf 0$; the role of nonzero offsets is required for translational invariance of the projector as will be discussed at the end of this subsection. In contrast to the Laplacian eigenvectors, each basis function carries a definite anchor site and a definite color, and we combine the two into the composite index $I=N_c\,n+a$, which runs over
\begin{equation}
  n_v = N_c\,(L/N)^3
\end{equation}
values. The dimension of the distillation space is therefore set by the coarse-grid spacing rather than by a spectral cutoff.

Equation~\eqref{eq:wavelet-basis-def} is an instance of the spectral-wavelet construction of Eq.~\eqref{eq:spectral-wavelet} with three identifications: the operator is the gauge-covariant Laplacian rather than a graph Laplacian, so the basis functions carry color and transform covariantly; the vertex set is subsampled to a coarse sublattice, the lattice analogue of the downsampling step of diffusion wavelets~\cite{CoifmanMaggioni2006}; and the kernel is a low-pass function
\begin{equation}
  h_{\tau,m}(\lambda) = {1\over {(1+\tau\lambda)}^{m}}, \qquad h_{\tau,m}(0)=1 .
  \label{eq:resolvent-kernel}
\end{equation}
The $\xi_I$ are thus scaling functions spanning an approximation space, in the sense of Ref.~\cite{Hammond2011}, whose projector $V(t)$ (defined below) plays the role of the distillation operator, while $\mathbb I-V(t)$ projects onto the complementary detail space. Unlike the sharp spectral cutoff $\Theta(\lambda_{\max}-\lambda)$ of conventional distillation, $h_{\tau,m}$ decays as a power, $(\tau\lambda)^{-m}$, for $\lambda\gg\tau^{-1}$.

The parameters $\tau$ and $m$ control the spatial width of the basis functions. For the free Laplacian, the kernel has a mean-square radius of $\langle r^2\rangle=6m\tau$, corresponding to an r.m.s.\ width of $\sqrt{2m\tau}$ along each spatial direction. We match this width to the coarse-grid spacing,
\begin{equation}
  N=\sqrt{2m\tau},
  \label{eq:width-matching}
\end{equation}
so that neighboring basis functions overlap sufficiently to cover the spatial volume without gaps. While we fix $m$ and $\tau$ using this free-field relation in the present work, their values could be further optimized by scanning over the parameter space to maximize the overlap with the standard distillation subspace spanned by the low-lying Laplacian eigenvectors. Constructing the basis requires $m$ sparse solves per basis vector and time slice. The operator $\mathrm I-\tau\nabla^2(t)$ is Hermitian and positive definite with condition number at most $1+12\tau$, independently of the volume, so conjugate-gradient solves converge in a small, essentially volume-independent number of iterations.

The functions $\xi_I$ are localized but not mutually orthogonal. Collecting them into the rectangular matrix $\Phi_{\mathbf x,I}(t)=\xi_I(\mathbf x,t)$, the Gram matrix and the orthogonal projector onto their span are,
\begin{equation}
  G(t)=\Phi^\dagger(t)\,\Phi(t), \qquad V(t)=\Phi(t)\,G^{-1}(t)\,\Phi^\dagger(t).
  \label{eq:gram-matrix1}
\end{equation}
For linearly independent $\xi_I$, $V(t)$ is Hermitian and idempotent, and it depends only on the span of the basis, not on the choice of basis within it. It is nevertheless convenient to work with an orthonormal basis, which we obtain through symmetric L\"owdin orthogonalization~\cite{Lowdin1950},
\begin{equation}
  \tilde\xi_I(\mathbf x,t)=\sum_J\xi_J(\mathbf x,t)\,\big[G^{-1/2}(t)\big]_{JI},
  \qquad
  V(\mathbf x,\mathbf y;t)=\sum_I\tilde\xi_I(\mathbf x,t)\,\tilde\xi_I^\dagger(\mathbf y,t).
  \label{eq:lowdin}
\end{equation}
Among all orthonormal sets spanning the same space, the L\"owdin basis is the one closest to the original functions in the least-squares sense; it therefore retains their localization, which we verify numerically in Sec.~\ref{sec:wavelet-benchmark}. Sequential Gram--Schmidt orthogonalization would instead single out an ordering of the coarse sites and progressively delocalize the later basis vectors. A flow-based alternative was proposed in Ref.~\cite{mv6l-m8rq}; for our purposes the L\"owdin construction \cite{Lowdin1950, Lowdin1970} is found to be considerably cheaper. The orthonormal functions $\tilde\xi_I$ are used throughout the remainder of this work, in the elementals, the perambulators, and the stochastic sources of Sec.~\ref{sec:wavelet-stochastic}. With Eq.~\eqref{eq:lowdin}, $V(t)$ replaces the eigenvector-built distillation operator $\sum_k\xi_k(t)\xi_k^\dagger(t)$, constructed using  Eq.~\eqref{eq:laplacian-eigenproblem}.

For the smeared field $\tilde q=Vq$ to transform like $q(\mathbf x,t)\to\Omega(\mathbf x,t)q(\mathbf x,t)$, the projector must transform as $V(\mathbf x,\mathbf y;t)\to\Omega(\mathbf x,t)V(\mathbf x,\mathbf y;t)\Omega^\dagger(\mathbf y,t)$. Since $\nabla^2(t)\to\Omega(t)\nabla^2(t)\Omega^\dagger(t)$, any function of it, in particular the resolvent kernel, transforms in the same way. Acting on the fixed color sources, this yields
\begin{equation}
  \Phi(t)\to\Omega(t)\,\Phi(t)\,R^\dagger(t), \qquad R(t)\equiv\bigoplus_n\Omega(\mathbf x_n+\mathbf s,t),
  \label{eq:phi-transform}
\end{equation}
where $R(t)$ is block diagonal in the anchor sites and rotates only the color index of each source. Consequently,
\begin{equation}
  G(t)\to R(t)\,G(t)\,R^\dagger(t), \qquad
  \tilde\Phi(t)\to\Omega(t)\,\tilde\Phi(t)\,R^\dagger(t), \qquad
  V(t)\to\Omega(t)\,V(t)\,\Omega^\dagger(t),
  \label{eq:gaugetransform}
\end{equation}
where $\tilde\Phi=\Phi G^{-1/2}$, and we use $G^{-1/2}\to RG^{-1/2}R^\dagger$. The rotations $R(t)$ cancel in the projector, as required. The L\"owdin basis inherits the block structure of Eq.~\eqref{eq:phi-transform}: under a gauge transformation, each orthonormal basis vector is only rotated in color at its own anchor site. This property allows gauge-invariant measures of the elementals to be defined site by site, and we will describe that method in Sec.~\ref{sec:wavelet-stochastic}.

Note the Laplacian eigenbasis makes no reference to any particular lattice site, whereas the wavelet basis is anchored to a coarse grid. On a given configuration, $V(t)$ is therefore covariant only under translations by $N\mathbb Z^3$. A momentum-projected operator constructed from the smeared fields, $\widetilde{\mathcal O}(\mathbf p,t)=\sum_{\mathbf x}e^{-i\mathbf p\cdot\mathbf x}\,\mathcal O[\tilde q](\mathbf x,t)$, will contain, besides the target momentum $\mathbf p$, other components with the aliased momenta $\mathbf p+\mathbf G_{\mathbf m}$, where,
\begin{equation}
  \mathbf G_{\mathbf m}=\frac{2\pi}{Na}\,\mathbf m=\frac{2\pi}{L}\,\frac{L}{Na}\,\mathbf m,\qquad \mathbf m\in\mathbb Z_N^3\setminus\{\mathbf 0\},
  \label{eq:reciprocal-vectors}
\end{equation}
are the reciprocal vectors of the coarse lattice. In the ensemble average, displacing the grid by $\mathbf s$ is equivalent to translating the operator by $\mathbf s$, so the component $\widetilde{\mathcal O}^{(\mathbf m)}$ with momentum $\mathbf p+\mathbf G_{\mathbf m}$ acquires a phase $e^{i\mathbf G_{\mathbf m}\cdot\mathbf s}$. 
Since components of different total momentum do not mix in the ensemble average,
with independent offsets $\mathbf s_f$ at the sink and $\mathbf s_i$ at the source, the two-point correlation function is given by,
\begin{equation}
  \big\langle\widetilde{\mathcal O}_{\mathbf s_f}(\mathbf p,t)\,\widetilde{\mathcal O}^\dagger_{\mathbf s_i}(\mathbf p,0)\big\rangle
  =\sum_{\mathbf m}e^{i\mathbf G_{\mathbf m}\cdot(\mathbf s_f-\mathbf s_i)}\,C^{(\mathbf m)}(t),
  \qquad
  C^{(\mathbf m)}(t)=\sum_\alpha\big|Z_\alpha^{(\mathbf m)}\big|^2e^{-E_\alpha(\mathbf p+\mathbf G_{\mathbf m})\,t}.
  \label{eq:corr-alias}
\end{equation}
On a fixed grid all phases equal one and every alias contributes. If instead $\mathbf s_i$ and $\mathbf s_f$ are drawn uniformly and independently from $\mathbb Z_N^3$, then $N^{-3}\sum_{\mathbf s}e^{i\mathbf G_{\mathbf m}\cdot\mathbf s}=\delta_{\mathbf m,\mathbf 0}$ yields the estimation:
\begin{equation}
  \mathbb E_{\mathbf s_f,\mathbf s_i}\big[C(t;\mathbf s_f,\mathbf s_i)\big]=C^{(\mathbf 0)}(t),
  \label{eq:shift-average}
\end{equation}
ensuring that only the target momentum survives. The independence of the two offsets is essential.
On individual configurations, aliased contributions remain as
fluctuations that vanish upon averaging over the random shifts,
rather than as a systematic bias, analogous to the random
sublattice shifts discussed in Subsec.~\ref{sec:wavelet:aliasing}.
In practice, we draw one source offset $\mathbf s_i^{(c)}$ and one sink offset $\mathbf s_f^{(c)}$ per configuration $c$, use the latter for all sink times to retain the temporal correlations needed in correlated fits, and get the average,
\begin{equation}
  \bar C(t)=\frac{1}{N_{\rm cfg}}\sum_{c=1}^{N_{\rm cfg}}C^{(c)}\big(t;\mathbf s_f^{(c)},\mathbf s_i^{(c)}\big),
  \qquad
  S_{IJ}(t_f,t_i)=\tilde\xi_I^{\,\mathbf s_f\dagger}(t_f)\,D^{-1}\,\tilde\xi_J^{\,\mathbf s_i}(t_i).
  \label{eq:shifted-estimator}
\end{equation}

All wavelet-distillation results reported here use random coarse-grid
shifts.  Aliasing of the momenta of the individual hadrons inside a multi-hadron operator changes only its overlaps with the states of a given total momentum; the energies can be affected through aliasing of the total momentum. At zero total momentum the aliased sectors carry $|\mathbf P|\geq 2\pi/(Na)$ and lie well above the low-lying levels, so on a fixed grid they would act only as additional higher excited-state contamination, suppressed at large Euclidean time. At nonzero momentum, however, an alias can be lighter than the target state and hence it can make a fitted correlator dominated by a lighter state from another sector. We illustrate this in  Subsec.~\ref{sec:wavelet-highmom} by comparison with a fixed-grid calculation.

\subsection{Numerical setup and smearing profile}
\label{sec:wavelet-benchmark}

All calculations in this section use an $N_f=3$ CLS ensemble~\cite{CLS} with $L^3\times T=24^3\times48$, $a\approx0.098$~fm and $m_\pi\approx429$~MeV, with Wilson-clover valence quarks at the sea quark mass. We analyze 100 gauge configurations. The scaling functions of spectral wavelet basis as described in Subsec.~\ref{sec:wavelet-construction} are constructed on a coarse grid with $N=8$, i.e.\ $3^3=27$ anchor sites, giving $n_v=3\times27=81$. Eq.~\eqref{eq:width-matching} then requires $m\tau=32$, and we use $m=8$, $\tau=4$. The Laplacian is constructed from gauge links smeared with 15 steps of APE smearing~\cite{APESMEARING} with $\alpha=2.5$, and the resolvent solves use even-odd preconditioned conjugate gradient with a relative residual of $10^{-10}$.
At each time slice, wavelet distillation requires $m n_v$
sparse linear solves with $\mathrm{I}-\tau\nabla^2(t)$,
each taking approximately $I_{\rm cg}$ conjugate-gradient
(CG) iterations~\cite{HestenesStiefel1952,Saad2003}.
The resulting basis is L\"owdin-orthogonalized, with costs
scaling as $\mathcal{O}(n_v^2V_3)$ for Gram-matrix construction
and basis rotation, and $\mathcal{O}(n_v^3)$ for computing
the inverse square root of the Gram matrix.  On each gauge
configuration, we independently sample source and sink
coarse-grid offsets and form the estimator of
Eq.~\eqref{eq:shifted-estimator}, which is then averaged
over configurations.

For the conventional distillation, we compute the lowest $n_v$
eigenmodes of the same Laplacian, either using PRIMME in
\texttt{PRIMME\_DYNAMIC} mode~\cite{PRIMME,StathopoulosMcCombs2007}
or with Chebyshev-filtered subspace
iteration~\cite{10.1016/j.jcp.2006.03.017, doi:10.1137/050630404}, to
convergence tolerance $\epsilon_{\rm eig}=10^{-8}$. The wavelet
filter is applied either through nested even--odd-preconditioned
CG solves or as a single Chebyshev expansion with truncation error
below $10^{-13}$~\cite[Ch.~8]{Trefethen2019}. Dirac inversions are
performed using the CG solver in the MILC codebase~\cite{MILC},
with tolerance $\epsilon_D=10^{-12}$. Both constructions use
distillation spaces of equal dimension. On the lattice studied
here, constructing the wavelet basis is approximately five times
faster than constructing the Laplacian eigenbasis, both with
PRIMME and CG, and with the Chebyshev methods.
Appendix~\ref{sec:wavelet-cost} provides the solver details,
parameters, and timings.

\begin{figure}[htbp]
  \centering
  \includegraphics[width=0.7\textwidth]{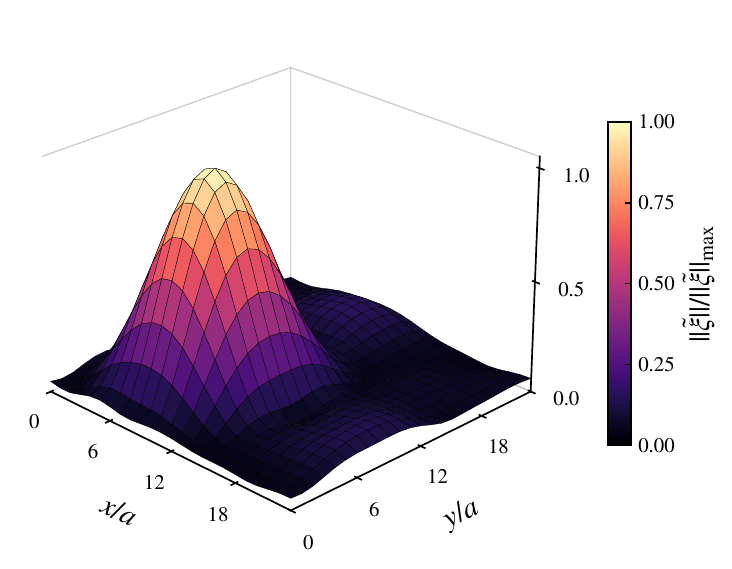}
  \caption{Spatial profile of an orthonormalized wavelet basis function anchored at the coarse-lattice site $\mathbf x_n=(8,8,0)$, in the $z=0$ plane. Here we illustrate the gauge-invariant magnitude of the orthonormal wavelet basis function $||\tilde{\xi}(\mathbf x)||=\big[\sum_a\tilde\xi^\dagger_{n,a}(\mathbf x)\,\tilde\xi_{n,a}(\mathbf x)\big]^{1/2}$, summed over the $N_c=3$ color partners and normalized to its maximum.}
  \label{fig:wavelet-basis-construction}
\end{figure}

We next examine the spatial profile of the wavelet basis functions
constructed above.
Figure~\ref{fig:wavelet-basis-construction} shows an orthonormalized basis function anchored at $\mathbf x_n=(8,8,0)$, through the gauge-invariant magnitude $||\tilde{\xi}(\mathbf x)||=[\sum_a\tilde\xi^\dagger_{n,a}(\mathbf x)\,\tilde\xi_{n,a}(\mathbf x)]^{1/2}$ in the $z=0$ plane. As can be observed, the function is peaked at its anchor site and decays smoothly with distance; L\"owdin orthogonalization does not noticeably delocalize it. This ensures the locality of the wavelet bases in our construction.

\begin{figure}[htbp]
   \centering
   \includegraphics[width=0.55\textwidth]{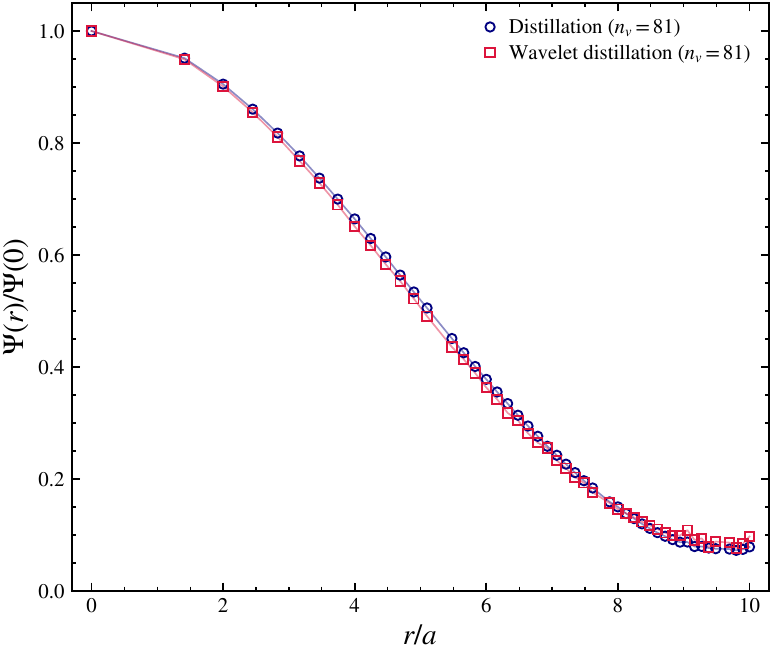}
   \caption{Smearing profile $\Psi(r)$, as defined in Eq.~\eqref{eq:smearing-profile}, for conventional distillation with the lowest 81 Laplacian eigenvectors and for wavelet distillation with $n_v=81$.}
  \label{fig:smearing-profile}
\end{figure}

Next, we compare the smearing profiles of conventional and wavelet
distillation using their full projectors. Following
Ref.~\cite{hadronspectrum2009}, we define
\begin{equation}
  \Psi(r) = \frac{1}{N(r)} \sum_{\mathbf x,t}\;\sum_{|\mathbf r|=r}\sqrt{\mathrm{Tr}\big[V_{\mathbf x,\mathbf x+\mathbf r}(t)\,V_{\mathbf x+\mathbf r,\mathbf x}(t)\big]},
  \label{eq:smearing-profile}
\end{equation}
where, $N(r) = \sum_{|\mathbf r|=r} 1$, is the number of lattice displacements of length $r$. Here, the trace is taken over the color indices. For wavelet distillation the average over $\mathbf x$ also averages over the position relative to the coarse grid. We show the normalized profiles in Fig.~\ref{fig:smearing-profile},  obtained with the same number of basis functions ($n_v = 81$). The two profiles agree quite closely, indicating that the wavelet
projector produces a smearing range comparable to that of
conventional distillation at the same $n_v$, despite the
power-law decay of the resolvent filter in the Laplacian
spectrum, in contrast to the sharp spectral cutoff of
conventional distillation. With our given choice of parameters, the normalized subspace overlap with the standard distillation space spanned by the lowest $n_v=81$ Laplacian eigenvectors $\{\chi_\ell\}$,
\begin{equation}
\frac{1}{n_v}\sum_{k,\ell=1}^{n_v} \big|\langle \tilde{\xi}_k \mid \chi_\ell \rangle\big|^2 \approx 0.88,\quad n_v = 81,
  \label{eq:subspace-overlap}
\end{equation}
where $\{\tilde{\xi}_k\}$ denotes the orthonormalized wavelet basis, is about $88\%$.

The principal gain at this stage is the reduced cost of basis
construction. As detailed in Appendix~\ref{sec:wavelet-cost}, the wavelet basis
is approximately five times cheaper to construct than the
Laplacian eigenbasis on this lattice, at comparable numerical
precision and with closely matched smearing profiles.
The construction requires no Laplacian eigensolve.
Together with locality-based contraction acceleration, this
offers the prospect of greater computational savings on
larger volumes and in GEVP analyses with large operator bases,
subject to maintaining comparable statistical precision.

\subsection{Finite-volume spectra at zero momentum}
\label{sec:wavelet-fv-spectra}

A matching smearing profile is necessary but not sufficient: what matters is whether the wavelet space provides the same overlap with the low-lying states, and their energy spectra. We test this by comparing finite-volume energies obtained with the two distillation spaces for a single baryon, a two-meson system analyzed using a GEVP ~\cite{MICHAEL198558,luscherwolff1990} setup, and the two-baryon deuteron channel with both dibaryon and local hexaquark operators.

\subsubsection{Operators and contractions}
For the nucleon, we use the interpolating operators defined in Eq.~\eqref{eq:protonOps}. For the two-nucleon system in the deuteron channel, we construct the following dibaryon type operator, 
\begin{eqnarray}
  \tilde{\mathcal N}_I^{\alpha}(\mathbf p,t)&=&\sum_{\mathbf x}e^{-i\mathbf p\cdot\mathbf x}\,\mathcal N_I^{\alpha}(\mathbf x, t)\\
  \mathcal O_D^{j}(\mathbf P=\mathbf 0,t)&=&\epsilon_{IJ}\,\tilde{\mathcal N}_I^{\alpha}(\mathbf 0,t)\,(C\gamma_jP_+)_{\alpha\beta}\,\tilde{\mathcal N}_J^{\beta}(\mathbf 0,t),\label{eq:DeuteronOps}
\end{eqnarray}
where, individual nucleons are projected to zero momenta. Here $I,J\in\{0,1\}$ label proton and neutron, $j$ is the polarization of the deuteron, $\epsilon_{IJ}$ is the antisymmetric tensor, and $P_+=\tfrac12(1+\gamma_4)$. Within distillation, the correlators factorize into the nucleon elemental \cite{hadronspectrum2009}
\begin{equation}
  (\phi^\omega)^{\alpha\beta\eta}_{IJK}(t)\equiv\sum_{\mathbf x}\delta_{\omega\alpha}\,\epsilon_{abc}\,(C\gamma_5)_{\beta\eta}\,\tilde\xi_I^a(\mathbf x,t)\,\tilde\xi_J^b(\mathbf x,t)\,\tilde\xi_K^c(\mathbf x,t) ,
  \label{elemental_def}
\end{equation}
and the perambulators $(S_{u,d})_{IJ}(t_f,t_i)=\tilde\xi_I^\dagger(t_f)\,D_{u,d}^{-1}\,\tilde\xi_J(t_i)$, where $\tilde\xi_I$ denotes either the Laplacian eigenvectors or the orthonormal wavelet functions of Eq.~\eqref{eq:lowdin}.  The positive-parity nucleon correlator is given by,
\begin{align}
  C_N(t_f,t_i)
  &=(P_+)_{\omega\omega'}\,(\phi^\omega)^{\alpha\beta\eta}_{IJK}(t_f)\,\big[(\phi^{\omega'})^{\alpha'\beta'\eta'}_{I'J'K'}(t_i)\big]^*\,(S_d)^{\eta\eta'}_{KK'}(t_f,t_i)\nonumber\\
  &\quad\times\Big[(S_u)^{\alpha\alpha'}_{II'}(t_f,t_i)\,(S_u)^{\beta\beta'}_{JJ'}(t_f,t_i)-(S_u)^{\alpha\beta'}_{IJ'}(t_f,t_i)\,(S_u)^{\beta\alpha'}_{JI'}(t_f,t_i)\Big],
  \label{eq:DistillationNCorr}
\end{align}
and the deuteron correlator,
\begin{align}
  C_D(t_f,t_i)
  &=\Gamma^j_{l_pl_n}\,(\phi^{l_p})^{\alpha_1\alpha_2\beta_1}_{I_1I_2J_1}(t_f)\,(\phi^{l_n})^{\beta_2\beta_3\alpha_3}_{J_2J_3I_3}(t_f)\nonumber\\
  &\quad\times \sum_{\omega,\sigma\in S_3}\mathrm{sgn}(\omega)\,\mathrm{sgn}(\sigma)\prod_{n=1}^{3}(S_u)^{\alpha_n\alpha'_{\omega(n)}}_{I_nI'_{\omega(n)}}(t_f,t_i)\,(S_d)^{\beta_n\beta'_{\sigma(n)}}_{J_nJ'_{\sigma(n)}}(t_f,t_i)\nonumber\\
  &\quad\times\,\Big[\Gamma^j_{l'_pl'_n}\,(\phi^{l'_p})^{\alpha'_1\alpha'_2\beta'_1}_{I'_1I'_2J'_1}(t_i)\,(\phi^{l'_n})^{\beta'_2\beta'_3\alpha'_3}_{J'_2J'_3I'_3}(t_i)\Big]^*,
  \label{eq:DeuteronContraction}
\end{align}
where $\Gamma^j_{l_pl_n}$ projects the two nucleon spins onto the $J^P=1^+$ component $j$, the up-quark (down-quark) indices are $\{I,\alpha\}$ ($\{J,\beta\}$), and the sums over the permutations $\omega,\sigma$ of the three identical up and down quark lines generate all direct and exchange contractions. In the isospin limit $S_u=S_d$. Evaluated naively, Eq.~\eqref{eq:DeuteronContraction} costs $\mathcal O(n_v^{12})$, whereas contracting the perambulators with the elementals one at a time reduces this to a sequence of steps of cost at most $\mathcal O(n_v^4)$.

For the $I=2$ two-pion system we use the five bilocal operators,
\begin{equation}
  (\pi\pi)_{I=2}(\mathbf P=\mathbf 0)=\pi^+(\mathbf p)\,\pi^+(-\mathbf p),\qquad |\mathbf p|^2\le4,
  \label{eq:pipi-operator1}
\end{equation}
with $\mathbf p$ in units of $2\pi/L$, and extract the spectrum from the principal correlators of the resulting $5\times5$ correlation matrix.
\subsubsection{Results on finite volume spectra}
\begin{figure}[htbp]
  \centering
  \includegraphics[width=0.8\linewidth]{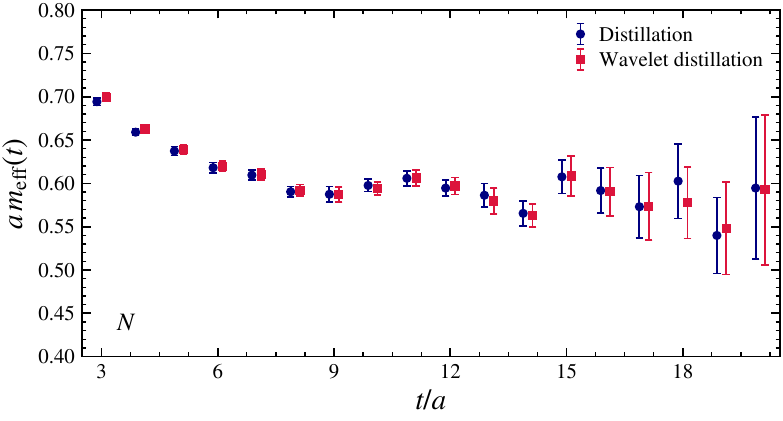}
  \caption{Effective mass corresponding to the  ground state of nucleon obtained from conventional distillation (blue) and wavelet distillation (red). In both cases we use  $n_v=81$ basis functions to construct the respective distillation spaces.}
  \label{fig:fv-spectra_nu}
\end{figure}

\paragraph{Single nucleon:}
In Fig.~\ref{fig:fv-spectra_nu} we present the usual effective mass ($m_{\mathrm{eff}} = \mathrm{log[C(t)/C(t+1)]}$) in lattice units obtained from
correlators ($C(t)$) computed on the ensemble described above using
the preceding constructions. The wavelet results are shown by the red-square marker whereas blue circles represent results obtained with regular distillation. We use the same number of basis vectors ($n_v = 81$).
In both cases the results are indistinguishable within statistical uncertainties at large Euclidean times, and the uncertainties are of similar size. 

\begin{figure}[htbp]
  \centering  
  \includegraphics[width=0.8\linewidth]{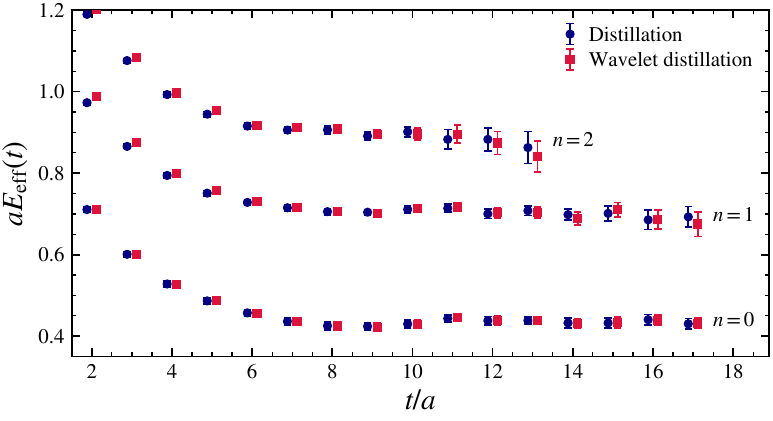} 
  \caption{Comparison of effective energies of the three lowest-lying finite-volume states
in the $I=2$ two-pion channel, extracted from a GEVP constructed using the
five operators defined in Eq.~\eqref{eq:pipi-operator1}, for conventional and wavelet distillations.}
  \label{fig:fv-spectra_pipi}
\end{figure}

\paragraph{Two pions:} Fig.~\ref{fig:fv-spectra_pipi} shows the corresponding results
for the two-pion system, obtained from a GEVP using the five
operators defined in Eq.~\eqref{eq:pipi-operator1}.
The two distillation methods agree very well within statistical
uncertainties for both the ground state and the extracted
excited levels. The latter provide a more demanding test of
the operator overlaps supported by the distillation space.
This observation clearly shows that the localized wavelet basis is as effective as the Laplacian eigenvectors in removing the ultraviolet modes while retaining the low-energy content of the quark fields.

\begin{figure}[htbp]
  \centering
  \includegraphics[width=0.8\linewidth]{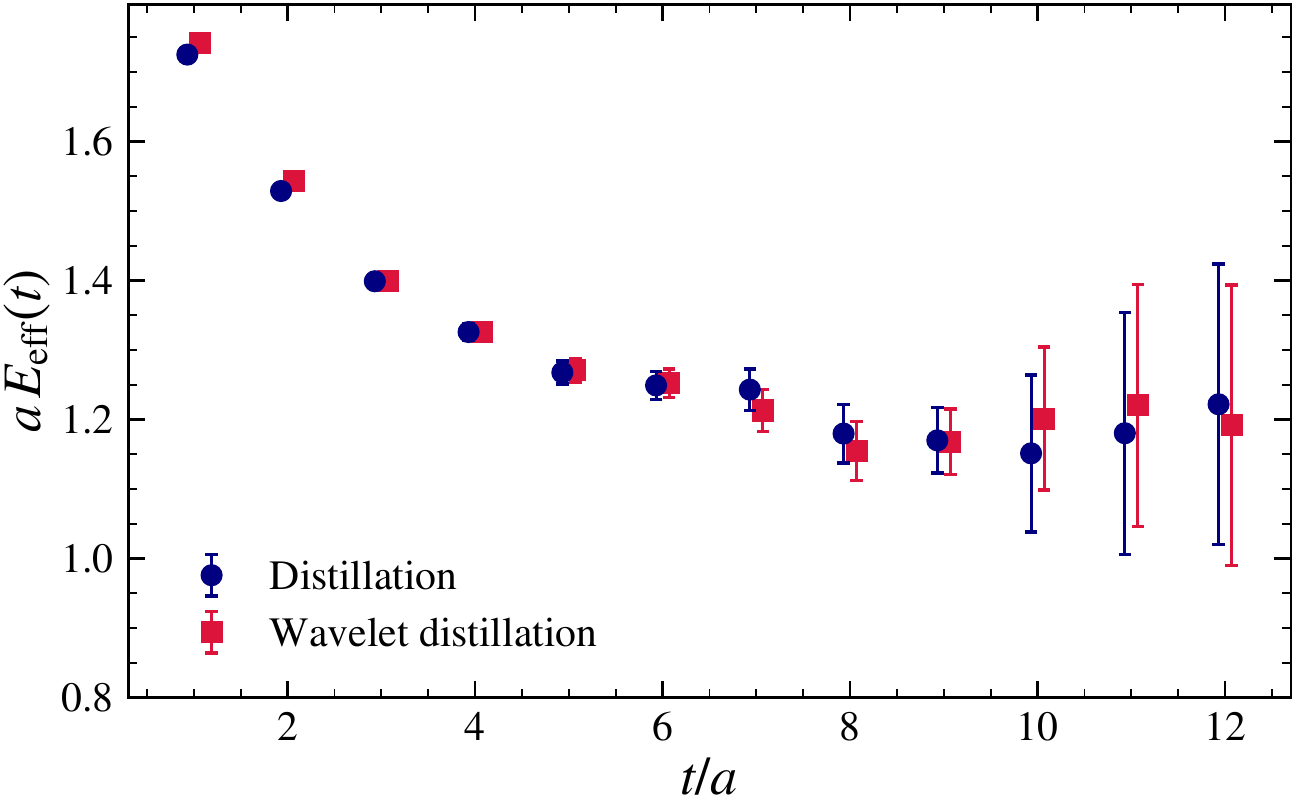}
  \caption{Effective mass of the ground state of Deuteron from conventional (navy) and wavelet (crimson) distillations with the dibaryon operator of Eq.~\eqref{eq:DeuteronOps}.}
  \label{fig:wavelet_NP_bilocal}
\end{figure}

\paragraph{Deuteron with dibaryon operators:}
The preceding examples demonstrate the efficacy of wavelet
distillation relative to conventional distillation in comparatively
simple hadronic channels. We next test its performance in the more
demanding deuteron channel, extending the comparison to a two-baryon system of considerable interest in contemporary lattice QCD. 
 Figure~\ref{fig:wavelet_NP_bilocal} shows the ground-state effective energy in the deuteron channel corresponding to the dibaryon operator of Eq.~\eqref{eq:DeuteronOps}. 
The two methods again agree very well within
statistical uncertainties throughout the Euclidean-time range
shown, supporting the ability of the wavelet space to retain
the low-energy two-baryon information resolved by conventional
distillation.

\begin{figure}[htbp]
  \centering
  \includegraphics[width=0.8\linewidth]{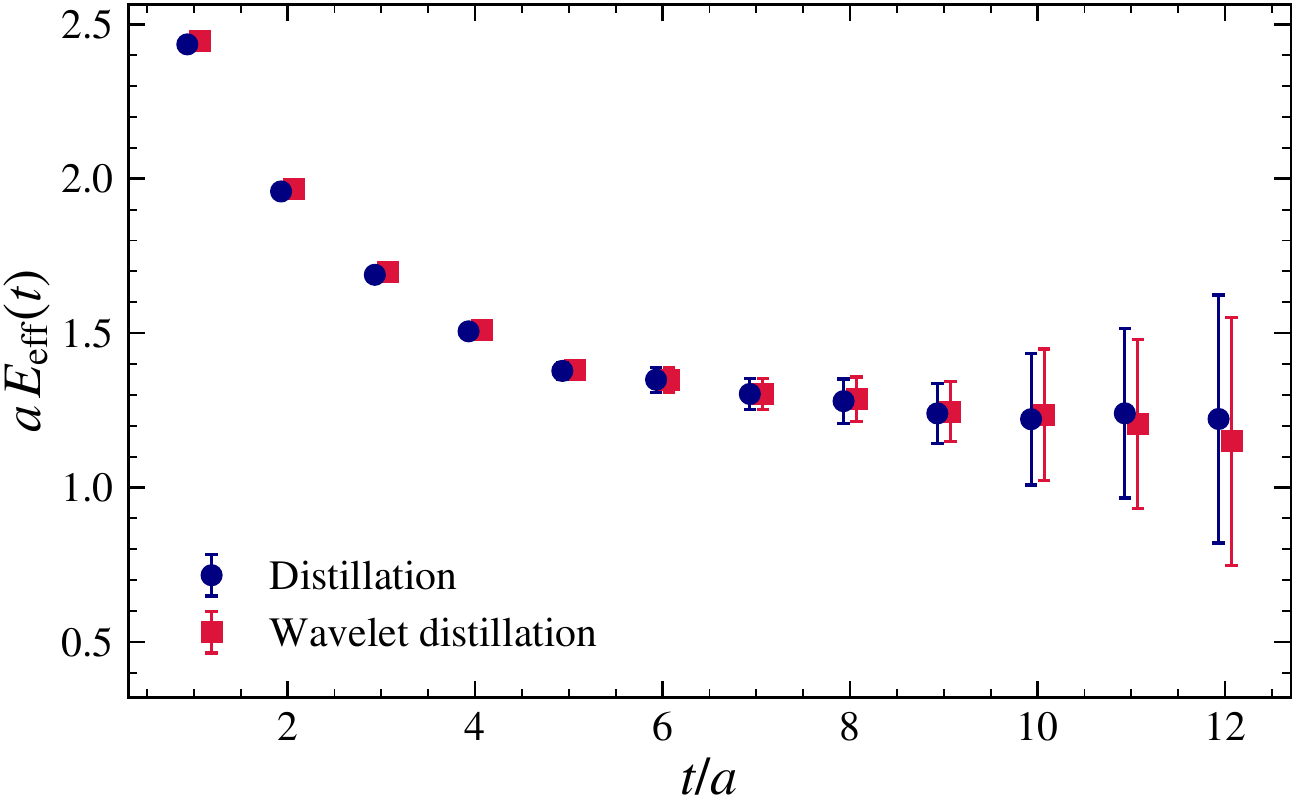}
  \caption{Effective mass of the ground state of Deuteron from conventional (navy) and wavelet (crimson) distillations  with the hexaquark operator of Eq.~\eqref{eq:NPHexaQuark}. Here we use a sink which is coarse-grained by two levels of the gauge-covariant fast wavelet transform (see text for explanation).}
  \label{fig:wavelet_NP_Hexaquark}
\end{figure}

\paragraph{Deuteron with hexaquark operators:}
Alongside dibaryon interpolating fields, local multi-hadron operators are also commonly included in GEVP bases. In general computation of correlation functions for such operators are relatively more computationally expensive in GEVP setup. It is thus also interesting to compute those in wavelet distillation. We consider the following hexaquark-type operator,
\begin{equation}
  \mathcal O_H^{j}(\mathbf x,t)=\epsilon_{IJ}\,\mathcal N_I^{\alpha}(\mathbf x,t)\,(C\gamma_jP_+)_{\alpha\beta}\,\mathcal N_J^{\beta}(\mathbf x,t).
  \label{eq:NPHexaQuark}
\end{equation}
with $\epsilon_{IJ}$ as the antisymmetric tensor. Since all six quarks share a single site, its elemental does not factorize into two nucleon elementals but forms a rank-6 tensor. Consequently the corresponding contraction scales as $\mathcal O(n_v^7)$. Following Ref.~\cite{khhk-qw9n}, we avoid this cost by evaluating the sink contraction in position space. Rather than the spatial decimation or stochastic site sampling used there~\cite{Detmold_2021,PhysRevD.103.014514}, we reconstruct the smeared quark propagators in position space from the perambulators and coarse-grain the sink with two levels of the gauge-covariant fast wavelet transform, using Eq.~\eqref{eq:GaugeCovariantFWT1} of Sec.~\ref{sec:wavelet:Sparsening}. That reduces the number of sink sites by a factor of 64. The same procedure is applied in both distillation spaces. Figure~\ref{fig:wavelet_NP_Hexaquark} shows the resulting effective energies for both the setups. Here again results from both the distillation setups agree within statistical uncertainties. This also demonstrates that the two wavelet constructions can be combined: wavelet distillation for the quark smearing and the fast wavelet transform for sparsening the contraction.  This approach can also be applied to local tetraquark operators,
offering the potential for substantial reductions in contraction costs.

\subsection{Nucleons at nonzero momentum}
\label{sec:wavelet-highmom}

In Lattice QCD calculations, computation of finite-momentum hadron correlators are essential for probing internal structures of hadrons, for example, for determining
form factors, scattering amplitudes, and parton distributions.  At large momenta, however, deteriorating
signal-to-noise ratios and reduced energy gaps make ground-state
isolation increasingly difficult, while discretization effects
become more pronounced. Although momentum
smearing improves overlaps with boosted states~\cite{PhysRevD.93.094515},
precise calculations remain computationally demanding,
motivating more efficient operator construction and contraction
methods. It is thus important to assess the efficacy of wavelet distillation in computing boosted correlators.

Momentum-dependent distillation bases have been developed to
improve the statistical precision of boosted-hadron
correlators~\cite{PhysRevD.103.034502}. We incorporate the same
idea into wavelet distillation by modifying the gauge links
entering the spatial Laplacian according to the
momentum-smearing prescription of
Ref.~\cite{PhysRevD.93.094515},
\begin{equation}
  \tilde U(\mathbf k)_{\mathbf x,\mu}=U_{\mathbf x,\mu}\,e^{ia\mathbf k\cdot\hat\mu},\qquad \mathbf k=\zeta\,\mathbf p,
  \label{eq:momentumphases}
\end{equation}
where $\mathbf k$ is the momentum carried by each quark. The naive value for a baryon is $\zeta=1/3$; following Ref.~\cite{PhysRevD.93.094515} we use $\zeta=0.4$. With the resulting Laplacian $\nabla^2(\mathbf k,t)$, the momentum-dressed basis is
\begin{equation}
  \xi^{\mathbf k;\mathbf s}_{n,a}(\mathbf x,t)=\big(\mathrm I-\tau\,\nabla^2(\mathbf k,t)\big)^{-m}\,\delta_{\mathbf x,\,\mathbf x_n+\mathbf s}\;e_a,
  \label{eq:wavelet-basis-momentum}
\end{equation}
followed by the same orthonormalization as in Subsec. \ref{sec:wavelet-construction}. For conventional distillation we compute the eigenvectors of the same dressed Laplacian, as in Refs.~\cite{PhysRevD.93.094515,PhysRevD.103.034502}. 
Although conventional and wavelet distillation use the same
contraction strategy, wavelet distillation reduces the cost
of constructing momentum-dependent bases by replacing a
separate Laplacian eigensolve at each momentum with sparse
linear solves. The benchmark in
Subsec.~\ref{sec:wavelet-benchmark} shows an approximately
fivefold reduction in basis-construction time on the
$24^3\times48$ lattice. Even greater savings are expected 
on larger volumes.

\begin{figure}[htbp]
  \centering
  \includegraphics[width=0.9\linewidth]{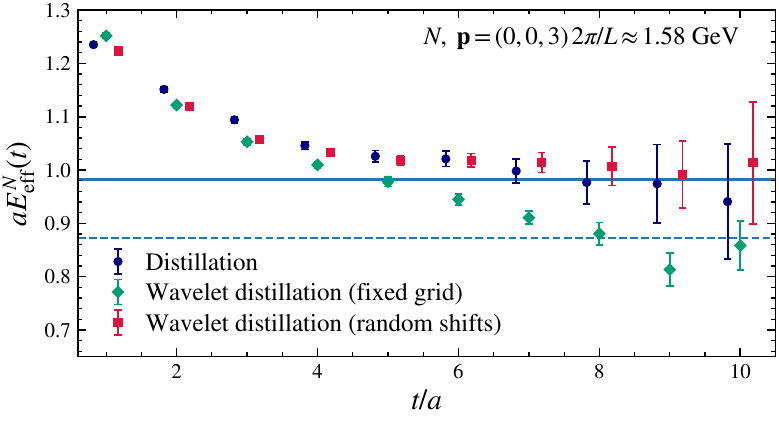}
  \caption{Nucleon effective energy at $\mathbf p=(0,0,3)\,2\pi/L\simeq1.58$~GeV, obtained using conventional distillation (blue), wavelet distillation on a fixed coarse grid (green), and wavelet distillation with randomly shifted coarse grids (red). For the fixed coarse grid, the nucleon operator also couples to an aliased zero-momentum state, leading to an apparently lower energy with reduced statistical uncertainty. In contrast, randomizing the coarse-grid shifts removes this aliasing effect, and the resulting effective energy is consistent with that obtained from conventional distillation. The horizontal lines the continuum (solid) and free-Wilson (dashed) dispersion relations obtained using  Eqs.~\eqref{eq:continuum-dispersion} and \eqref{eq:lattice-dispersion}, respectively.}
  \label{fig:nucleon-highmom-shift}
\end{figure}

As discussed in Subsec.~\ref{sec:wavelet-construction}, at nonzero momentum the fixed coarse grid can couple the operator to lighter aliased momenta. On our lattice, $L/N=3$, so the momenta $\mathbf n$ and $\mathbf n+3\mathbf m$ (in units of $2\pi/L$) are aliases of each other. In Fig.~\ref{fig:nucleon-highmom-shift} we illustrate effective mass of boosted nucleon correlators at $\mathbf n=(0,0,3)$, $|\mathbf p|\simeq1.58$~GeV, for regular distillation, wavelet distillation and wavelet distillation with random shifts as discussed above. Here the alias is the nucleon at rest. As can be observed, on the fixed grid the effective energy lies significantly below that from conventional distillation and has smaller uncertainties, as expected from an admixture of the lighter and less noisy zero-momentum nucleon. With random grid shifts, Eq.~\eqref{eq:shifted-estimator}, as used for all other results in this work, the wavelet result agrees with conventional distillation, with error bars of comparable size. The fixed-grid calculation is shown only to illustrate the effect of aliasing.
 
\begin{figure}[htbp]
  \centering
  \includegraphics[width=0.9\linewidth]{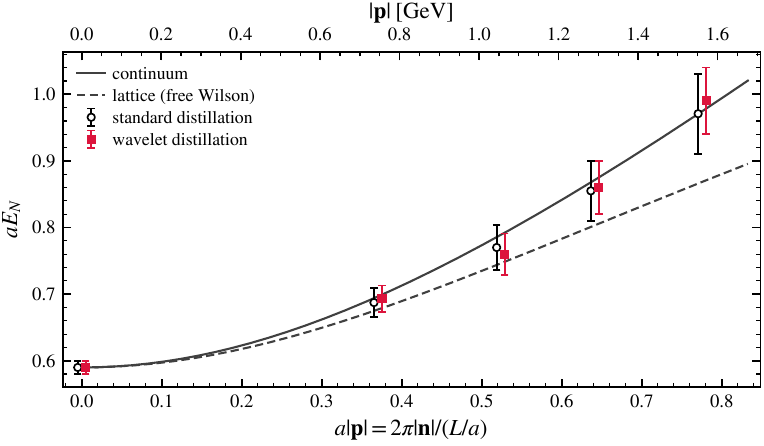}
  \caption{Nucleon energies at lattice momenta $\mathbf p=2\pi\mathbf n/L$, obtained using conventional distillation (black) and wavelet distillation with randomly shifted coarse grids (red), compared with the continuum (solid) and free-Wilson (dashed) dispersion relations given in Eqs.~\eqref{eq:continuum-dispersion} and \eqref{eq:lattice-dispersion}.}
  \label{fig:nucleon-dispersion}
\end{figure}

To further assess discretization effects, we compute boosted-nucleon
correlators at several momenta up to $|\mathbf p|\simeq1.6$~GeV. For each
setup and momentum, we extract the lowest energy from a
single-exponential fit at large Euclidean times. The fitted
energies are listed in Table~\ref{tab:nucleon_dispersion} and
plotted in Fig.~\ref{fig:nucleon-dispersion}, together with
the continuum dispersion relation,
\begin{equation}
  aE_N=a\sqrt{m_N^2+\mathbf p^2},
  \label{eq:continuum-dispersion}
\end{equation}
and the dispersion relation of a free Wilson fermion~\cite{PhysRevD.93.094515},
\begin{equation}
  \cosh(aE_N)=1+\frac{\big(e^{am_N}-1+a^2\hat{\mathbf p}^2/2\big)^2+a^2\bar{\mathbf p}^2}{2\big(e^{am_N}+a^2\hat{\mathbf p}^2/2\big)},
  \qquad
  \hat p_j=\frac2a\sin\frac{ap_j}{2},\quad \bar p_j=\frac1a\sin(ap_j).
  \label{eq:lattice-dispersion}
\end{equation}
The latter describes a point-like fermion and is not expected to reproduce the interacting nucleon; the difference between the two curves serves only as a rough indication of the size of lattice artifacts at a given momentum. The energies from wavelet distillation with random shifts and from conventional distillation agree within statistical uncertainties at all momenta, with comparable errors, and both follow the continuum dispersion relation. Once translation invariance is restored in the ensemble average, we thus find no systematic difference between the two methods at nonzero momentum, while the wavelet basis requires only sparse solves for each momentum instead of a new eigensolve and remains compatible with the localized contraction estimators of Sec.~\ref{sec:wavelet-stochastic}.

\begin{table}[htbp]
  \caption{Nucleon energies in lattice units at momenta $\mathbf p=2\pi\mathbf n/L$, with $L/a=24$ and $a=0.098$~fm, from conventional and wavelet distillation (with random coarse-grid shifts), together with the continuum and free-Wilson dispersion relations evaluated at $am_N=0.59$, corresponding to  $\mathbf p = 0$.}
  \label{tab:nucleon_dispersion}
  \centering
  \vspace*{0.2in}
  \setlength{\tabcolsep}{4pt}
  \begin{tabular}{cccccccc}
    \toprule
    & & & \multicolumn{2}{c}{Dispersion relation} & \multicolumn{2}{c}{Distillation} & \\
    \cmidrule(lr){4-5}\cmidrule(lr){6-7}
    $\mathbf n$ & $a|\mathbf p|$ & $|\mathbf p|$ [GeV] & Continuum & Lattice & Conventional & Wavelet & Fit range\\
    \midrule
    $(0,0,0)$ & 0     & 0     & --    & --    & 0.589(12) & 0.588(12) & [9--20]\\
    $(0,1,1)$ & 0.370 & 0.746 & 0.697 & 0.678 & 0.687(22) & 0.693(20) & [8--18]\\
    $(0,0,2)$ & 0.524 & 1.054 & 0.789 & 0.746 & 0.767(34) & 0.761(28) & [8--15]\\
    $(1,1,2)$ & 0.641 & 1.291 & 0.871 & 0.812 & 0.855(45) & 0.861(39) & [7--13]\\
    $(0,0,3)$ & 0.785 & 1.581 & 0.982 & 0.873 & 0.971(60) & 0.989(53) & [6--12]\\
    \bottomrule
  \end{tabular}
\end{table}

\section{Stochastic estimators}\label{sec:wavelet-stochastic}

By exploiting a low dimensional subspace, the distillation method \cite{hadronspectrum2009} enables the construction of large correlation matrices for
subsequent GEVP analyses, facilitating the extraction of tens of
finite-volume energy levels. However, at a fixed physical smearing scale, the
dimension $n_v$ of the distillation space grows approximately linearly with
the spatial volume of the lattice, and two costs grow with it. Construction of exact
perambulators requires $4n_v$ Dirac inversions per quark flavor and source
time, and the Wick contractions of hadronic correlation functions scale
polynomially with the dimension of the distillation space.

The method of stochastic LapH~\cite{PhysRevD.83.114505} addresses the first bottleneck by
estimating quark propagation with diluted noise vectors within the
distillation subspace. The number of Dirac inversions is then set by the
number of noise vectors and dilution projectors rather than directly by $n_v$,
and the Wick contractions at the source end are organized in the typically
smaller dilution space. The method exploits the reduced dimensionality of the
distillation subspace to enable stochastic estimation, with the variance
controlled through suitable dilution.

More recently, it has been shown that a distillation-space basis, localized
around the sites of a coarse grid, could make hadron elementals approximately
sparse~\cite{mv6l-m8rq}. A Hansen-Hurwitz estimator~\cite{HansenHurwitz}
with sampling probabilities built from the elementals can then evaluate the
contractions using a small fraction of their terms. In that formulation, quark
lines are represented by exact perambulators, and the indices of both source
and sink elementals are sampled. The sampled contributions are connected
through the $n_v\times n_v$ perambulators via a sequence of intermediate
tensors. However, as the authors noted in Ref.~\cite{mv6l-m8rq}, the central
intermediate depends on index sets obtained by marginalizing over one or two
elemental indices. This marginalization substantially reduces the sparsity
gained from localization, and the retained index sets become nearly fully
populated even at moderate sample sizes. 
Consequently, constructing this intermediate becomes nearly as
costly as the exact contraction, limiting the computational
savings from HH sampling in this implementation.

In this section, we combine the stochastic LapH and
contraction-sampling strategies of
Refs.~\cite{PhysRevD.83.114505,mv6l-m8rq} within wavelet
distillation. Following stochastic LapH, we estimate quark
propagation using diluted source-noise vectors in the localized
wavelet basis, discussed in Sec.~\ref{sec:wavelet-construction}.
We refer to this formulation as \emph{Stochastic Wavelets}.
Subsection~\ref{sec:stoch:source} describes its construction
and compares nucleon and deuteron correlators with those
obtained using stochastic LapH. In subsection~\ref{sec:stoch:hh} we introduce HH
sampling of the sink contractions, which are ordinarily
evaluated exactly in distillation calculations. We explain
how combining stochastic source estimation with sink-only
sampling improves the contraction-cost scaling relative
to sampling source and sink elementals connected by full
perambulators. Although developed here for the wavelet
basis, this strategy is also applicable in principle to
other localized distillation bases, including that of
Ref.~\cite{mv6l-m8rq}. In subsection~\ref{sec:stoch:results} we present results for the stochastic distillation for nucleon and deuteron
correlators and compare them with those obtained without HH sink sampling,
quantifying the computational costs and additional sampling
variance. Finally, in Subsec.~\ref{sec:stoch:scaling} we evaluate the operation
counts and perform a statistical analysis  to assess how the contraction
savings may increase with the dimension of the distillation
subspace.

Throughout this section, the wavelet basis carries the composite index
$I=N_c n+a$, where $n=0,\ldots,n_{\rm site}-1$ labels the coarse-grid anchor
site and $a=0,\ldots,N_c-1$ labels color. We write $(na)\equiv N_c n+a$ when
the coarse-site structure is relevant. For the ensemble studied here,
$n_{\rm site}=27$ and $n_v=N_c n_{\rm site}=81$, while $N_s=4$ denotes the
number of spin components.

\subsection{Stochastic wavelets}\label{sec:stoch:source}

Following Ref.~\cite{PhysRevD.83.114505}, we introduce noise vectors in the space spanned by the wavelet
basis vectors of Sec.~\ref{sec:wavelet-construction} and solve the Dirac equation with these noise
vectors as sources. Dilution reduces the variance introduced by the noise. The noise vectors $\boldsymbol{\zeta}\in\mathbb{C}^{n_v}$ are independent, with components drawn uniformly from
$\{\pm 1, \pm i\}$, so that
\begin{equation}
    \langle \zeta_i \rangle_{\zeta}=0, \qquad
    \langle \zeta_i\zeta_j^* \rangle_{\zeta}=\delta_{ij}.
    \label{eq:z4-noise-properties}
\end{equation}
We use full dilution in time and spin, so that every diluted source has support on a single time slice
and spin component. Within the $n_v$-dimensional wavelet space we use interlace dilution with $n_d$
projectors~\cite{PhysRevD.83.114505},
\begin{equation}
    \big(\mathcal{P}^{b}\big)_{ij} = \delta_{ij}\,\delta_{b,\,i\bmod n_d},
    \qquad b = 0,\ldots,n_d-1 ,
    \label{eq:interlace-dilution1}
\end{equation}
and diluted noise vectors $\boldsymbol{\zeta}^{\,b} = \mathcal{P}^{b}\boldsymbol{\zeta}$.

\begin{figure*}[tbp]
  \centering
  \includegraphics[width=0.95\textwidth]{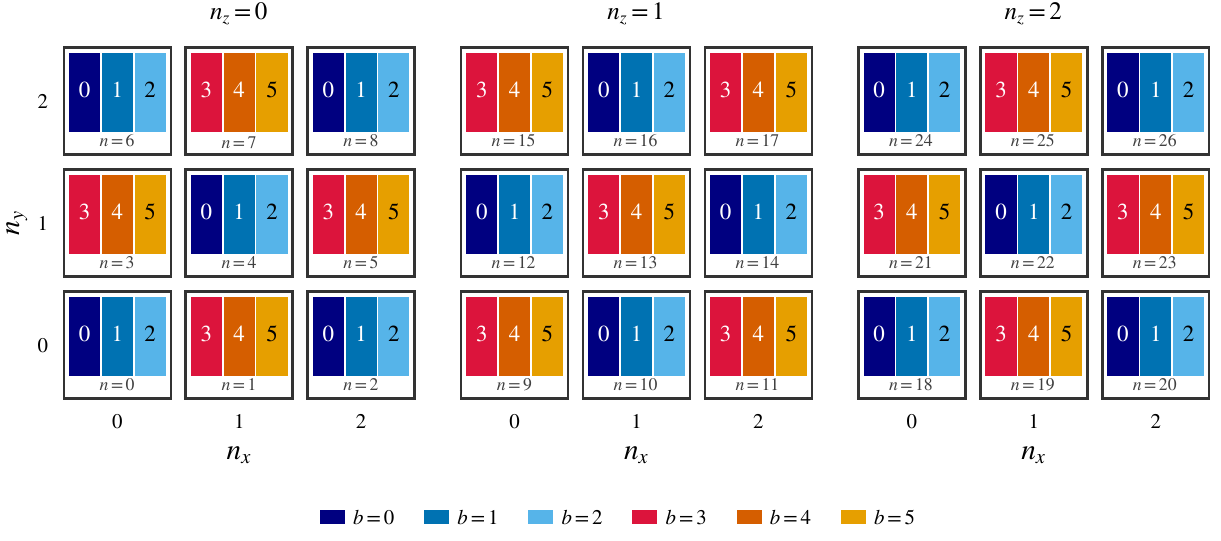} 
  \caption{Illustration of interlace dilution with $n_d = 6$ for the wavelet basis on the $3^3$ coarse lattice. Each
  square is a coarse site $n = n_x + 3n_y + 9n_z$, divided into its three color partners $a = 0,1,2$
  (basis index $I = 3n + a$) and labelled by its dilution projector
  $b = I \bmod 6 = 3\,(n \bmod 2) + a$: even sites use $b = 0$--$2$, odd sites $b = 3$--$5$.}
  \label{fig:interlace-dilution}
\end{figure*}

In stochastic LapH, each interlaced projector contains eigenmodes separated by a fixed number of
positions in the eigenvalue ordering, so that each projector samples the whole spectrum. 
In the
wavelet basis, the same prescription instead partitions
localized basis functions.
For the $3^3$ coarse grid and our ordering of the basis vectors,
interlace dilution with $n_d = 6$ combines full color dilution with a checkerboard of the coarse sites:
each projector contains one color at every second coarse site, up to the periodic boundary.  We depict this in
Fig.~\ref{fig:interlace-dilution}. Since the basis is localized, probing and coloring methods, such
as hierarchical probing~\cite{HProbing1, HProbing2} or multiplier-based
coloring~\cite{MultiplierBasedColoring}, could exploit its locality to construct more effective dilution
projectors on larger volumes.

Instead of the full perambulator, here we compute its action on the diluted noise,
\begin{equation}
    \mathcal{Q}_F[\zeta^b]_I(t_f,t_i) = (S_F)_{II'}(t_f,t_i)\,(\zeta^b)_{I'},
    \label{eq:diluted-solution-vectors}
\end{equation}
which requires $4n_d$ Dirac inversions per noise vector, flavor and source time, independently of
$n_v$. For brevity we suppress the spin indices here and below. With this setup we discuss the nucleon and deuteron correlators below.

\subsubsection{Nucleon} The nucleon correlator of Eq.~\eqref{eq:DistillationNCorr} 
has the stochastic representation,
\begin{equation}
    C_N(t_f,t_i) = \Big\langle (P_+)_{\omega\omega'}\,
        \mathcal{B}[\rho_1,\rho_2,\eta_1,t_f]^{\omega}_{b_1b_2a_1}
        \Big(\tilde{\mathcal{B}}[\rho_1,\rho_2,\eta_1,t_i]^{\omega'}_{b_1b_2a_1}
        - \tilde{\mathcal{B}}[\rho_2,\rho_1,\eta_1,t_i]^{\omega'}_{b_2b_1a_1}\Big)^{*}\Big\rangle_{\rho,\eta},
    \label{eq:StochDistillNCorr}
\end{equation}
where the sink and source blocks are defined as,
\begin{align}
    \mathcal{B}[\rho_1,\rho_2,\eta_1,t_f]^{\omega}_{b_1b_2a_1} &=
        (\phi^{\omega})_{IJK}(t_f)\,\mathcal{Q}_u[\rho_1^{b_1}]_I\,
        \mathcal{Q}_u[\rho_2^{b_2}]_J\,\mathcal{Q}_d[\eta_1^{a_1}]_K,
        \label{eq:SinkBlock}\\
    \tilde{\mathcal{B}}[\rho_1,\rho_2,\eta_1,t_i]^{\omega}_{b_1b_2a_1} &=
        (\phi^{\omega})_{IJK}(t_i)\,[\rho_1^{b_1}]_I\,[\rho_2^{b_2}]_J\,[\eta_1^{a_1}]_K.
        \label{eq:SourceBlock}
\end{align}
Here $P_+$ is the positive-parity projector,
$(\phi^\omega)_{IJK}$ is the nucleon elemental defined
in Eq.~\eqref{elemental_def}, with open spin index $\omega$,
and $\mathcal{Q}_q$ denotes the diluted solution vectors
of Eq.~\eqref{eq:diluted-solution-vectors}, evaluated
at $(t_f,t_i)$. 
Spin contractions are implicit, and
repeated basis, spin, and dilution indices are summed.
The noise vectors $\rho_1$, $\rho_2$, and $\eta_1$
are assigned to the two $u$-quark lines and the
$d$-quark line, respectively. Exchanging the two
$u$-quark noise vectors and their associated dilution
indices in the second source block generates the exchange
contraction, with the relative minus sign required
by fermionic antisymmetry.

Here three independent noise vectors are the minimum that are needed for 
the factorized unbiased estimator used here. In practice we draw triples of distinct vectors from a larger pool of $N_\rho$ noise vectors, and average over all such choices and their permutations.

Both the source and sink blocks are required for every noise permutation, but only the sink block involves
quark propagators. 
The source blocks can therefore
be computed once per gauge configuration and source
time and reused for all sink times. However, the sink blocks
must be evaluated at each sink time and they will dominate the contraction cost.
Their leading cost is
$\mathcal{O}(n_v^3 n_d)$ per noise triple and sink
time, up to spin factors, and hence grows cubically
with the distillation-space dimension $n_v$ at fixed
dilution number $n_d$. The sampling strategy introduced
in Subsec.~\ref{sec:stoch:hh} targets this dominant cost.

\subsubsection{Deuteron} 
Using the proton and neutron source and sink blocks, the
deuteron correlator in Eq. 
\eqref{eq:DeuteronContraction} can be rewritten as,
\begin{align}
    C_D(t_f,t_i) = \Big\langle\, & \Gamma^{j}_{l_pl_n}\,
        \mathcal{B}[\rho_1,\rho_2,\eta_1,t_f]^{l_p}_{b_1b_2a_1}\,
        \mathcal{B}[\eta_2,\eta_3,\rho_3,t_f]^{l_n}_{a_2a_3b_3}
        \sum_{\pi,\sigma\in S_3} \operatorname{sgn}(\pi)\operatorname{sgn}(\sigma) \nonumber\\
    & \hspace*{-0.1in} \times \Big(\Gamma^{j}_{l'_pl'_n}\,
        \tilde{\mathcal{B}}[\rho_{\pi(1)},\rho_{\pi(2)},\eta_{\sigma(1)}]^{l'_p}_{b_{\pi(1)}b_{\pi(2)}a_{\sigma(1)}}\,
        \tilde{\mathcal{B}}[\eta_{\sigma(2)},\eta_{\sigma(3)},\rho_{\pi(3)}]^{l'_n}_{a_{\sigma(2)}a_{\sigma(3)}b_{\pi(3)}}
        \Big)^{*}\Big\rangle_{\rho,\eta},
    \label{eq:DeuteronContractStoch}
\end{align}
where the source blocks are evaluated at $t_i$. 
The first and second sink blocks correspond to the proton
and neutron, respectively, with the appropriate quark
flavors in each block. The tensor $\Gamma^j$ couples the
nucleon spin indices $l_p$ and $l_n$ to the spin-triplet
component $j$. The six independent noise vectors
$\rho_i$ and $\eta_i$, $i=1,2,3$, are assigned to the
three $u$- and three $d$-quark lines, with dilution
indices $b_i$ and $a_i$, respectively.
Repeated spin
and dilution indices are summed, with $j$ held fixed.
The sums over the permutations $\pi$
and $\sigma$ of the identical $u$ and $d$ lines, with their signs, include all direct and exchange
contractions. We also average over the permutation of the noise assignment to the quark lines as in the case for single nucleons. 
Once the source and sink blocks are available, their
assembly in Eq.~\eqref{eq:DeuteronContractStoch} costs
at most $\mathcal{O}(n_d^4)$ per Wick term, up to spin
factors, and is independent of $n_v$ at fixed dilution. An ordering of the contractions can always be found in which each sink block is contracted with a
source block sharing at least two of its three quark lines, so the sum runs over two dilution indices, costing $\mathcal{O}(n_d^{2\times3-2}) = \mathcal{O}(n_d^4)$. The remaining
contraction does not increase this leading cost.

\subsubsection{Comparison with stochastic LapH} 
Following Sec.~\ref{sec:wavelet-fv-spectra}, we use $n_v=81$ basis
vectors for both Stochastic LapH and Stochastic Wavelet distillation. In each
case, interlace dilution with $n_d=6$ in the distillation space
and full spin dilution yield $24$ diluted sources per noise
vector. We use one source time on each of $100$ gauge
configurations of the CLS A653 ensemble~\cite{CLS}. On each configuration we generate six independent noise vectors.
For the
deuteron, we average over all permutations of these six vectors. For the nucleon, we average
over every three-element subset from the six noise vectors
and all of its permutations.

\begin{figure*}[tbp]
  \centering
  \includegraphics[width=0.7\textwidth]{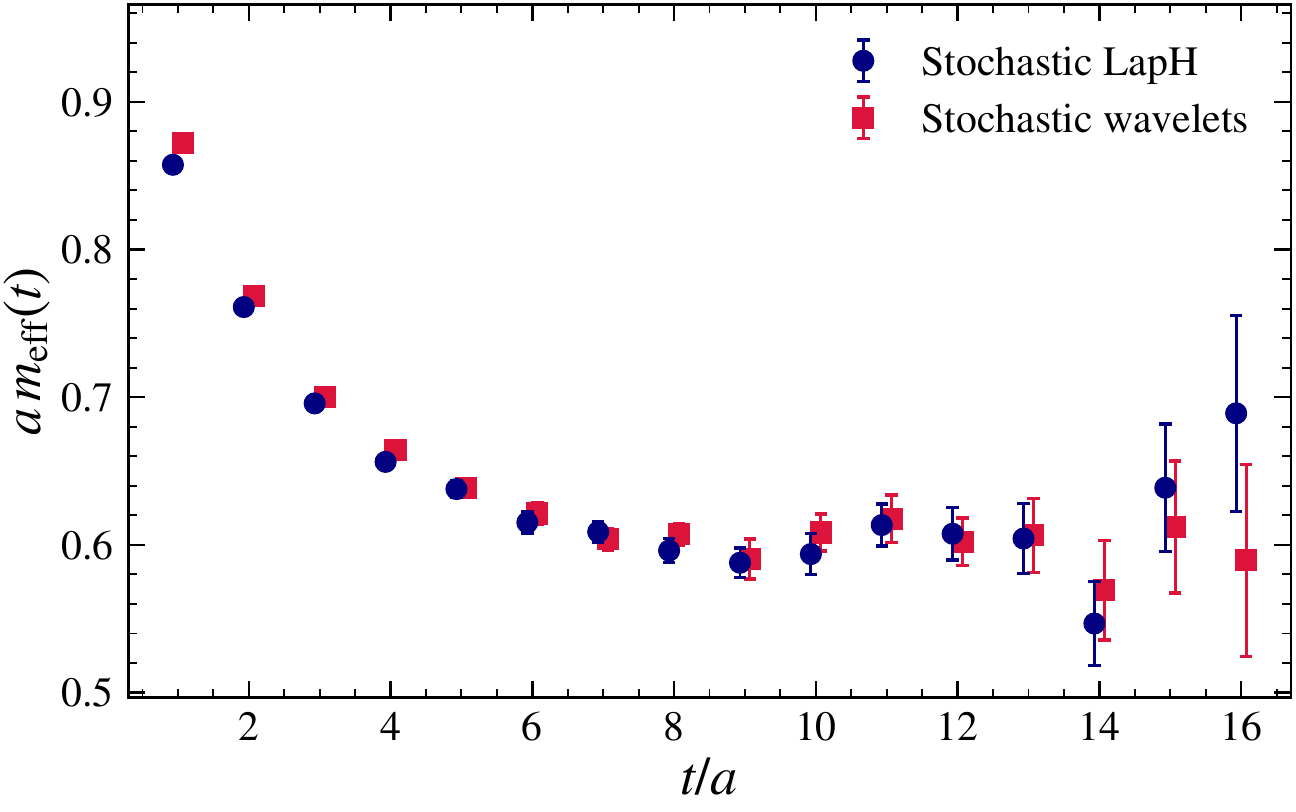}
  \includegraphics[width=0.7\textwidth]{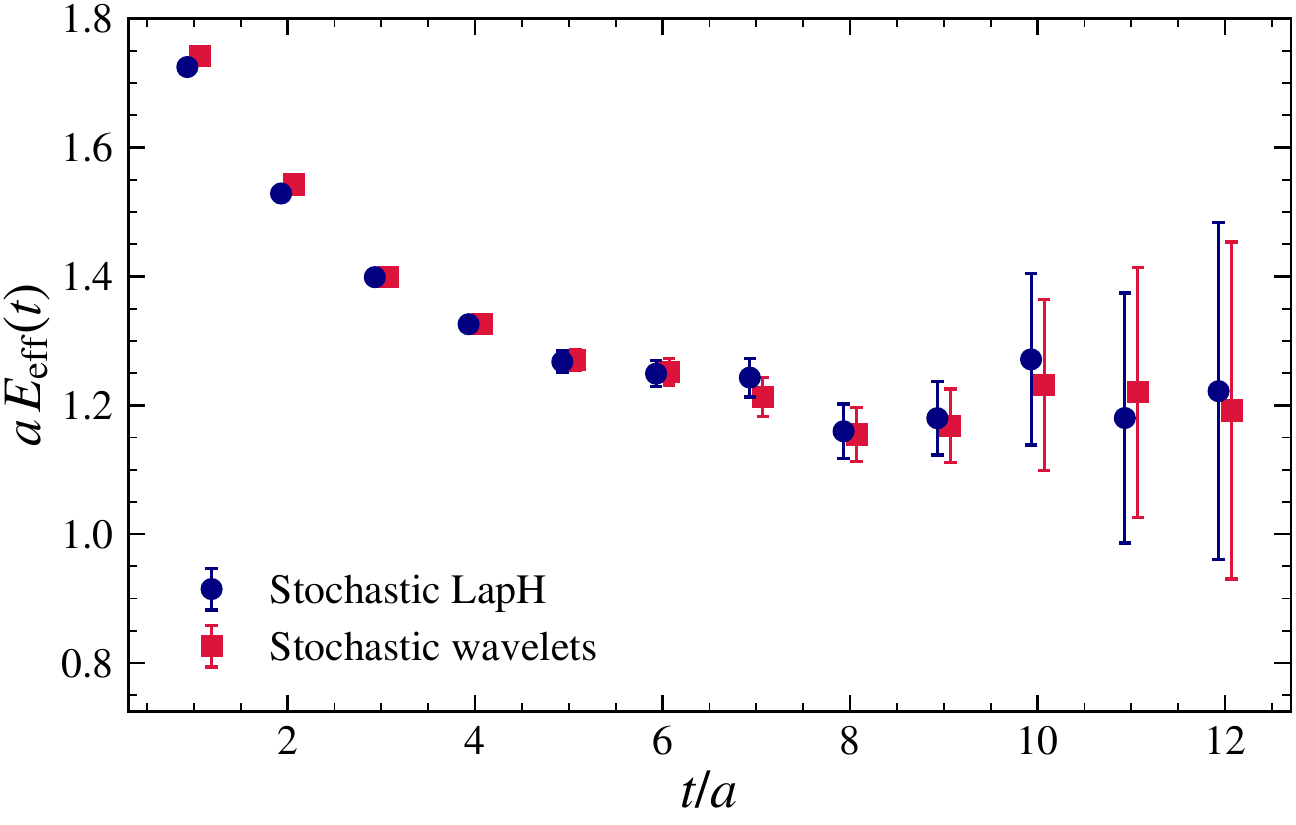}
  \caption{Nucleon (top) and deuteron (bottom) effective masses from stochastic LapH (blue) and
  stochastic wavelet distillation (red), both with $\mathbb{Z}_4$ noise and interlace-6 dilution in the
  distillation space and full spin dilution.}
  \label{fig:stoch-effective-mass}
\end{figure*}

Figure~\ref{fig:stoch-effective-mass} compares the nucleon and
deuteron effective masses obtained with the two setups. The results agree very well within statistical
uncertainties over the full range of Euclidean times, showing comparable statistical errors.
These results show that for the systems and parameters studied, stochastic estimation
in the localized wavelet basis  achieves overall precision
comparable to that in the Laplacian eigenbasis. 
Further studies
are of course required to assess this behavior for other values of $n_v$
and $n_d$, lattice spacings and volumes, and interpolating
operators. 

We next use Hansen--Hurwitz sampling~\cite{HansenHurwitz} to exploit
wavelet localization in the sink-side Wick contractions and assess
the resulting computational savings and statistical precision.

\subsection{Hansen--Hurwitz estimation at the sink}\label{sec:stoch:hh}
The preceding sections showed that conventional and wavelet distillation yield statistically consistent finite-volume energies for both single- and multihadron systems, and additionally  comparable agreement between their stochastic implementations. Despite similar spectra and smearing profiles, the underlying bases have qualitatively different spatial structures:  Laplacian eigenvectors are delocalized over the spatial volume, whereas the wavelet basis vectors are localized around their coarse sites. As a result, the nucleon elemental entering the
sink block, as expressed in Eq.~\eqref{eq:SinkBlock}, is dense in the Laplacian eigenbasis but sparse in the
wavelet basis. To quantify this, and to construct the sampling distribution, we need a
gauge-invariant measure of how the magnitude of the elemental is distributed over its indices.

\begin{figure*}[tbp]
  \centering
  \includegraphics[width=0.48\textwidth]{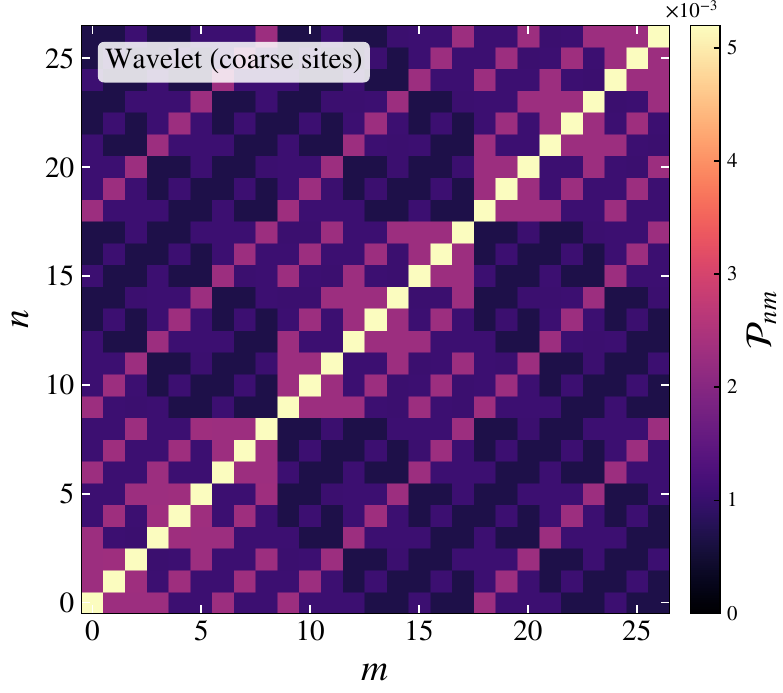}
  \hfill
  \includegraphics[width=0.48\textwidth]{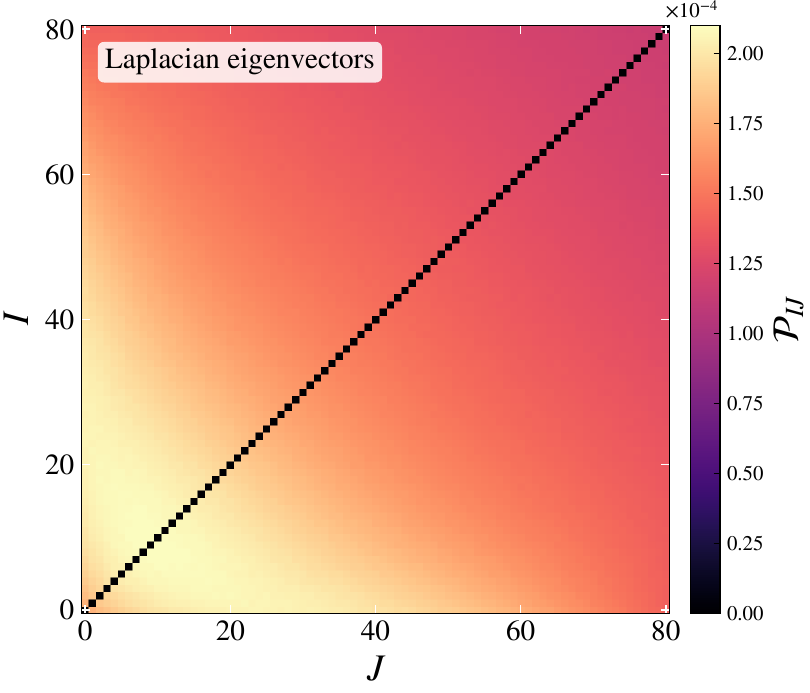}
  \caption{Left: coarse-site marginalized weight $P_{nm}$ of the elemental in the wavelet basis, as in 
  Eq.~\eqref{eq:site_pdf}. Right: marginalized weight $P_{IJ}$ in the Laplacian eigenbasis, as in 
  Eq.~\eqref{eq:hh-probability}. The localization of the wavelet basis makes the elemental sparse.}
  \label{fig:sparsity-comparison}
\end{figure*}

In the wavelet basis of Sec.~\ref{sec:wavelet-construction},
the composite index $I=(n,a)$ labels a coarse-grid site $n$
and a color component $a$ at its anchor point. Under a gauge
transformation $G$, the basis vectors associated with each
coarse site undergo a color rotation determined by $G$ at
that anchor point. Accordingly, the elemental transforms as,
\begin{equation}
    (\phi^\omega)_{(na)(mb)(lc)} \;\to\; \sum_{a'b'c'} (\phi^\omega)_{(na')(mb')(lc')}\,
        G^\dagger(n)_{a'a}\, G^\dagger(m)_{b'b}\, G^\dagger(l)_{c'c} ,
    \label{eq:elemental-gauge}
\end{equation}
where $G(n)$ denotes the gauge transformation at the anchor
point of coarse site $n$. Although the magnitudes of individual
elemental entries are gauge dependent, unitarity ensures that
the Frobenius norm over all three color indices at a fixed
coarse-site triple is gauge invariant. We therefore define
the elemental weight at the level of coarse-site triples as,
\begin{equation}
    P_{nml} = \bigg\langle \frac{|\phi|_{nml}(t)}{\sum_{n'm'l'} |\phi|_{n'm'l'}(t)} \bigg\rangle_{U,t} , \quad {\mathrm{with}} \quad
    |\phi|_{nml}(t) = \bigg[\sum_{abc} \big|(\phi)_{(na)(mb)(lc)}(t)\big|^2\bigg]^{1/2}  .
    \label{eq:site_pdf}
\end{equation}
In the Laplacian eigenbasis, the entries of the elemental are themselves gauge invariant, and the
weight can also similarly be  defined entry by entry as,
\begin{equation}
    P_{IJK} = \bigg\langle \frac{|\phi(t)_{IJK}|}{\sum_{I'J'K'} |\phi(t)_{I'J'K'}|}\bigg\rangle_{U,t} .
    \label{eq:hh-probability}
\end{equation}
Figure~\ref{fig:sparsity-comparison} compares the marginalized weights $P_{nm} = \sum_l P_{nml}$ in
the wavelet basis and $P_{IJ} = \sum_K P_{IJK}$ in the Laplacian eigenbasis. In the wavelet basis,
the weight is concentrated on pairs of nearby coarse sites, whereas in the Laplacian eigenbasis it is spread over all index pairs with a dense distribution reflecting the delocalized Laplacian
eigenvectors. The spatial localization of the wavelet basis can therefore be
exploited to accelerate hadronic contractions through suitable
stochastic estimators, as discussed below.

\subsubsection{Hansen-Hurwitz (HH) Estimator} 

Consider a tensor contraction with a sum,
\begin{equation}
  C=\sum_{\mathbf X}T(\mathbf X).
\end{equation}
Let $\mathbf X_s$, with $s=1,\ldots,N_{hh}$, be independent samples drawn with
replacement from a normalized probability distribution $P(\mathbf X)$.  
The HH estimator \cite{HansenHurwitz} replaces the sum $C$ by,
\begin{equation}
    \widehat C = \frac{1}{N_{\rm hh}} \sum_{s=1}^{N_{\rm hh}} \frac{T(\mathbf X_s)}{P(\mathbf X_s)},
    \qquad \mathbf X_s \sim P .
    \label{eq:HH-general1}
\end{equation}
The estimator is unbiased, $\mathbb{E}[\widehat C] = C$, for any $P$ that is non-zero wherever
$T$ is. Its variance,
\begin{equation}
    \operatorname{Var}(\widehat C) = \frac{1}{N_{\rm hh}}
        \bigg[\sum_{\mathbf X} \frac{|T(\mathbf X)|^2}{P(\mathbf X)} - |C|^2\bigg],
    \label{eq:HH-variance1}
\end{equation}
is minimized by $P \propto |T|$. The variance-minimizing distribution is
\begin{equation}
  P_{\mathrm{opt}}(\mathbf X)
  =\frac{|T(\mathbf X)|}{\sum_{\mathbf X'}|T(\mathbf X')|}.
  \label{eq:HH-optimal-probability}
\end{equation}
Constructing $P_{\mathrm{opt}}$ from the complete hadronic contraction, however,  would
eliminate the computational advantage, since it requires evaluating all tensor
elements.  A practical alternative is to construct the
sampling distribution from the elementals alone and importance-sample
the relevant indices. This strategy was investigated for exact
distillation in Ref.~\cite{mv6l-m8rq}; however, its benefit was limited
by the sequential contractions and the cost of storing or contracting
intermediate tensors. In stochastic wavelet distillation, the localization-induced sparsity
of the elementals makes HH sampling more effective for
evaluating the sink block $\mathcal{B}$ in
Eq.~\eqref{eq:SinkBlock}. We therefore use the elemental itself as a computationally
inexpensive proxy, namely the distribution $P_{nml}$ of Eq.~\eqref{eq:site_pdf}.

\subsubsection{Coarse-site sampling} Rather than sampling the composite distillation-space indices, denoted by $\{I,J,K\}$ (see Eq.~\ref{elemental_def} for their entries), we sample only
their coarse sites, drawing site triples $(n,m,l)$ from $P_{nml}$, and evaluate the color sums exactly.
For fixed dilution indices $(b_1,b_2,a_1)$ we define
\begin{equation}
    F^{\omega}_{nml}(t_f) = \sum_{abc} (\phi^\omega)_{(na)(mb)(lc)}(t_f)\,
        \mathcal{Q}_u[\rho_1^{b_1}]_{(na)}\,\mathcal{Q}_u[\rho_2^{b_2}]_{(mb)}\,\mathcal{Q}_d[\eta_1^{a_1}]_{(lc)},
    \label{eq:HH-integrand}
\end{equation}
where the dilution labels of $F^\omega_{nml}$ and the time arguments of the solution vectors are
suppressed. To reduce the sampling variance, a set $\mathcal{D}$ of high-probability triples is evaluated
exactly and only its complement $\mathcal{R}$ is sampled. For a sampling budget of $N_{\rm hh}$ draws, we
take $N_R (= N_{\rm hh} - |\mathcal{D}|)$ independent draws with replacement from
\begin{equation}
    P^R_{nml} = \frac{P_{nml}}{1-\sum_{(n',m',l')\in\mathcal{D}} P_{n'm'l'}},
    \qquad (n,m,l)\in\mathcal{R}.
    \label{eq:HH-residual-probability}
\end{equation}
The HH estimator of the sink block of Eq.~\eqref{eq:SinkBlock} is then given by,
\begin{equation}
    \widehat{\mathcal{B}}^{\,\omega}_{b_1b_2a_1}(t_f)
    = \sum_{(n,m,l)\in\mathcal{D}} F^\omega_{nml}(t_f)
    + \sum_{\substack{(n,m,l)\in\mathcal{R}\\ f_{nml}>0}} \frac{f_{nml}}{N_R P^R_{nml}}\,F^\omega_{nml}(t_f),
    \label{eq:HH-sink-block}
\end{equation}
where $f_{nml}$ is the number of times a triple was drawn, so that the second sum runs over distinct
drawn triples. Since $\mathbb{E}_{\rm HH}[f_{nml}] = N_R P^R_{nml}$, the estimator is unbiased at fixed
gauge configuration and noise vectors, provided $P^R_{nml} > 0$ wherever the integrand is non-zero.

All color components of a drawn site triple enter with the same weight, allowing the color contraction to be evaluated exactly. Only the sum over coarse sites is stochastic. The set $\mathcal{D}$ contains the triples with an expected number of draws of at least one ($N_{\mathrm{hh}}P_{nml} \geq 1$). Since these triples would be evaluated at least once regardless, computing them exactly removes their contribution to the variance at essentially no computational cost. On the current ensemble, this selects the 27 diagonal triples ($n = m = l$), which have significantly higher probabilities than all other triples and carry approximately 7\% of the total weight. The optimal strategy for selecting the deterministic index set
could be investigated in future work.

On each gauge configuration, HH samples are generated
independently of the source-noise vectors, with a separate
sample set for each source-noise permutation. Each sample set
is reused across sink times to retain temporal correlations.
We find that independent sampling across source noise permutations
substantially reduces the additional HH variance compared with reusing a common sample set. The remaining sampling
fluctuations are included in the statistical uncertainties of the
correlators. Below we describe the construction for the deuteron with a dibaryon operator.

\subsubsection{HH estimator for Deuteron} The deuteron correlator in Eq.~\eqref{eq:DeuteronContractStoch} contains products
of two sink blocks. If both blocks were estimated from the same HH draws, their sampling fluctuations
would be correlated and the estimate of the product biased. With the sampling strategy used here, this
is avoided automatically: every choice of noise triple entering a sink block has its own independent set
of site draws, and in each term of Eq.~\eqref{eq:DeuteronContractStoch}, the proton and neutron blocks
carry complementary sets of three noise vectors. The two factors of every product are therefore always
estimated from independent draws, so that the product estimator is unbiased, while each sink block is
still computed only once and reused in all terms in which it appears.

The antisymmetry of the elemental allows two further reductions. First, antisymmetrizing the sink
blocks under exchange in their two identical-flavor lines combines the 36 Wick-contraction terms into 9. Second,
averaging over permutations of the noise vectors among quark lines of the same flavor is equivalent to
antisymmetrizing the source operator. With the source blocks antisymmetrized within each nucleon, only 180
of the 720 orderings of the six noise vectors therefore are needed to reproduce the average over all 720 exactly. 

We now estimate the contraction cost and its dependence on
the dimensions of the basis and dilution spaces.

\subsubsection{Contraction order and cost.} 
\begin{figure*}[tbp]
  \centering
  \includegraphics[width=\textwidth]{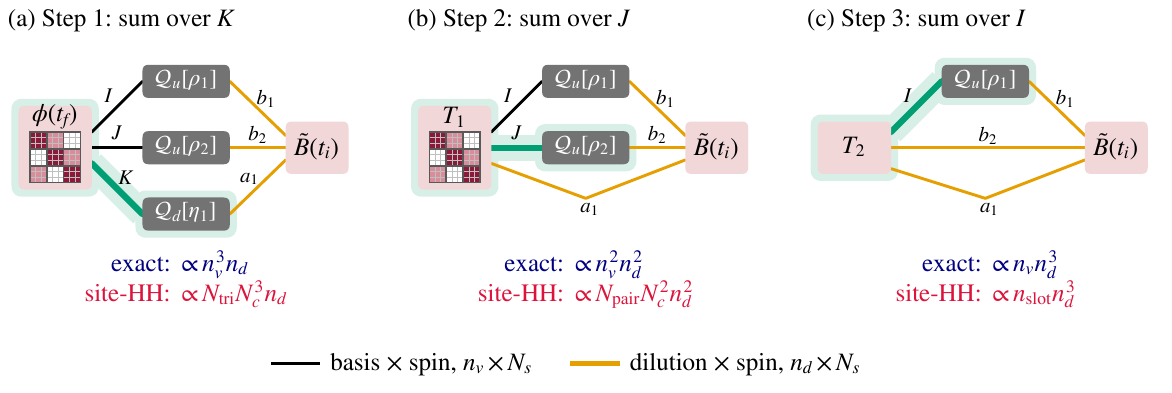}   
  \caption{A schematic illustration of the order of contractions for the nucleon sink block corresponding to Eq.~\eqref{eq:SinkBlock}, for one noise
  triple with the HH sink. The sampled sink elemental is contracted in turn over three steps $K$ (a), $J$ (b) and
  $I$ (c), each time with one quark line. In (c) the spin projection is applied and the result is
  contracted with the source block. Below each panel: the theoretical cost estimation of the step for the exact (Eq.~\eqref{eq:cost_exact})   and HH (Eq.~\eqref{eq:cost_hh}) sinks, are  shown by navy and 
  crimson colors, respectively. See text for details.}
  \label{fig:hh-contraction-steps}
\end{figure*}

We first compare the cost of evaluating the sink block of
Eq.~\eqref{eq:SinkBlock} exactly and with the HH estimator. The block is evaluated by contracting the
sink elemental with one quark line at a time and storing each result as an intermediate tensor. We illustrate the procedure schematically in Fig.~\ref{fig:hh-contraction-steps}. First, the index $K$ of the sink elemental $\phi(t_f)$, a rank-3
tensor, is contracted with the $d$-quark line $\mathcal{Q}_d[\eta_1]$, which gives the intermediate
$T_1$ at a cost of $\mathcal{O}(n_v^3 n_d)$ up to spin factors. Next, $T_1$ is contracted over $J$ with
the $u$-quark line $\mathcal{Q}_u[\rho_2]$, which gives $T_2$ at a cost of $\mathcal{O}(n_v^2 n_d^2)$.
Finally, $T_2$ is contracted over $I$ with the remaining $u$-quark line $\mathcal{Q}_u[\rho_1]$. For the
nucleon this last step is combined with the spin structure of the correlator, the parity projection and
the source diquark, so that only the dilution indices remain open; it costs $\mathcal{O}(n_v n_d^3)$ up
to spin factors. With the spin factors restored, the exact evaluation cost is,
\begin{equation}
    \mathrm{Ops}_{\rm exact} = n_v^3 N_s^2 n_d + n_v^2 N_s^3 n_d^2 + 2\,n_v N_s^2 n_d\,(N_s^2 + n_d^2)
    \label{eq:cost_exact}
\end{equation}
complex multiply-adds per noise triple and sink time. The remaining contraction with the source block,
$\mathcal{O}(n_d^3)$ operations, is negligible and the same for both methods.

For the HH estimator, let $N_{\rm tri}$ be the number of distinct site triples $(n,m,l)$ among the
$N_{\rm hh}$ draws, $N_{\rm pair}$ the number of distinct site pairs $(n,m)$, obtained by marginalizing
over $l$, and $n_{\rm slot}$ the number of distinct first indices $(na)$, i.e.\ $N_c$ times the number
of distinct first sites $n$. The same sequence of contractions then costs
\begin{equation}
    \mathrm{Ops}_{\rm HH} = N_{\rm tri} N_c^3 N_s^2 n_d + N_{\rm pair} N_c^2 N_s^3 n_d^2
        + 2\,n_{\rm slot} N_s^2 n_d\,(N_s^2 + n_d^2).
    \label{eq:cost_hh}
\end{equation}
Sampling replaces the $n_v^3$ terms of the dominant first step by $N_{\rm tri}N_c^3$ terms. The cost of
the second step grows as more site pairs are drawn, but since $N_c^2 N_{\rm pair} \leq n_v^2$ it never
exceeds that of the corresponding exact step, which it reaches only when all site pairs are present.

That the saturating step is a subleading one is a consequence of the stochastic source. In the
localized-distillation formulation of Ref.~\cite{mv6l-m8rq}, both the source and the sink tensors are sampled and joined through exact $n_v\times n_v$ perambulators, so that the saturating intermediate is the dominant one and can approach the $\mathcal{O}(n_v^4)$ cost of the exact contraction. In the present formulation the quark lines are $n_v\times  n_d$ matrices of diluted solution vectors and only the sink elemental is sampled. The saturating intermediate is then that of the second step, which costs a fraction $N_s n_d/n_v$ of the dominant first step, about a quarter of Eq.~\eqref{eq:cost_exact} for $n_v = 81$ and should be less for larger $n_v$. The reduction of the dominant $n_v^3$ term is therefore retained even when all site pairs have been drawn. A nested Hansen–Hurwitz estimator, in which instead of sampling all the sink indices together, the set of sink indices are partitioned and each partitions are independently sampled using relevant conditional probability, could possibly soften the saturation of second contraction step further.

The stochastic source offers a second advantage. As in stochastic LapH, the correlator is averaged over
all permutations of the noise vectors, which reduces the variance introduced by the noise. If each
permutation is given its own independent set of site draws, at negligible extra cost, the same average
also reduces the variance introduced by the HH estimator, since the sampling errors of different
permutations are then independent. The variance due to the noise vectors is unchanged with respect to
stochastic LapH, but the HH contribution is reduced substantially. On the present ensemble, at
$N_{\rm hh} = 1600$, it falls from about 18\% of the variance of the nucleon correlator with a single set of draws for all permutations to about 3.5\% with independent draws, so that a moderate number of draws suffices.

For the deuteron, the assembly of the source and sink blocks in Eq.~\eqref{eq:DeuteronContractStoch}
contracts tensors that carry only dilution and spin indices. Choosing the contraction order separately
for each Wick term bounds the cost of each assembly step by $\mathcal{O}(n_d^4)$, up to spin factors, so
that at fixed dilution the assembly cost is independent of $n_v$. On the present ensemble it takes about
a tenth of the time of the sink blocks.

\subsection{Results using stochastic wavelet distillation}\label{sec:stoch:results}
In this section, we compare stochastic wavelet distillation using
HH sampling at the sink with the corresponding calculation using
exact sink contractions. Both methods use the same gauge configurations, the same $N_\rho = 6$ noise vectors and interlace dilution with $n_d = 6$. We have averaged over all permutations of noise triples  with its own independent set of site draws, which are kept fixed for all sink times of a configuration. Since everything except the site draws is shared, the difference between the two calculations on each configuration is the HH sampling error alone. This allows us to determine the variance added by the sampling directly, under bootstrap.
 
\begin{figure}[tbp]
  \centering
  \includegraphics[width=0.6\columnwidth]{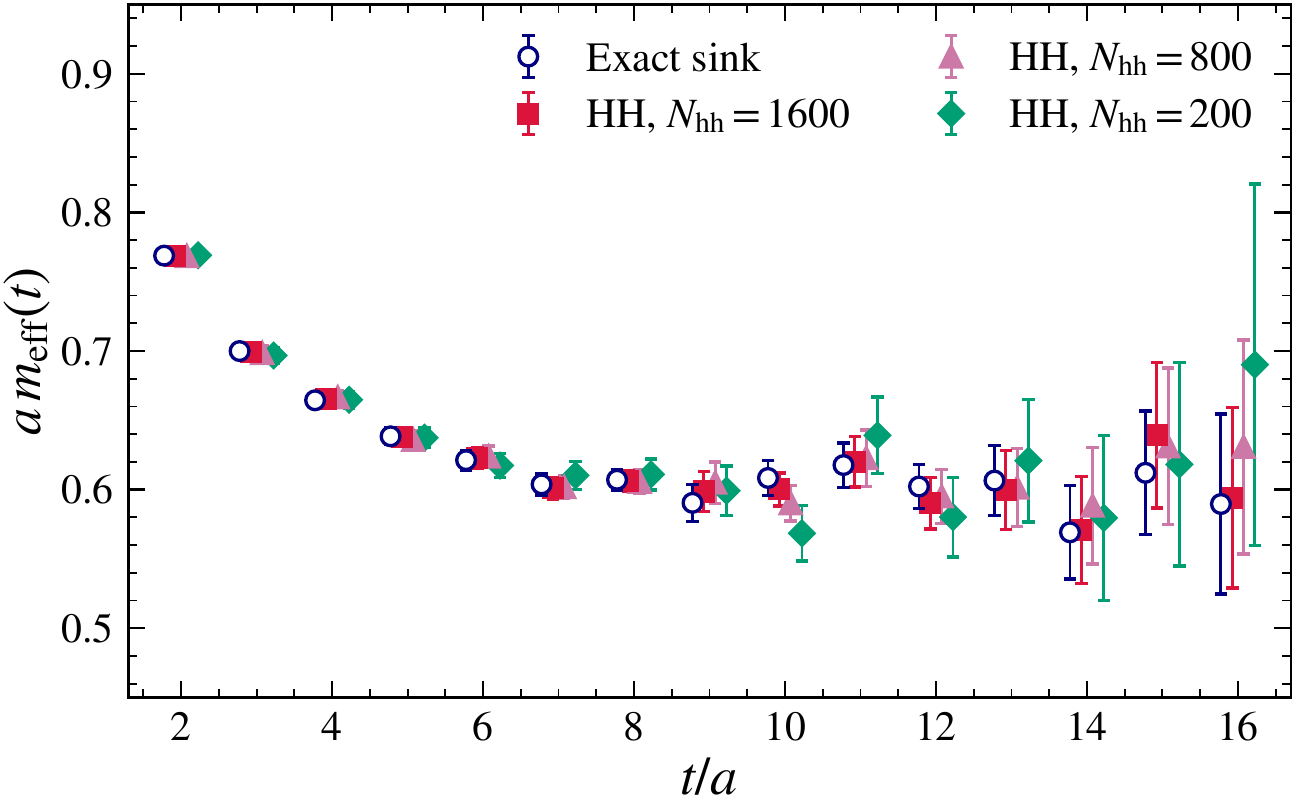}
  \caption{Nucleon effective mass obtained using stochastic wavelet distillation with the exact sink contraction (open circles) and the HH-sink estimator with $N_{\rm hh} = 1600$, $800$ and $200$ site draws,  using independent site draws for each noise triple.}
  \label{fig:wavelet_nucl_stoch_HH}
\end{figure}

Fig.~\ref{fig:wavelet_nucl_stoch_HH} shows the nucleon effective mass with the
exact sink and with the HH sink for $N_{\rm hh} = 200$, $800$ and $1600$ site draws. The HH results agree
with the exact ones at all Euclidean times, for every number of draws. At early times the sampling
error is negligible. The error bars of all four calculations are indistinguishable, since the gauge and
noise-vector fluctuations dominate. The sampling error becomes visible only at later times, where the
signal is small, and decreases with the number of draws. The important point is that even with $N_{\rm hh}=200$, corresponding to approximately $1\%$ of
the $19\,683$ site triples, the effective mass agrees with that
obtained from exact sink contractions within statistical uncertainties.
 
\begin{figure}[tbp]
  \centering
  \includegraphics[width=0.6\columnwidth]{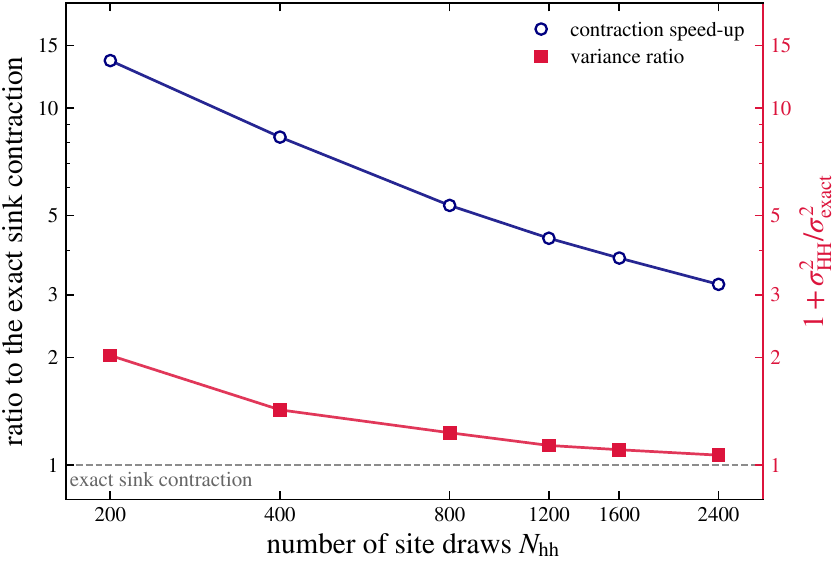}
  \caption{Plot illustrating the speed-up of the nucleon sink contraction relative to the exact contraction (circles, left
  axis) and the corresponding variance ratio of the fitted ground-state energy (squares, right axis), shown as functions of the
  number of site draws $N_{\rm hh}$. 
  The variance ratio (VR) is defined as $1 + \mathrm{Var}(E_{\rm HH} -
  E_{\rm exact})/\mathrm{Var}(E_{\rm exact})$,
  using paired bootstrap samples from correlated single-exponential fits over $t/a=8\text{--}16$. The speed-up is given by the ratio of wall-clock times for a single configuration. Both vertical axes use the same logarithmic scale, and the dashed line denotes the exact sink contraction.}
  \label{fig:hh-cost-variance}
\end{figure}

Fig.~\ref{fig:hh-cost-variance} compares the computational savings
from HH sink sampling with the resulting statistical precision of
the nucleon correlator. The speedup is defined as the ratio of
wall-clock times for exact and HH sink contractions.
The
precision is measured by the variance of the fitted ground-state energy, obtained from a correlated
single-exponential fit to the correlator in the range $t/a = 8$--$16$. Since the two calculations share
the configurations and noise vectors, the difference of their fitted energies on each bootstrap sample,
$E_{\rm HH} - E_{\rm exact}$, contains only the site sampling error, and its variance is the contribution of
the HH estimator to the variance of the result. We therefore define the variance ratio (VR) as,
\begin{equation}
    \mathrm{VR} = 1+\frac{\sigma^2_{\rm HH}}{\sigma^2_{\rm exact}}
    = 1+\frac{\mathrm{Var}(E_{\rm HH} - E_{\rm exact})}{\mathrm{Var}(E_{\rm exact})},
    \label{eq:hh-variance-ratio}
\end{equation}
with both variances from the same bootstrap samples. 
As $N_{\rm hh}$ increases from $200$ to $2400$, we observe that the speedup decreases
from $13.6$ to $3.2$, while the variance ratio approaches unity.
At $N_{\rm hh}=1600$, HH sampling accelerates the sink contraction
by a factor of $3.8$, with only a $5\%$ increase in the statistical
uncertainty of the fitted energy.

\begin{figure}[tbp]
  \centering
  \includegraphics[width=0.6\columnwidth]{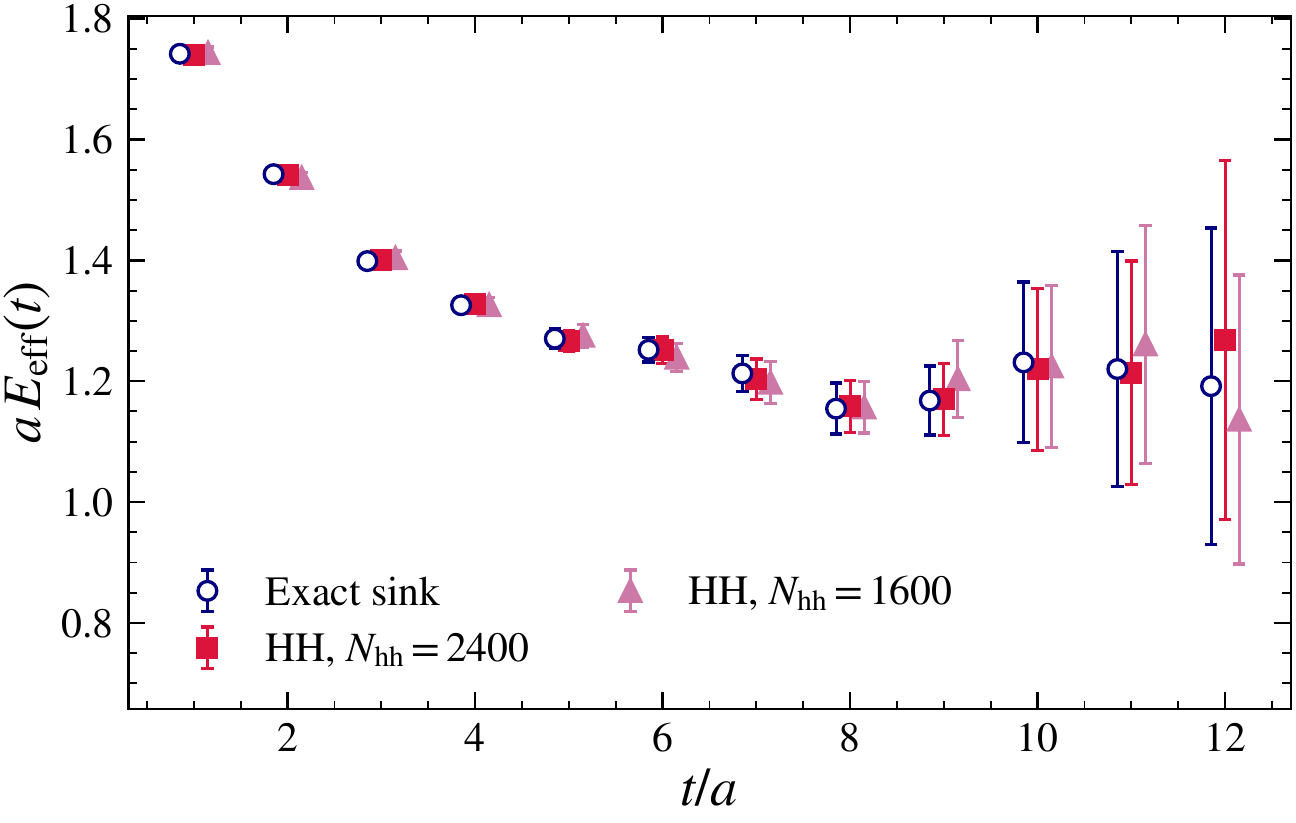}
  \caption{Deuteron effective energy obtained using stochastic wavelet distillation with the exact sink
  contraction (open circles) and with the HH sink for $N_{\rm hh} = 2400$ and $1600$ site draws. For each term, the proton and neutron sink blocks are estimated using independent site draws.}
  \label{fig:wavelet_NP_stoch_HH}
\end{figure}
 
\paragraph*{Deuteron} 
Figure~\ref{fig:wavelet_NP_stoch_HH} shows the deuteron effective energy with
exact sink and with the HH sink at $N_{\rm hh} = 2400$ and $1600$,  using the same $100$
gauge configurations of the CLS ensemble~\cite{CLS}.  The two blocks in each product in
Eq.~\eqref{eq:DeuteronContractStoch} are estimated using independent draws, ensuring that the product remains
unbiased. The results agree within
statistical uncertainties throughout the Euclidean-time range
shown, with only a modest increase in errors. The relative statistical uncertainty of the deuteron correlator
grows more rapidly with Euclidean time than that of the nucleon.
In this calculation, the additional HH fluctuations contribute
a smaller fraction of the total variance, which is dominated
by gauge and source-noise fluctuations. The contraction cost
is dominated by sink-block evaluation; assembling the source
and sink blocks requires approximately one tenth of the
sink-block evaluation time. At $N_{\rm hh}=1600$, we find that HH sampling
reduces the total deuteron contraction cost by a factor of
approximately $3.0$, with comparable statistical precision.

These gains are obtained on a small coarse grid that limits
the benefit of localization. With only three coarse sites per
spatial direction, every site triple lies within the range of the basis functions, producing a broad sampling distribution.
Consequently, a large fraction of the site pairs is drawn
even at moderate $N_{\rm hh}$, and the second step in
Eq.~\eqref{eq:cost_hh} approaches its exact-contraction cost. For bigger lattices the gain is expected to be larger, and the next subsection examines how the computational gain is
expected to evolve with the dimension of the distillation space.

\subsection{Dependence on the dimension of the distillation space}\label{sec:stoch:scaling}
The results discussed in Subsec.~\ref{sec:stoch:results} are obtained with a $3^3$ coarse sites geometry, where
every site triple lies within the range of the basis functions. 
To examine the dependence on the distillation-space
dimension, we repeat the analysis on the same $24^3$ spatial
lattice using a $4^3$ coarse grid, corresponding to $n_v=192$,
with resolvent parameters $\tau=2.25$ and $m=8$. The nominal
localization scale, $\sqrt{2m\tau}=6$ in lattice units, equals
the coarse-grid spacing, as in the $n_v=81$ construction. 
We perform the comparison on a single gauge configuration,
computing the quark lines separately in each wavelet basis.
For each fixed set of source-noise and solution vectors, we
vary only the HH samples, using independent draws for each
noise triple. The same procedure is used for both values of
$n_v$.
This comparison probes the dependence on the basis dimension
at fixed lattice volume; it is not a direct test of
physical-volume scaling.

\begin{figure*}[tbp]
  \centering
  \includegraphics[width=\textwidth]{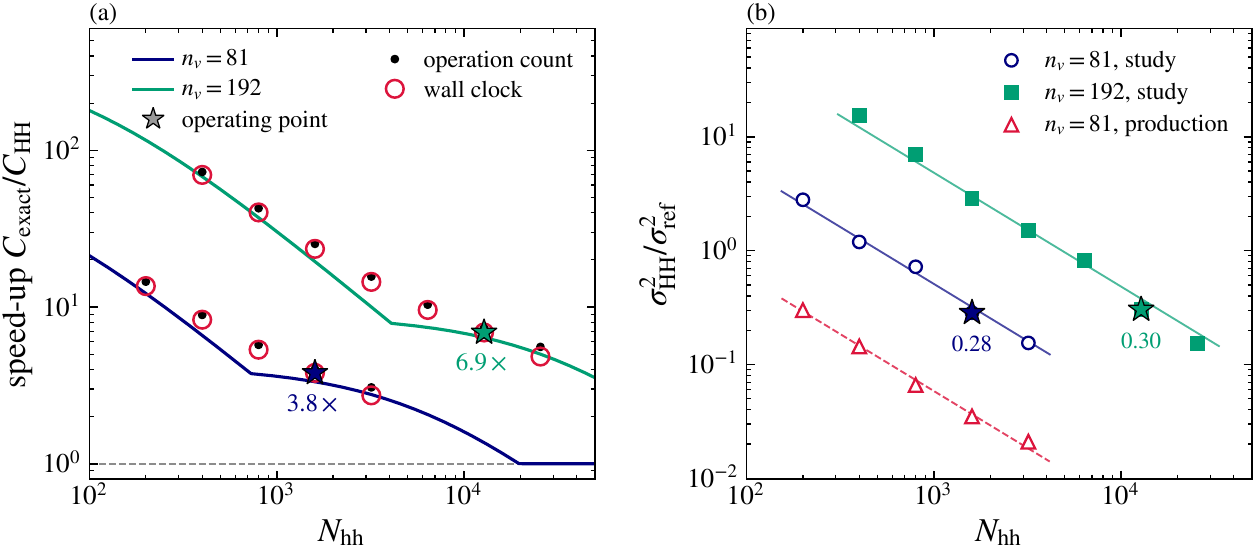}
  \caption{(a) Plot representing the nucleon sink-contraction speed-up, $C_{\rm exact}/C_{\rm HH}$,
versus the number of site draws $N_{\rm hh}$ for $n_v=81$ and $192$.
Lines show lower bounds on the operation-count speed-up from
Eqs.~\eqref{eq:cost_exact} and~\eqref{eq:cost_hh}, assuming each
draw yields a new site triple and pair until saturation.
Dots show speed-ups calculated from the operation counts of the
actual sampling plans; circles show measured wall-clock speed-ups.
 (b) Relative sampling variance,
$\sigma_{\rm HH}^2/\sigma_{\rm ref}^2$, of the correlator.
In the single-configuration study, the HH variance is evaluated
at fixed source noise and solution vectors, while
$\sigma_{\rm ref}^2=\sigma_{\rm noise}^2$ is determined from
independent noise sets. For the production ensemble
($n_v=81$, $t/a=8$--$16$),
$\sigma_{\rm ref}^2=\sigma_{\rm gauge}^2+\sigma_{\rm noise}^2$.
Lines show fits proportional to $1/N_{\rm hh}$.
Stars mark the operating points at
$\sigma_{\rm HH}^2/\sigma_{\rm noise}^2\approx0.3$
with the corresponding speed-ups in (a).
  }
  \label{fig:hh-projection-variance}
\end{figure*}

Fig.~\ref{fig:hh-projection-variance}(a) shows the speed-up of the sink
contraction as a function of $N_{\rm hh}$. The lines give lower
bounds on the operation-count speedup,  obtained from
Eqs.~\eqref{eq:cost_exact} and~\eqref{eq:cost_hh} by assuming that every draw yields a new site triple
and a new site pair, until all of them are exhausted, using $N_{\rm tri} \leq \min(N_{\rm hh}, n_{\rm site}^3)$ and
$N_{\rm pair} \leq \min(N_{\rm hh}, n_{\rm site}^2)$. 
Repeated sampling of high-probability
triples and pairs therefore yields operation-count speedups
above these bounds. The measured wall-clock speedups agree
with those predicted from the actual sampled contraction plans
to within approximately $10\%$ for both basis dimensions,
supporting the cost estimates.
The operation counts of Eqs.~\eqref{eq:cost_exact} and~\eqref{eq:cost_hh} therefore describe the actual
cost of the implementation. 
At fixed $N_{\rm hh}$, the speedup is substantially larger for
$n_v=192$: the leading term in the exact contraction cost
increases by $(192/81)^3\simeq13$, whereas the HH cost is
controlled primarily by the sampled index sets. A meaningful
efficiency comparison, however, must also account for the
number of samples required to maintain comparable precision.

\subsubsection{Sampling variance}
The variance of the correlator computed with the HH sink has three sources of error: the fluctuations of the gauge field, $\sigma^2_{\rm gauge}$, those of the noise vectors at the source, $\sigma^2_{\rm noise}$,
and the sampling at the sink, $\sigma^2_{\rm HH}$. 
On a fixed gauge configuration,  we 
compute the noise-averaged correlator with the exact and with the HH sink. 
Repeating the exact
calculation with independent sets of noise vectors gives $\sigma^2_{\rm noise}$, and repeating the HH
calculation with independent site draws at fixed noise vectors gives $\sigma^2_{\rm HH}$; their ratio
$\sigma^2_{\rm HH}/\sigma^2_{\rm noise}$ is shown in Fig.~\ref{fig:hh-projection-variance}(b). 

In our reported final calculation with $n_v = 81$, the correlator is averaged over 100 gauge configurations. There, the paired difference
between HH and exact sink contractions, as in
Eq.~\eqref{eq:hh-variance-ratio}, measures the additional
sampling contribution relative to the full baseline variance,
$\sigma^2_{\rm HH}/(\sigma^2_{\rm gauge}
+\sigma^2_{\rm noise})$, over $t/a=8$--$16$.
All three data sets exhibit approximately $1/N_{\rm hh}$
behavior, consistent with independent sampling.
The smaller ratio in the production calculation is consistent
with a substantial contribution from gauge fluctuations. Since both ratios refer to the same quantity, the noise-averaged correlator, their quotient at equal
$N_{\rm hh}$ is $1 + \sigma^2_{\rm gauge}/\sigma^2_{\rm noise}$. At $n_v = 81$, the
gauge fluctuations exceed those of the noise vectors by almost an order of magnitude and dominate the
variance of the ensemble average.
A quantitative separation of gauge and source-noise variances
requires averaging the conditional variance estimates over
representative gauge configurations.

\subsubsection{Comparison at matched sampling variance}
This motivates a sampling criterion analogous to that used in
stochastic LapH: the source-noise contribution should remain
subdominant to gauge fluctuations, while the variance from site
sampling should remain subdominant compared with that of the noise vectors. 
As a practical criterion, we choose
$\sigma^2_{\rm HH}/\sigma^2_{\rm noise}\simeq0.3$,
which can be estimated on a small set of configurations before a
production run. In the present test, this corresponds to
$N_{\rm hh}=1600$ for $n_v=81$ and $N_{\rm hh}=12\,800$
for $n_v=192$, with variance ratios of $0.28$ and $0.30$,
respectively. At the former operating point, the production
analysis gives an approximately $10\%$ increase in the
variance of the fitted nucleon energy, equivalent to about
a $5\%$ increase in its statistical uncertainty
(discussed in Subsec.~\ref{sec:stoch:results}).

At these matched operating points the speed-up of the sink contraction increases from $3.8$ for
$n_v = 81$ to $6.9$ for $n_v = 192$. We illustrate those by star symbol in Fig.~\ref{fig:hh-projection-variance}(a). The gain thus
grows with the dimension of the distillation space. Its rate is set by how the required number of draws
grows with $n_v$. The cost of the exact contraction grows as $n_v^3$, by a factor of about $13$ between
the two cases, whereas the cost of the HH sink at the operating point grows by about $7$, and the
required number of draws by about $8$.  For large distillation spaces, the required number of draws is expected to
grow linearly with $n_v$, at fixed localization and
fixed $n_d$, and hence the operation-count bounds imply a speedup growing
at least as $\mathcal{O}(n_v/(N_s n_d))$. 
The results for $n_v = 81$ and $192$
indicate that, if this expectation holds, considerably larger values of $n_v$
are needed to reach the asymptotic regime. 
Nevertheless, both cases demonstrate
substantial reductions in sink-contraction cost with a
controlled additional sampling variance.

\section{Summary and outlook}
\label{sec:summary-outlook}
In this work we have introduced a novel multiresolution analysis of
lattice QCD based on wavelets. The results presented here demonstrate
that spatial localization can be exploited for efficient
computation of lattice-QCD correlation-functions. A distillation space
constructed directly from localized, resolvent-filtered sources
reproduces the energies obtained with the conventional Laplacian
eigenbasis within statistical uncertainties. This agreement extends from
single-hadron correlators to excited energy levels, the deuteron
channel, and boosted nucleons. These comparisons support the use of
localized approximation spaces, that can be generated cheaply, for
variational spectroscopy without requiring a Laplacian
eigensolve. Moreover the computational benefits of this formulation
arise at several stages. Wavelet filtering provides a systematic
alternative to direct decimation, with reduced early-time
excited-state contamination at equal spatial reduction. In wavelet
distillation, the basis construction is measured to be approximately five times computationally cheaper on
a moderately sized lattice studied here, making on-demand regeneration a
practical alternative to persistent basis storage and disk
input/output. Moreover this advantage is expected to grow even further at larger volumes.  Spatial localization also yields hadron
elementals whose weight is dominated by a small subset of entries,
enabling Hansen-Hurwitz importance sampling to accelerate sink contractions within stochastic distillation. On the ensemble studied here, nucleon and deuteron contractions are accelerated approximately by factors of 3.8 and 3.0, respectively, at comparable precision.

Controlling the approximations and stochastic contributions is
essential to realizing these benefits. In particular, we find that
coarse-grid anchoring can introduce momentum aliases that can contaminate
boosted correlators. However, these unwanted contributions can be removed in expectation through independent random shifts at the source and sink, with the residual fluctuations absorbed into the statistical uncertainties. Likewise, the HH sampling budget needs to be chosen according to its contribution to the correlator variance.

The wider significance of this framework lies in its potential to
expand this formulation to large volumes and to a broad range of operators within GEVP setups.  The two wavelet approaches can also be combined,
linking quark-field smearing with the spatial sparsening of
contractions. Further applications include additional multihadron
channels, local multiquark operators, and nuclear systems beyond the
deuteron. Optimizing the filters, dilution projectors, and
importance-sampling distributions may provide additional
savings. Extensions to three-point functions would assess the utility
of the framework for hadron and nuclear matrix elements. Together, these
developments could extend the computational gains demonstrated here
into more extensive spectroscopy, scattering, and structure
calculations on large lattices that are becoming increasingly accessible.

\vspace{-0.2in}
\section{Acknowledgment} 
 This work is supported by the Department of Atomic Energy, Government of India,  under Project Identification Numbers RTI-4012 and RTI-4013. M. B. acknowledges support from the visitor program of the Department of Theoretical Physics, TIFR. Computations were carried out on the computing clusters at the Department of Theoretical Physics, TIFR, Mumbai, and with ILGTI's computing resources. We thank Piyush Srivastava, Gunnar Bali and Navdeep Singh Dhindsa for valuable discussions at various stages. We thank Gunnar Bali and Wolfgang S\"oldner for providing the  CLS ensemble.  We would also like to thank  Ajay Salve and Kapil Ghadiali for computational support. 

We used Anthropic's Claude (Opus 5.5) to assist in optimizing
the Wick-contraction codes developed by us for stochastic LapH
and stochastic wavelet distillation. We also utilized this tool in implementing the
Chebyshev method for the Laplacian eigenvector and expanding the inverse-power of  resolvent operator in terms of Chebyshev polynomial. We validated all optimizedand new implementations against our original codes, confirming
agreement between their results. We also used OpenAI's ChatGPT
(GPT-5.6) to improve English phrasing. The authors take full responsibility for the accuracy and integrity of this work.

\clearpage
\appendix

\section{Computational Cost}
\label{sec:wavelet-cost}

Wavelet distillation reduces computational costs at two stages:
basis construction and Wick
contractions through Hansen--Hurwitz sampling. At matched
subspace dimension, dilution scheme, and number of noise vectors,
conventional and wavelet distillation require the same number of
Dirac inversions. 

\paragraph{Lattice setup:}
We compare the two methods on $100$ gauge configurations of the
CLS $24^3\times48$ ensemble, generated with the clover action at
$a\approx0.098$~fm and $m_\pi\approx429$~MeV. The valence quarks
use the Wilson--clover action with masses matched to the sea
quarks.

\paragraph{Basis construction:}
Following Sec.~\ref{sec:wavelet-construction}, the $n_v=81$ wavelet
basis uses a coarse grid of spacing $N=8$ ($3^3$ anchor sites, three
color sources each), and the width-matching condition,
Eq.~\eqref{eq:width-matching}, $m\tau=32$, is realized with $m=8$,
$\tau=4$. For $n_v=192$ ($4^3$ anchor sites, $N=6$) it gives
$m\tau=18$, and we use $m=8$, $\tau=2.25$. Both bases use the same
spatial covariant Laplacian, built from links with $15$
APE-smearing steps~\cite{APESMEARING} at $\alpha=2.5$.

For each basis we compare a standard iterative method, with a Chebyshev method based on a polynomial filter. The lowest $n_v$ eigenvectors of $-\nabla^2(t)$ are computed with
PRIMME~\cite{PRIMME,StathopoulosMcCombs2007} in
\texttt{PRIMME\_DYNAMIC} mode and with Chebyshev-filtered subspace
iteration
(ChFSI)~\cite{10.1016/j.jcp.2006.03.017, doi:10.1137/050630404}
with SVQB orthonormalization~\cite{doi:10.1137/S1064827500370883}, both to a
residual tolerance $\epsilon_{\rm eig}=10^{-8}$. The wavelet filter $(\mathrm{I}-\tau\nabla^2)^{-m}$
is applied to the $n_v$ point sources either through $m$ nested
solves with even--odd-preconditioned conjugate
gradient~\cite{HestenesStiefel1952,Saad2003} (relative residual
$10^{-10}$). We also employ a single Chebyshev expansion of degree
$K$~\cite{Trefethen2019, RevModPhys.78.275}, truncated such that the
operator-norm error is below $\eta=10^{-13}$.

Table~\ref{tab:wavelet-params} lists the wavelet parameters, the
Chebyshev degree, the condition number of the Gram matrix
$G=\Phi^\dagger\Phi$ of the unorthonormalized basis, and the
subspace overlap with the exact eigenvectors. 

\begin{table}[h!]
  \centering
  \caption{Wavelet-basis parameters, Chebyshev degree $K$
  ($\eta=10^{-13}$), Gram-matrix condition number, and subspace
  overlap $S=n_v^{-1}\|V_{\mathrm eig}^\dagger V_{\mathrm wav}\|_F^2$ between the
  L\"owdin-orthonormalized basis $V_{\mathrm wav}$ and the $n_v$ lowest
  eigenvectors $V_{\mathrm eig}$ of $-\nabla^2$.}
  \label{tab:wavelet-params}
  \vspace*{0.2in}
  \begin{tabular}{cccccccc}
    \toprule
    $n_v$ & $N$ & $m$ & $\tau$ & $m\tau$ & $K$ & $\mathrm{cond}(G)$ & $S$ \\
    \midrule
    81  & 8 & 8 & 4.00 & 32 & 153 & $7.9\times10^{3}$ & 0.883 \\
    192 & 6 & 8 & 2.25 & 18 & 115 & $3.2\times10^{4}$ & 0.881 \\
    \bottomrule
  \end{tabular}
\end{table}

\newpage
In Table~\ref{tab:cost-timings} we provide the basis-construction times per
configuration, averaged over $10$ independent runs. With both Krylov and Chebyshev methods the wavelet basis is about five times cheaper than the eigenbasis.
 
\begin{table}[htbp]
  \centering
  \caption{Wall-clock basis-construction times in minutes per
  configuration on the CLS $24^3\times48$ ensemble, using 32 cores
  of an Intel Xeon Platinum 8358 processor. The last column gives
  the ratio of eigenbasis to wavelet times.}
  \label{tab:cost-timings}
  \vspace*{0.2in}
  \begin{tabular}{clccc}
    \toprule
    $n_v$ & Method & Eigenbasis & Wavelet & Ratio \\
    \midrule
    81  & PRIMME / EO-CG        & 9.36  & 1.82  & 5.14  \\
        & Chebyshev (ChFSI / expansion)  & 1.53  & 0.32  & 4.78  \\
    \midrule
    192 & PRIMME / EO-CG        & 36.5  & 2.91  & 12.54  \\
        & Chebyshev (ChFSI / expansion)  & 3.07  & 0.56  & 5.49  \\
    \bottomrule
  \end{tabular}
\end{table}

\paragraph{Dirac inversions:}
Exact perambulators require $4n_v=324$ Dirac inversions per
flavor and source time slice. With full spin dilution and
$n_d=6$ distillation-space dilution projectors, stochastic
estimation requires $4n_d=24$ inversions per noise vector.
Our choice of $N_\rho=6$ noise vectors therefore requires
$144$ inversions per flavor and source time slice in both
stochastic LapH and stochastic wavelet distillation.
The inversion counts are identical, although the measured
runtimes can differ because the source vectors differ. Dirac inversions are performed using the MILC
codebase~\cite{MILC}, with residual tolerance
$\epsilon_D=10^{-12}$.

The following two appendices describe the HH sink-contraction
algorithm and its computational performance.

\section{Pseudocode for the HH sink contraction}\label{app:pseudocode}

We describe the evaluation of the nucleon sink block, using HH-estimator as detailed in Sec.\ref{sec:stoch:hh}, and provide pseudocode for each stages. We defined the nucleon sink block in Eq.~\eqref{eq:SinkBlock}, for one noise triple
and sink time and the HH estimator in Eq.~\eqref{eq:HH-sink-block}. We suppressed the spin indices. Each
quark line $\mathcal{Q}_i \equiv \mathcal{Q}[\zeta_{r_i}]$ is then an $n_v\times n_d$ matrix, with rows
labelled by the composite index $(na)$ of coarse site $n$ and color $a$, and columns by the dilution
index. In the implementation, the lines also carry their spins, and the spin structure of the nucleon
operator is applied in the last step. This changes the cost of each step by spin factors only, as shown by
Eqs.~\eqref{eq:cost_exact} and~\eqref{eq:cost_hh}.

For one plan of site draws we define the sets of distinct triples, pairs and first sites as,
\begin{equation}
\begin{gathered}
    \mathcal{X}_3 = \{(n,m,l)\ \text{drawn or in}\ \mathcal{D}\},\\
    \mathcal{X}_2 = \{(n,m)\,|\,(n,m,\cdot)\in\mathcal{X}_3\},\qquad
    \mathcal{X}_1 = \{n\,|\,(n,\cdot,\cdot)\in\mathcal{X}_3\},
\end{gathered}
\end{equation}
which determine the cost of the three contraction steps.

Below we provide pseudocode for each stages:

\begin{algorithm}[H]
\caption{Site draws for one noise triple}\label{alg:draw}
\begin{algorithmic}[1]
\Require site distribution $P_{nml}$, number of draws $N_{\rm hh}$, seed of the noise triple
\State $\mathcal{D}\equiv$ triples with $N_{\rm hh}P_{nml}\ge1$; $N_R\equiv N_{\rm hh}-|\mathcal{D}|$
\State draw $N_R$ triples from $P^R_{nml}$, Eq.~\eqref{eq:HH-residual-probability}, with replacement; $f_{nml}$ is the multiplicity of $(n,m,l)$
\State $w_{nml}= 1$ for $(n,m,l)\in\mathcal{D}$, \quad $w_{nml}= f_{nml}/(N_RP^R_{nml})$ otherwise
\State \Return $\mathcal{X}_3$ with weights $w_{nml}$
\end{algorithmic}
\end{algorithm}

\begin{algorithm}[H]
\caption{First contraction (sum over $K$)}\label{alg:T1}
\begin{algorithmic}[1]
\Require weighted triples $\mathcal{X}_3$; elemental $\phi(t_f)$; line $\mathcal{Q}_3$
\State $T_1 = 0$ (Initialize)
\For{$(n,m,l)\in\mathcal{X}_3$} \Comment{$|\mathcal{X}_3|\,N_c^3\,n_d$}
    \State $T_1[(na),(mb)][b_3] \mathrel{+}= w_{nml}\sum_c \phi_{(na)(mb)(lc)}(t_f)\,\mathcal{Q}_{3,(lc)}[b_3]$ \quad for all $a,b,b_3$
\EndFor
\State \Return $T_1$
\end{algorithmic}
\end{algorithm}

\begin{algorithm}[H]
\caption{Second contraction (sum over $J$)}\label{alg:T2}
\begin{algorithmic}[1]
\Require $T_1$; pairs $\mathcal{X}_2$; line $\mathcal{Q}_2$
\State $T_2 = 0$ (Initialize)
\For{$(n,m)\in\mathcal{X}_2$} \Comment{$|\mathcal{X}_2|\,N_c^2\,n_d^2$}
    \State $T_2[(na)][b_2b_3] \mathrel{+}= \sum_b \mathcal{Q}_{2,(mb)}[b_2]\,T_1[(na),(mb)][b_3]$ \quad for all $a,b_2,b_3$
\EndFor
\State \Return $T_2$
\end{algorithmic}
\end{algorithm}

\begin{algorithm}[H]
\caption{Sink block and contraction with the source block (sum over $I$)}\label{alg:sink}
\begin{algorithmic}[1]
\Require $T_2$; first sites $\mathcal{X}_1$; line $\mathcal{Q}_1$; source block $\tilde{\mathcal{B}}$
\State $\widehat{\mathcal{B}}[b_1b_2b_3] = \sum_{n\in\mathcal{X}_1}\sum_a \mathcal{Q}_{1,(na)}[b_1]\,T_2[(na)][b_2b_3]$ \Comment{$|\mathcal{X}_1|\,N_c\,n_d^3$}
\State \Return $\hat c(t_f) = \sum_{b_1b_2b_3}(\tilde{\mathcal{B}}^*[b_1b_2b_3]-\tilde{\mathcal{B}}^*[b_2b_1b_3])\,\widehat{\mathcal{B}}[b_1b_2b_3]$
\end{algorithmic}
\end{algorithm}

\begin{algorithm}[H]
\caption{Nucleon correlator on one gauge configuration}\label{alg:config}
\begin{algorithmic}[1]
\For{each noise triple $p$, in parallel}
    \State $\mathcal{X}_3^{(p)}\gets$ Stage~\ref{alg:draw} with the seed of $p$ \Comment{independent draws per triple}
    \For{each sink time $t_f$} \Comment{same draws for all $t_f$}
        \State $\hat c_p(t_f)\gets$ Stages~\ref{alg:T1}--\ref{alg:sink}
    \EndFor
\EndFor
\State \Return $C(t_f) = N_p^{-1}\sum_p \hat c_p(t_f)$ \Comment{fixed summation order}
\end{algorithmic}
\end{algorithm}

The exact sink contraction is obtained by running algorithms mentioned in stages ~\ref{alg:T1}--\ref{alg:sink} over all site
triples with unit weight. For the deuteron, the proton and neutron sink blocks of each term carry
complementary noise triples and are therefore built from independent sets of draws, ensuring that their
product is estimated without bias.

\section{Timings and sampling statistics of the HH sink}\label{app:hh-timings}

Here we present the timings and sampling statistics of the HH sink for the results presented in  Subsecs.~\ref{sec:stoch:results} and \ref{sec:stoch:scaling}, particularly 
for the measurements behind Figs.~\ref{fig:hh-cost-variance} and~\ref{fig:hh-projection-variance}. 
All timings mentioned here are wall-clock times of the nucleon sink contraction
with 32 OpenMP threads on a single node of Intel Xeon Platinum 8358 processor, for the exact contraction and for the HH sink with independent
site draws for each of the $120$ noise triples. For each number of draws we list the average numbers of
distinct site triples, $N_{\rm tri}$, and site pairs, $N_{\rm pair}$, in the sampling plans, which
determine the cost of the first two contraction steps in Eq.~\eqref{eq:cost_hh}, and the speed-up
predicted by the operation counts of Eqs.~\eqref{eq:cost_exact} and~\eqref{eq:cost_hh} for these plans. All timings listed below represent the average over $10$ independent runs. 

 Table~\ref{tab:hh-timings-81} lists the results on the $24^3$ ensemble with
three coarse sites per direction ($n_{\rm site} = 27$, i.e.\ $19\,683$ site triples and $729$ site pairs).
The exact sink contraction takes $156.8$~s. The variance ratio is that of the fitted ground-state
energy, Eq.~\eqref{eq:hh-variance-ratio}, from approximately 100 configurations.

\begin{table}[htb]
  \caption{
  Performance of HH estimator for nucleon sink contractions at $n_v=81$.
Columns report the mean numbers of distinct site triples and pairs
per sampling plan, wall-clock time $t_{\rm HH}$, measured speed-up
$t_{\rm exact}/t_{\rm HH}$ with $t_{\rm exact}=156.8$~s,
speed-up predicted from operation counts, variance ratio
$\sigma_{\rm HH}^2/\sigma_{\rm exact}^2$ of the fitted nucleon
energy [Eq.~\eqref{eq:hh-variance-ratio}].}
  \label{tab:hh-timings-81}
\vspace*{0.2in}  
  \begin{ruledtabular}
  \begin{tabular}{rrrrrrr}
    $N_{\rm hh}$ & $N_{\rm tri}$ & $N_{\rm pair}$ & $t_{\rm HH}$ [s] & speed-up & op.\ count
      & var.\ ratio \\
    \hline
     200 &  193 & 159 & 11.53 & 13.60 & 14.51 & 2.028  \\
     400 &  379 & 276 & 18.89 &  8.30 &  8.90 & 1.428 \\
     800 &  733 & 432 & 29.36 &  5.34 &  5.73 & 1.232  \\
    1200 &  1108  & 522  & 36.29 &  4.32 &  4.62   & 1.135  \\
    1600 & 1405 & 589 & 41.23 &  3.80 &  4.02 & 1.104  \\
    2400 & 2033 & 651 & 48.86 &  3.21 &  3.42 & 1.066  \\
    3200 & 2605 & 687 & 57.4\phantom{0} &  2.73 &  3.07 &  --   \\
  \end{tabular}
  \end{ruledtabular}
\end{table}

The measured speed-ups are found to be about $90\%$ of those predicted by the operation counts, the difference
being due to overheads common to both contractions. At $N_{\rm hh} = 1600$ the HH estimator
adds about $10\%$ to the variance of the fitted energy, and about $3.5\%$ to that of the correlator in
the range $t/a = 8$--$16$.

In Table~\ref{tab:hh-timings-192} we list the results of the study on the same
lattice with four coarse sites per direction ($n_{\rm site} = 64$, i.e.\ $262\,144$ site triples and
$4096$ site pairs), with $n_v = 192$, on a single configuration and for two sink time slices. Here the exact sink
contraction takes $51.6$~s.

\begin{table}[htb]
  \caption{As Table~\ref{tab:hh-timings-81}, for $n_v = 192$ (single configuration, two sink time
  slices, $t_{\rm exact} = 51.6$~s).}
  \label{tab:hh-timings-192}
\vspace*{0.2in}  
  \begin{ruledtabular}
  \begin{tabular}{rrrrrr}
    $N_{\rm hh}$ & $N_{\rm tri}$ & $N_{\rm pair}$ & $t_{\rm HH}$ [s] & speed-up & op.\ count \\
    \hline
      400 &   396 &  365 & 0.740 & 69.70 & 72.77 \\
      800 &   786 &  674 & 1.284 & 40.15 & 42.70 \\
     1600 &  1550 & 1174 & 2.194 & 23.51 & 25.29 \\
     3200 &  3034 & 1890 & 3.562 & 14.47 & 15.64 \\
     6400 &  5880 & 2743 & 5.385 &  9.58 & 10.35 \\
    12800 & 11211 & 3511 & 7.491 &  6.88 &  7.38 \\
  \end{tabular}
  \end{ruledtabular}
\end{table}

The measured speed-ups are again found to be around $90\%$ of the operation-count predictions. Compared with
$n_v = 81$, here a much smaller fraction of the site triples and pairs is drawn at a given $N_{\rm hh}$. At
$N_{\rm hh} = 3200$, about $1\%$ of the triples and $46\%$ of the pairs  are drawn for $n_v = 192$, against $13\%$ and $94\%$ for
$n_v = 81$. The second contraction step therefore saturates much later, which is why the speed-up at a given number of draws is so much larger.

At fixed localization scale and dilution, the operation-count
bounds imply an asymptotically increasing speedup if the number
of samples required at fixed precision grows linearly with $n_v$.
The present comparison demonstrates an increased gain with basis
dimension at fixed lattice volume; tests on larger volumes are
needed to establish the corresponding physical-volume scaling.

\section{Fit results for fast-wavelet-transform sparsening}
\label{app:fwt-fits}
We extract the lowest energies from the correlators presented in
Subsec.~\ref{sec:wavelet:results} using correlated single-state fits
at large Euclidean times, with the cosh or single-exponential forms
specified in Subsec.~\ref{sec:fwt-setup}.
In Tables~\ref{tab:fwt_fits_mesons} and~\ref{tab:fwt_fits_baryons}
we summarize the results for the full correlators ($s=0$) and those
coarse-grained using the fast wavelet transform at levels $s=1,2,3$,
with either fixed coarse grids or random sublattice shifts.
The energy difference $\Delta E = E - E_{\rm full}$ is evaluated
on common bootstrap samples, preserving the correlations between
the coarse-grained and full results. The reported
$\chi^2/\mathrm{dof}$ corresponds to the fit to the central
correlator values. The fit windows are $t/a\in[10-31]$ for the pion, $[10-20]$ for
the $\rho$ meson, and $[10-18]$ for the nucleon and two-nucleon
system. For the latter two channels at $s=3$, the window is
$[12-18]$.

\begin{table}[htbp]
\caption{Extracted energies in lattice units for the pion and the $\rho$ meson extracted using correlated single-state fits to the
full ($s=0$) and coarse-grained ($s=1,2,3$) correlators. Results are shown for a fixed coarse grid and also with random
sublattice shifts. Here $\Delta E = E - E_{\rm full}$.}
\vspace*{0.2in}
\label{tab:fwt_fits_mesons}
\begin{ruledtabular}
\begin{tabular}{clcccccc}
 & & \multicolumn{3}{c}{Fixed grid} & \multicolumn{3}{c}{Random shifts}\\
\cline{3-5}\cline{6-8}
$s$ & Scheme & $aE$ & $\chi^2/\mathrm{dof}$ & $a\Delta E\times10^{3}$ & $aE$ & $\chi^2/\mathrm{dof}$ & $a\Delta E\times10^{3}$\\
\hline
\multicolumn{8}{l}{$\pi$}\\
0 & Full     & 0.42065(52) & 1.06 & --          & 0.42065(52) & 1.06 & --\\
1 & Bior 4,4 & 0.42067(53) & 1.10 & 0.021(50)   & 0.42071(53) & 1.13 & 0.058(46)\\
  & Bior 5,5 & 0.42067(54) & 1.25 & 0.02(13)    & 0.42074(54) & 1.29 & 0.09(13)\\
  & Decimate & 0.42064(52) & 1.04 & $-$0.011(47)& 0.42064(53) & 1.09 & $-$0.016(42)\\
2 & Bior 4,4 & 0.42060(55) & 1.23 & $-$0.05(21) & 0.42098(54) & 1.31 & 0.33(19)\\
  & Bior 5,5 & 0.42031(54) & 1.14 & $-$0.34(28) & 0.42052(54) & 1.32 & $-$0.13(28)\\
  & Decimate & 0.41995(59) & 1.00 & $-$0.70(29) & 0.42061(54) & 1.09 & $-$0.04(31)\\
3 & Bior 4,4 & 0.42087(61) & 1.10 & 0.22(39)    & 0.42048(55) & 1.11 & $-$0.18(41)\\
  & Bior 5,5 & 0.42058(57) & 0.82 & $-$0.07(46) & 0.42038(56) & 1.10 & $-$0.27(45)\\
  & Decimate & 0.42130(75) & 1.75 & 0.65(67)    & 0.42073(67) & 1.25 & 0.08(57)\\
\hline
\multicolumn{8}{l}{$\rho$}\\
0 & Full     & 0.6437(45) & 0.95 & --           & 0.6437(45) & 0.95 & --\\
1 & Bior 4,4 & 0.6437(45) & 0.88 & 0.00(22)     & 0.6435(45) & 0.88 & $-$0.13(24)\\
  & Bior 5,5 & 0.6433(46) & 0.75 & $-$0.33(57)  & 0.6432(45) & 0.74 & $-$0.48(60)\\
  & Decimate & 0.6445(45) & 0.87 & 0.83(22)     & 0.6435(45) & 0.96 & $-$0.20(30)\\
2 & Bior 4,4 & 0.6430(47) & 0.95 & $-$0.65(90)  & 0.6435(47) & 0.80 & $-$0.20(88)\\
  & Bior 5,5 & 0.6419(50) & 0.61 & $-$1.8(1.7)  & 0.6425(49) & 0.56 & $-$1.2(1.7)\\
  & Decimate & 0.6478(48) & 1.45 & 4.1(1.6)     & 0.6468(47) & 0.48 & 3.1(1.5)\\
3 & Bior 4,4 & 0.6492(58) & 0.62 & 5.5(3.2)     & 0.6409(57) & 0.33 & $-$2.8(3.4)\\
  & Bior 5,5 & 0.6441(68) & 0.45 & 0.5(4.2)     & 0.6385(65) & 0.97 & $-$5.2(4.2)\\
  & Decimate & 0.6965(90) & 3.92 & 52.8(6.8)    & 0.6577(77) & 0.65 & 14.0(6.0)\\
\end{tabular}
\end{ruledtabular}
\end{table}

\begin{table}[htbp]
\caption{Same as Table~\ref{tab:fwt_fits_mesons}, for the nucleon ($N$) and the two-nucleon system ($NP$).}
\vspace*{0.2in}
\label{tab:fwt_fits_baryons}
\begin{ruledtabular}
\begin{tabular}{clcccccc}
 & & \multicolumn{3}{c}{Fixed grid} & \multicolumn{3}{c}{Random shifts}\\
\cline{3-5}\cline{6-8}
$s$ & Scheme & $aE$ & $\chi^2/\mathrm{dof}$ & $a\Delta E\times10^{3}$ & $aE$ & $\chi^2/\mathrm{dof}$ & $a\Delta E\times10^{3}$\\
\hline
\multicolumn{8}{l}{$N$}\\
0 & Full     & 0.9238(80) & 0.44 & --           & 0.9238(80) & 0.44 & --\\
1 & Bior 4,4 & 0.9235(81) & 0.48 & $-$0.29(55)  & 0.9237(81) & 0.45 & $-$0.14(57)\\
  & Bior 5,5 & 0.9224(83) & 0.55 & $-$1.4(1.5)  & 0.9224(83) & 0.57 & $-$1.4(1.5)\\
  & Decimate & 0.9247(80) & 0.43 & 0.89(45)     & 0.9243(81) & 0.51 & 0.49(53)\\
2 & Bior 4,4 & 0.9237(84) & 0.77 & $-$0.1(2.1)  & 0.9227(86) & 0.44 & $-$1.1(2.4)\\
  & Bior 5,5 & 0.9221(89) & 0.49 & $-$1.7(3.6)  & 0.9204(90) & 0.52 & $-$3.4(3.9)\\
  & Decimate & 0.9290(87) & 1.10 & 5.2(3.8)     & 0.9256(91) & 1.21 & 1.8(3.8)\\
3 & Bior 4,4 & 0.928(18)  & 1.23 & 4(17)        & 0.931(17)  & 0.36 & 7(15)\\
  & Bior 5,5 & 0.919(18)  & 0.70 & $-$5(18)     & 0.926(18)  & 0.41 & 2(17)\\
  & Decimate & 0.956(28)  & 3.31 & 32(26)       & 0.933(20)  & 1.33 & 9(19)\\
\hline
\multicolumn{8}{l}{$NP$}\\
0 & Full     & 2.003(13) & 0.69 & --            & 2.003(13) & 0.69 & --\\
1 & Bior 4,4 & 2.002(14) & 0.70 & $-$0.94(79)   & 2.002(14) & 0.72 & $-$1.52(92)\\
  & Bior 5,5 & 1.999(14) & 0.71 & $-$2.8(2.4)   & 2.001(14) & 0.77 & $-$1.8(2.5)\\
  & Decimate & 2.004(14) & 0.72 & 1.3(1.3)      & 2.004(14) & 0.70 & 0.8(1.1)\\
2 & Bior 4,4 & 1.997(14) & 0.76 & $-$6.1(4.4)   & 1.993(14) & 0.72 & $-$8.8(9.3)\\
  & Bior 5,5 & 1.998(14) & 0.79 & $-$5.7(7.0)  & 1.991(15) & 0.96 & $-$10.8(9.8)\\
  & Decimate & 2.009(15) & 0.91 & 6.2(7.3)      & 1.994(16) & 1.20 & $-$9.8(11.9)\\
3 & Bior 4,4 & 1.992(34) & 1.45 & $-$21(32)     & 1.982(27) & 0.71 & $-$21(24)\\
  & Bior 5,5 & 1.989(30) & 1.44 & $-$40(29)     & 1.991(32) & 1.23 & $-$11(29)\\
  & Decimate & 2.021(68) & 2.95 &   48(64)     & 1.973(33) & 0.81 & $-$31(31)\\
\end{tabular}
\end{ruledtabular}
\end{table}

\clearpage
\bibliography{wavelet}

\begin{thebibliography}{61}%
\makeatletter
\providecommand \@ifxundefined [1]{%
 \@ifx{#1\undefined}
}%
\providecommand \@ifnum [1]{%
 \ifnum #1\expandafter \@firstoftwo
 \else \expandafter \@secondoftwo
 \fi
}%
\providecommand \@ifx [1]{%
 \ifx #1\expandafter \@firstoftwo
 \else \expandafter \@secondoftwo
 \fi
}%
\providecommand \natexlab [1]{#1}%
\providecommand \enquote  [1]{``#1''}%
\providecommand \bibnamefont  [1]{#1}%
\providecommand \bibfnamefont [1]{#1}%
\providecommand \citenamefont [1]{#1}%
\providecommand \href@noop [0]{\@secondoftwo}%
\providecommand \href [0]{\begingroup \@sanitize@url \@href}%
\providecommand \@href[1]{\@@startlink{#1}\@@href}%
\providecommand \@@href[1]{\endgroup#1\@@endlink}%
\providecommand \@sanitize@url [0]{\catcode `\\12\catcode `\$12\catcode
  `\&12\catcode `\#12\catcode `\^12\catcode `\_12\catcode `\%12\relax}%
\providecommand \@@startlink[1]{}%
\providecommand \@@endlink[0]{}%
\providecommand \url  [0]{\begingroup\@sanitize@url \@url }%
\providecommand \@url [1]{\endgroup\@href {#1}{\urlprefix }}%
\providecommand \urlprefix  [0]{URL }%
\providecommand \Eprint [0]{\href }%
\providecommand \doibase [0]{https://doi.org/}%
\providecommand \selectlanguage [0]{\@gobble}%
\providecommand \bibinfo  [0]{\@secondoftwo}%
\providecommand \bibfield  [0]{\@secondoftwo}%
\providecommand \translation [1]{[#1]}%
\providecommand \BibitemOpen [0]{}%
\providecommand \bibitemStop [0]{}%
\providecommand \bibitemNoStop [0]{.\EOS\space}%
\providecommand \EOS [0]{\spacefactor3000\relax}%
\providecommand \BibitemShut  [1]{\csname bibitem#1\endcsname}%
\let\auto@bib@innerbib\@empty
\bibitem [{\citenamefont {Michael}(1985)}]{MICHAEL198558}%
  \BibitemOpen
  \bibfield  {author} {\bibinfo {author} {\bibfnamefont {C.}~\bibnamefont
  {Michael}},\ }\bibfield  {title} {\bibinfo {title} {Adjoint sources in
  lattice gauge theory},\ }\href
  {https://doi.org/https://doi.org/10.1016/0550-3213(85)90297-4} {\bibfield
  {journal} {\bibinfo  {journal} {Nuclear Physics B}\ }\textbf {\bibinfo
  {volume} {259}},\ \bibinfo {pages} {58} (\bibinfo {year} {1985})}\BibitemShut
  {NoStop}%
\bibitem [{\citenamefont {L\"uscher}\ and\ \citenamefont
  {Wolff}(1990)}]{luscherwolff1990}%
  \BibitemOpen
  \bibfield  {author} {\bibinfo {author} {\bibfnamefont {M.}~\bibnamefont
  {L\"uscher}}\ and\ \bibinfo {author} {\bibfnamefont {U.}~\bibnamefont
  {Wolff}},\ }\bibfield  {title} {\bibinfo {title} {How to calculate the
  elastic scattering matrix in two-dimensional quantum field theories by
  numerical simulation},\ }\href@noop {} {\bibfield  {journal} {\bibinfo
  {journal} {Nuclear Physics B}\ }\textbf {\bibinfo {volume} {339}},\ \bibinfo
  {pages} {222} (\bibinfo {year} {1990})}\BibitemShut {NoStop}%
\bibitem [{\citenamefont {Gattringer}\ and\ \citenamefont
  {Lang}(2009)}]{gattringerlang2009}%
  \BibitemOpen
  \bibfield  {author} {\bibinfo {author} {\bibfnamefont {C.}~\bibnamefont
  {Gattringer}}\ and\ \bibinfo {author} {\bibfnamefont {C.~B.}\ \bibnamefont
  {Lang}},\ }\href@noop {} {\emph {\bibinfo {title} {Quantum Chromodynamics on
  the Lattice: An Introductory Presentation}}}\ (\bibinfo  {publisher}
  {Springer},\ \bibinfo {year} {2009})\BibitemShut {NoStop}%
\bibitem [{\citenamefont {Kronfeld}\ \emph {et~al.}(2022)\citenamefont
  {Kronfeld} \emph {et~al.}}]{kronfeld2022}%
  \BibitemOpen
  \bibfield  {author} {\bibinfo {author} {\bibfnamefont {A.~S.}\ \bibnamefont
  {Kronfeld}} \emph {et~al.},\ }\href@noop {} {\bibinfo {title} {Lattice {QCD}
  and particle physics}} (\bibinfo {year} {2022}),\ \Eprint
  {https://arxiv.org/abs/arXiv:2207.07641} {arXiv:2207.07641} \BibitemShut
  {NoStop}%
\bibitem [{\citenamefont {Peardon}\ \emph {et~al.}(2009)\citenamefont
  {Peardon}, \citenamefont {Bulava}, \citenamefont {Foley}, \citenamefont
  {Morningstar}, \citenamefont {Dudek}, \citenamefont {Edwards}, \citenamefont
  {Jo{\'o}}, \citenamefont {Lin}, \citenamefont {Richards},\ and\ \citenamefont
  {Juge}}]{hadronspectrum2009}%
  \BibitemOpen
  \bibfield  {author} {\bibinfo {author} {\bibfnamefont {M.}~\bibnamefont
  {Peardon}}, \bibinfo {author} {\bibfnamefont {J.}~\bibnamefont {Bulava}},
  \bibinfo {author} {\bibfnamefont {J.}~\bibnamefont {Foley}}, \bibinfo
  {author} {\bibfnamefont {C.}~\bibnamefont {Morningstar}}, \bibinfo {author}
  {\bibfnamefont {J.}~\bibnamefont {Dudek}}, \bibinfo {author} {\bibfnamefont
  {R.~G.}\ \bibnamefont {Edwards}}, \bibinfo {author} {\bibfnamefont
  {B.}~\bibnamefont {Jo{\'o}}}, \bibinfo {author} {\bibfnamefont {H.-W.}\
  \bibnamefont {Lin}}, \bibinfo {author} {\bibfnamefont {D.~G.}\ \bibnamefont
  {Richards}},\ and\ \bibinfo {author} {\bibfnamefont {K.~J.}\ \bibnamefont
  {Juge}} (\bibinfo {collaboration} {Hadron Spectrum Collaboration}),\
  }\bibfield  {title} {\bibinfo {title} {Novel quark-field creation operator
  construction for hadronic physics in lattice qcd},\ }\href
  {https://doi.org/10.1103/PhysRevD.80.054506} {\bibfield  {journal} {\bibinfo
  {journal} {Phys. Rev. D}\ }\textbf {\bibinfo {volume} {80}},\ \bibinfo
  {pages} {054506} (\bibinfo {year} {2009})}\BibitemShut {NoStop}%
\bibitem [{\citenamefont {Morningstar}\ \emph {et~al.}(2011)\citenamefont
  {Morningstar}, \citenamefont {Bulava}, \citenamefont {Foley}, \citenamefont
  {Juge}, \citenamefont {Lenkner}, \citenamefont {Peardon},\ and\ \citenamefont
  {Wong}}]{PhysRevD.83.114505}%
  \BibitemOpen
  \bibfield  {author} {\bibinfo {author} {\bibfnamefont {C.}~\bibnamefont
  {Morningstar}}, \bibinfo {author} {\bibfnamefont {J.}~\bibnamefont {Bulava}},
  \bibinfo {author} {\bibfnamefont {J.}~\bibnamefont {Foley}}, \bibinfo
  {author} {\bibfnamefont {K.~J.}\ \bibnamefont {Juge}}, \bibinfo {author}
  {\bibfnamefont {D.}~\bibnamefont {Lenkner}}, \bibinfo {author} {\bibfnamefont
  {M.}~\bibnamefont {Peardon}},\ and\ \bibinfo {author} {\bibfnamefont {C.~H.}\
  \bibnamefont {Wong}},\ }\bibfield  {title} {\bibinfo {title} {Improved
  stochastic estimation of quark propagation with laplacian heaviside smearing
  in lattice qcd},\ }\href {https://doi.org/10.1103/PhysRevD.83.114505}
  {\bibfield  {journal} {\bibinfo  {journal} {Phys. Rev. D}\ }\textbf {\bibinfo
  {volume} {83}},\ \bibinfo {pages} {114505} (\bibinfo {year}
  {2011})}\BibitemShut {NoStop}%
\bibitem [{\citenamefont {Knechtli}\ \emph {et~al.}(2022)\citenamefont
  {Knechtli}, \citenamefont {Korzec}, \citenamefont {Peardon},\ and\
  \citenamefont {Urrea-Ni{\~n}o}}]{Knechtli:2022bji}%
  \BibitemOpen
  \bibfield  {author} {\bibinfo {author} {\bibfnamefont {F.}~\bibnamefont
  {Knechtli}}, \bibinfo {author} {\bibfnamefont {T.}~\bibnamefont {Korzec}},
  \bibinfo {author} {\bibfnamefont {M.}~\bibnamefont {Peardon}},\ and\ \bibinfo
  {author} {\bibfnamefont {J.~A.}\ \bibnamefont {Urrea-Ni{\~n}o}},\ }\bibfield
  {title} {\bibinfo {title} {Optimizing creation operators for charmonium
  spectroscopy on the lattice},\ }\href
  {https://doi.org/10.1103/PhysRevD.106.034501} {\bibfield  {journal} {\bibinfo
   {journal} {Phys. Rev. D}\ }\textbf {\bibinfo {volume} {106}},\ \bibinfo
  {pages} {034501} (\bibinfo {year} {2022})}\BibitemShut {NoStop}%
\bibitem [{\citenamefont {H{\"o}llwieser}\ \emph {et~al.}(2023)\citenamefont
  {H{\"o}llwieser}, \citenamefont {Knechtli}, \citenamefont {Korzec},
  \citenamefont {Peardon},\ and\ \citenamefont
  {Urrea-Ni{\~n}o}}]{Hollwieser:2022doy}%
  \BibitemOpen
  \bibfield  {author} {\bibinfo {author} {\bibfnamefont {R.}~\bibnamefont
  {H{\"o}llwieser}}, \bibinfo {author} {\bibfnamefont {F.}~\bibnamefont
  {Knechtli}}, \bibinfo {author} {\bibfnamefont {T.}~\bibnamefont {Korzec}},
  \bibinfo {author} {\bibfnamefont {M.}~\bibnamefont {Peardon}},\ and\ \bibinfo
  {author} {\bibfnamefont {J.~A.}\ \bibnamefont {Urrea-Ni{\~n}o}},\ }\bibfield
  {title} {\bibinfo {title} {Constructing static quark-antiquark creation
  operators from laplacian eigenmodes},\ }\href
  {https://doi.org/10.1103/PhysRevD.107.034511} {\bibfield  {journal} {\bibinfo
   {journal} {Phys. Rev. D}\ }\textbf {\bibinfo {volume} {107}},\ \bibinfo
  {pages} {034511} (\bibinfo {year} {2023})}\BibitemShut {NoStop}%
\bibitem [{\citenamefont {Bali}\ \emph {et~al.}(2016)\citenamefont {Bali},
  \citenamefont {Lang}, \citenamefont {Musch},\ and\ \citenamefont
  {Sch\"afer}}]{PhysRevD.93.094515}%
  \BibitemOpen
  \bibfield  {author} {\bibinfo {author} {\bibfnamefont {G.~S.}\ \bibnamefont
  {Bali}}, \bibinfo {author} {\bibfnamefont {B.}~\bibnamefont {Lang}}, \bibinfo
  {author} {\bibfnamefont {B.~U.}\ \bibnamefont {Musch}},\ and\ \bibinfo
  {author} {\bibfnamefont {A.}~\bibnamefont {Sch\"afer}} (\bibinfo
  {collaboration} {RQCD Collaboration}),\ }\bibfield  {title} {\bibinfo {title}
  {Novel quark smearing for hadrons with high momenta in lattice qcd},\ }\href
  {https://doi.org/10.1103/PhysRevD.93.094515} {\bibfield  {journal} {\bibinfo
  {journal} {Phys. Rev. D}\ }\textbf {\bibinfo {volume} {93}},\ \bibinfo
  {pages} {094515} (\bibinfo {year} {2016})}\BibitemShut {NoStop}%
\bibitem [{\citenamefont {Egerer}\ \emph {et~al.}(2021)\citenamefont {Egerer},
  \citenamefont {Edwards}, \citenamefont {Orginos},\ and\ \citenamefont
  {Richards}}]{PhysRevD.103.034502}%
  \BibitemOpen
  \bibfield  {author} {\bibinfo {author} {\bibfnamefont {C.}~\bibnamefont
  {Egerer}}, \bibinfo {author} {\bibfnamefont {R.~G.}\ \bibnamefont {Edwards}},
  \bibinfo {author} {\bibfnamefont {K.}~\bibnamefont {Orginos}},\ and\ \bibinfo
  {author} {\bibfnamefont {D.~G.}\ \bibnamefont {Richards}} (\bibinfo
  {collaboration} {On behalf of the HadStruc Collaboration}),\ }\bibfield
  {title} {\bibinfo {title} {Distillation at high momentum},\ }\href
  {https://doi.org/10.1103/PhysRevD.103.034502} {\bibfield  {journal} {\bibinfo
   {journal} {Phys. Rev. D}\ }\textbf {\bibinfo {volume} {103}},\ \bibinfo
  {pages} {034502} (\bibinfo {year} {2021})}\BibitemShut {NoStop}%
\bibitem [{\citenamefont {Lang}\ \emph {et~al.}(2025)\citenamefont {Lang},
  \citenamefont {Edwards},\ and\ \citenamefont {Peardon}}]{mv6l-m8rq}%
  \BibitemOpen
  \bibfield  {author} {\bibinfo {author} {\bibfnamefont {N.}~\bibnamefont
  {Lang}}, \bibinfo {author} {\bibfnamefont {R.~G.}\ \bibnamefont {Edwards}},\
  and\ \bibinfo {author} {\bibfnamefont {M.~J.}\ \bibnamefont {Peardon}}
  (\bibinfo {collaboration} {for the Hadron Spectrum Collaboration}),\
  }\bibfield  {title} {\bibinfo {title} {Optimizing stochastic algorithms for
  hadron correlation function computations in lattice qcd using a localized
  distillation basis},\ }\href {https://doi.org/10.1103/mv6l-m8rq} {\bibfield
  {journal} {\bibinfo  {journal} {Phys. Rev. D}\ }\textbf {\bibinfo {volume}
  {111}},\ \bibinfo {pages} {114510} (\bibinfo {year} {2025})}\BibitemShut
  {NoStop}%
\bibitem [{\citenamefont {Detmold}\ \emph {et~al.}(2021)\citenamefont
  {Detmold}, \citenamefont {Murphy}, \citenamefont {Pochinsky}, \citenamefont
  {Savage}, \citenamefont {Shanahan},\ and\ \citenamefont
  {Wagman}}]{Detmold_2021}%
  \BibitemOpen
  \bibfield  {author} {\bibinfo {author} {\bibfnamefont {W.}~\bibnamefont
  {Detmold}}, \bibinfo {author} {\bibfnamefont {D.~J.}\ \bibnamefont {Murphy}},
  \bibinfo {author} {\bibfnamefont {A.~V.}\ \bibnamefont {Pochinsky}}, \bibinfo
  {author} {\bibfnamefont {M.~J.}\ \bibnamefont {Savage}}, \bibinfo {author}
  {\bibfnamefont {P.~E.}\ \bibnamefont {Shanahan}},\ and\ \bibinfo {author}
  {\bibfnamefont {M.~L.}\ \bibnamefont {Wagman}},\ }\bibfield  {title}
  {\bibinfo {title} {Sparsening algorithm for multihadron lattice qcd
  correlation functions},\ }\href {https://doi.org/10.1103/PhysRevD.104.034502}
  {\bibfield  {journal} {\bibinfo  {journal} {Phys. Rev. D}\ }\textbf {\bibinfo
  {volume} {104}},\ \bibinfo {pages} {034502} (\bibinfo {year}
  {2021})}\BibitemShut {NoStop}%
\bibitem [{\citenamefont {Christian}\ and\ \citenamefont
  {Detmold}(2025)}]{xfmp-c2d4}%
  \BibitemOpen
  \bibfield  {author} {\bibinfo {author} {\bibfnamefont {S.}~\bibnamefont
  {Christian}}\ and\ \bibinfo {author} {\bibfnamefont {W.}~\bibnamefont
  {Detmold}},\ }\bibfield  {title} {\bibinfo {title} {Aspects of propagator
  sparsening in lattice qcd},\ }\href {https://doi.org/10.1103/xfmp-c2d4}
  {\bibfield  {journal} {\bibinfo  {journal} {Phys. Rev. D}\ }\textbf {\bibinfo
  {volume} {111}},\ \bibinfo {pages} {114514} (\bibinfo {year}
  {2025})}\BibitemShut {NoStop}%
\bibitem [{\citenamefont {Stump}\ and\ \citenamefont
  {Green}(2026)}]{khhk-qw9n}%
  \BibitemOpen
  \bibfield  {author} {\bibinfo {author} {\bibfnamefont {A.}~\bibnamefont
  {Stump}}\ and\ \bibinfo {author} {\bibfnamefont {J.~R.}\ \bibnamefont
  {Green}},\ }\bibfield  {title} {\bibinfo {title} {Position-space sampling for
  local multiquark operators in lattice qcd using distillation and the
  importance of tetraquark operators for ${T}_{cc}(3875{)}^{+}$},\ }\href
  {https://doi.org/10.1103/khhk-qw9n} {\bibfield  {journal} {\bibinfo
  {journal} {Phys. Rev. D}\ }\textbf {\bibinfo {volume} {113}},\ \bibinfo
  {pages} {034513} (\bibinfo {year} {2026})}\BibitemShut {NoStop}%
\bibitem [{\citenamefont {Daubechies}(1988)}]{daubechies1988}%
  \BibitemOpen
  \bibfield  {author} {\bibinfo {author} {\bibfnamefont {I.}~\bibnamefont
  {Daubechies}},\ }\bibfield  {title} {\bibinfo {title} {Orthonormal bases of
  compactly supported wavelets},\ }\href
  {https://doi.org/https://doi.org/10.1002/cpa.3160410705} {\bibfield
  {journal} {\bibinfo  {journal} {Communications on Pure and Applied
  Mathematics}\ }\textbf {\bibinfo {volume} {41}},\ \bibinfo {pages} {909}
  (\bibinfo {year} {1988})}\BibitemShut {NoStop}%
\bibitem [{\citenamefont {Mallat}(1989{\natexlab{a}})}]{mallat1989}%
  \BibitemOpen
  \bibfield  {author} {\bibinfo {author} {\bibfnamefont {S.}~\bibnamefont
  {Mallat}},\ }\bibfield  {title} {\bibinfo {title} {A theory for
  multiresolution signal decomposition: the wavelet representation},\ }\href
  {https://doi.org/10.1109/34.192463} {\bibfield  {journal} {\bibinfo
  {journal} {IEEE Transactions on Pattern Analysis and Machine Intelligence}\
  }\textbf {\bibinfo {volume} {11}},\ \bibinfo {pages} {674} (\bibinfo {year}
  {1989}{\natexlab{a}})}\BibitemShut {NoStop}%
\bibitem [{\citenamefont {Mallat}(2009)}]{mallat2009}%
  \BibitemOpen
  \bibfield  {author} {\bibinfo {author} {\bibfnamefont {S.}~\bibnamefont
  {Mallat}},\ }\href@noop {} {\emph {\bibinfo {title} {A Wavelet Tour of Signal
  Processing}}},\ \bibinfo {edition} {3rd}\ ed.\ (\bibinfo  {publisher}
  {Academic Press},\ \bibinfo {address} {Boston},\ \bibinfo {year}
  {2009})\BibitemShut {NoStop}%
\bibitem [{\citenamefont {Wilson}(1965)}]{wilson1965}%
  \BibitemOpen
  \bibfield  {author} {\bibinfo {author} {\bibfnamefont {K.~G.}\ \bibnamefont
  {Wilson}},\ }\bibfield  {title} {\bibinfo {title} {Model hamiltonians for
  local quantum field theory},\ }\href
  {https://doi.org/10.1103/PhysRev.140.B445} {\bibfield  {journal} {\bibinfo
  {journal} {Phys. Rev.}\ }\textbf {\bibinfo {volume} {140}},\ \bibinfo {pages}
  {B445} (\bibinfo {year} {1965})}\BibitemShut {NoStop}%
\bibitem [{\citenamefont {Bulut}\ and\ \citenamefont
  {Polyzou}(2013)}]{bulutpolyzou2013}%
  \BibitemOpen
  \bibfield  {author} {\bibinfo {author} {\bibfnamefont {F.}~\bibnamefont
  {Bulut}}\ and\ \bibinfo {author} {\bibfnamefont {W.~N.}\ \bibnamefont
  {Polyzou}},\ }\bibfield  {title} {\bibinfo {title} {Wavelets in field
  theory},\ }\href {https://doi.org/10.1103/PhysRevD.87.116011} {\bibfield
  {journal} {\bibinfo  {journal} {Phys. Rev. D}\ }\textbf {\bibinfo {volume}
  {87}},\ \bibinfo {pages} {116011} (\bibinfo {year} {2013})}\BibitemShut
  {NoStop}%
\bibitem [{\citenamefont {Evenbly}\ and\ \citenamefont
  {White}(2016)}]{PhysRevLett.116.140403}%
  \BibitemOpen
  \bibfield  {author} {\bibinfo {author} {\bibfnamefont {G.}~\bibnamefont
  {Evenbly}}\ and\ \bibinfo {author} {\bibfnamefont {S.~R.}\ \bibnamefont
  {White}},\ }\bibfield  {title} {\bibinfo {title} {Entanglement
  renormalization and wavelets},\ }\href
  {https://doi.org/10.1103/PhysRevLett.116.140403} {\bibfield  {journal}
  {\bibinfo  {journal} {Phys. Rev. Lett.}\ }\textbf {\bibinfo {volume} {116}},\
  \bibinfo {pages} {140403} (\bibinfo {year} {2016})}\BibitemShut {NoStop}%
\bibitem [{\citenamefont {George}\ \emph {et~al.}(2022)\citenamefont {George},
  \citenamefont {Sanders}, \citenamefont {Bagherimehrab}, \citenamefont
  {Sanders},\ and\ \citenamefont {Brennen}}]{george2022}%
  \BibitemOpen
  \bibfield  {author} {\bibinfo {author} {\bibfnamefont {D.~J.}\ \bibnamefont
  {George}}, \bibinfo {author} {\bibfnamefont {Y.~R.}\ \bibnamefont {Sanders}},
  \bibinfo {author} {\bibfnamefont {M.}~\bibnamefont {Bagherimehrab}}, \bibinfo
  {author} {\bibfnamefont {B.~C.}\ \bibnamefont {Sanders}},\ and\ \bibinfo
  {author} {\bibfnamefont {G.~K.}\ \bibnamefont {Brennen}},\ }\bibfield
  {title} {\bibinfo {title} {Entanglement in quantum field theory via wavelet
  representations},\ }\href {https://doi.org/10.1103/PhysRevD.106.036025}
  {\bibfield  {journal} {\bibinfo  {journal} {Phys. Rev. D}\ }\textbf {\bibinfo
  {volume} {106}},\ \bibinfo {pages} {036025} (\bibinfo {year}
  {2022})}\BibitemShut {NoStop}%
\bibitem [{\citenamefont {Basak}\ \emph
  {et~al.}(2026{\natexlab{a}})\citenamefont {Basak}, \citenamefont
  {Chakraborty}, \citenamefont {Mathur},\ and\ \citenamefont
  {Ratabole}}]{basak2026a}%
  \BibitemOpen
  \bibfield  {author} {\bibinfo {author} {\bibfnamefont {M.}~\bibnamefont
  {Basak}}, \bibinfo {author} {\bibfnamefont {D.}~\bibnamefont {Chakraborty}},
  \bibinfo {author} {\bibfnamefont {N.}~\bibnamefont {Mathur}},\ and\ \bibinfo
  {author} {\bibfnamefont {R.}~\bibnamefont {Ratabole}},\ }\bibfield  {title}
  {\bibinfo {title} {Hamiltonian formulation of the ($1+1$)-dimensional
  ${\ensuremath{\phi}}^{4}$ theory in a momentum-space daubechies wavelet
  basis},\ }\href {https://doi.org/10.1103/74c7-y1gp} {\bibfield  {journal}
  {\bibinfo  {journal} {Phys. Rev. D}\ }\textbf {\bibinfo {volume} {113}},\
  \bibinfo {pages} {085021} (\bibinfo {year} {2026}{\natexlab{a}})}\BibitemShut
  {NoStop}%
\bibitem [{\citenamefont {Basak}\ \emph
  {et~al.}(2026{\natexlab{b}})\citenamefont {Basak}, \citenamefont
  {Chakraborty},\ and\ \citenamefont {Mathur}}]{basak2026b}%
  \BibitemOpen
  \bibfield  {author} {\bibinfo {author} {\bibfnamefont {M.}~\bibnamefont
  {Basak}}, \bibinfo {author} {\bibfnamefont {D.}~\bibnamefont {Chakraborty}},\
  and\ \bibinfo {author} {\bibfnamefont {N.}~\bibnamefont {Mathur}},\
  }\bibfield  {title} {\bibinfo {title} {Efficient wavelet-based hamiltonian
  formulation of quantum field theories using flow equations},\ }\href
  {https://doi.org/10.1103/krjm-7h9c} {\bibfield  {journal} {\bibinfo
  {journal} {Phys. Rev. D}\ }\textbf {\bibinfo {volume} {114}},\ \bibinfo
  {pages} {014502} (\bibinfo {year} {2026}{\natexlab{b}})}\BibitemShut
  {NoStop}%
\bibitem [{\citenamefont {Federbush}(1995)}]{10.1143/PTP.94.1135}%
  \BibitemOpen
  \bibfield  {author} {\bibinfo {author} {\bibfnamefont {P.}~\bibnamefont
  {Federbush}},\ }\bibfield  {title} {\bibinfo {title} {A new formulation and
  regularization of gauge theories using a non-linear wavelet expansion},\
  }\href {https://doi.org/10.1143/PTP.94.1135} {\bibfield  {journal} {\bibinfo
  {journal} {Progress of Theoretical Physics}\ }\textbf {\bibinfo {volume}
  {94}},\ \bibinfo {pages} {1135} (\bibinfo {year} {1995})},\ \Eprint
  {https://arxiv.org/abs/https://academic.oup.com/ptp/article-pdf/94/6/1135/5441674/94-6-1135.pdf}
  {https://academic.oup.com/ptp/article-pdf/94/6/1135/5441674/94-6-1135.pdf}
  \BibitemShut {NoStop}%
\bibitem [{\citenamefont {Neuberger}(2018)}]{neuberger2018}%
  \BibitemOpen
  \bibfield  {author} {\bibinfo {author} {\bibfnamefont {H.}~\bibnamefont
  {Neuberger}},\ }\bibfield  {title} {\bibinfo {title} {Wavelets and lattice
  field theory},\ }\href@noop {} {\bibfield  {journal} {\bibinfo  {journal}
  {EPJ Web of Conferences}\ }\textbf {\bibinfo {volume} {175}},\ \bibinfo
  {pages} {11002} (\bibinfo {year} {2018})},\ \Eprint
  {https://arxiv.org/abs/arXiv:1707.09623} {arXiv:1707.09623} \BibitemShut
  {NoStop}%
\bibitem [{\citenamefont {Kaplan}\ and\ \citenamefont
  {Tilloy}(2026)}]{kaplantilloy2026}%
  \BibitemOpen
  \bibfield  {author} {\bibinfo {author} {\bibfnamefont {M.}~\bibnamefont
  {Kaplan}}\ and\ \bibinfo {author} {\bibfnamefont {A.}~\bibnamefont
  {Tilloy}},\ }\href@noop {} {\bibinfo {title} {Wavelet matrix product states
  for quantum fields}} (\bibinfo {year} {2026}),\ \Eprint
  {https://arxiv.org/abs/arXiv:2606.23823} {arXiv:2606.23823} \BibitemShut
  {NoStop}%
\bibitem [{\citenamefont {Lüscher}(2007)}]{luscher2007}%
  \BibitemOpen
  \bibfield  {author} {\bibinfo {author} {\bibfnamefont {M.}~\bibnamefont
  {Lüscher}},\ }\bibfield  {title} {\bibinfo {title} {Local coherence and
  deflation of the low quark modes in lattice qcd},\ }\href
  {https://doi.org/10.1088/1126-6708/2007/07/081} {\bibfield  {journal}
  {\bibinfo  {journal} {Journal of High Energy Physics}\ }\textbf {\bibinfo
  {volume} {2007}},\ \bibinfo {pages} {081} (\bibinfo {year}
  {2007})}\BibitemShut {NoStop}%
\bibitem [{\citenamefont {Babich}\ \emph {et~al.}(2010)\citenamefont {Babich},
  \citenamefont {Brannick}, \citenamefont {Brower}, \citenamefont {Clark},
  \citenamefont {Manteuffel}, \citenamefont {McCormick}, \citenamefont
  {Osborn},\ and\ \citenamefont {Rebbi}}]{babich2010}%
  \BibitemOpen
  \bibfield  {author} {\bibinfo {author} {\bibfnamefont {R.}~\bibnamefont
  {Babich}}, \bibinfo {author} {\bibfnamefont {J.}~\bibnamefont {Brannick}},
  \bibinfo {author} {\bibfnamefont {R.~C.}\ \bibnamefont {Brower}}, \bibinfo
  {author} {\bibfnamefont {M.~A.}\ \bibnamefont {Clark}}, \bibinfo {author}
  {\bibfnamefont {T.~A.}\ \bibnamefont {Manteuffel}}, \bibinfo {author}
  {\bibfnamefont {S.~F.}\ \bibnamefont {McCormick}}, \bibinfo {author}
  {\bibfnamefont {J.~C.}\ \bibnamefont {Osborn}},\ and\ \bibinfo {author}
  {\bibfnamefont {C.}~\bibnamefont {Rebbi}},\ }\bibfield  {title} {\bibinfo
  {title} {Adaptive multigrid algorithm for the lattice wilson-dirac
  operator},\ }\href {https://doi.org/10.1103/PhysRevLett.105.201602}
  {\bibfield  {journal} {\bibinfo  {journal} {Phys. Rev. Lett.}\ }\textbf
  {\bibinfo {volume} {105}},\ \bibinfo {pages} {201602} (\bibinfo {year}
  {2010})}\BibitemShut {NoStop}%
\bibitem [{\citenamefont {Cohen}\ \emph {et~al.}(1992)\citenamefont {Cohen},
  \citenamefont {Daubechies},\ and\ \citenamefont {Feauveau}}]{cdf1992}%
  \BibitemOpen
  \bibfield  {author} {\bibinfo {author} {\bibfnamefont {A.}~\bibnamefont
  {Cohen}}, \bibinfo {author} {\bibfnamefont {I.}~\bibnamefont {Daubechies}},\
  and\ \bibinfo {author} {\bibfnamefont {J.-C.}\ \bibnamefont {Feauveau}},\
  }\bibfield  {title} {\bibinfo {title} {Biorthogonal bases of compactly
  supported wavelets},\ }\href {https://doi.org/10.1002/cpa.3160450502}
  {\bibfield  {journal} {\bibinfo  {journal} {Communications on Pure and
  Applied Mathematics}\ }\textbf {\bibinfo {volume} {45}},\ \bibinfo {pages}
  {485} (\bibinfo {year} {1992})}\BibitemShut {NoStop}%
\bibitem [{\citenamefont {L{\"o}wdin}(1950)}]{Lowdin1950}%
  \BibitemOpen
  \bibfield  {author} {\bibinfo {author} {\bibfnamefont {P.-O.}\ \bibnamefont
  {L{\"o}wdin}},\ }\bibfield  {title} {\bibinfo {title} {On the
  non-orthogonality problem connected with the use of atomic wave functions in
  the theory of molecules and crystals},\ }\href
  {https://doi.org/10.1063/1.1747632} {\bibfield  {journal} {\bibinfo
  {journal} {J. Chem. Phys.}\ }\textbf {\bibinfo {volume} {18}},\ \bibinfo
  {pages} {365} (\bibinfo {year} {1950})}\BibitemShut {NoStop}%
\bibitem [{\citenamefont {L{\"o}wdin}(1970)}]{Lowdin1970}%
  \BibitemOpen
  \bibfield  {author} {\bibinfo {author} {\bibfnamefont {P.-O.}\ \bibnamefont
  {L{\"o}wdin}},\ }\bibfield  {title} {\bibinfo {title} {On the
  nonorthogonality problem},\ }\href
  {https://doi.org/10.1016/S0065-3276(08)60339-1} {\bibfield  {journal}
  {\bibinfo  {journal} {Adv. Quantum Chem.}\ }\textbf {\bibinfo {volume} {5}},\
  \bibinfo {pages} {185} (\bibinfo {year} {1970})}\BibitemShut {NoStop}%
\bibitem [{\citenamefont {Hansen}\ and\ \citenamefont
  {Hurwitz}(1943)}]{HansenHurwitz}%
  \BibitemOpen
  \bibfield  {author} {\bibinfo {author} {\bibfnamefont {M.~H.}\ \bibnamefont
  {Hansen}}\ and\ \bibinfo {author} {\bibfnamefont {W.~N.}\ \bibnamefont
  {Hurwitz}},\ }\bibfield  {title} {\bibinfo {title} {{On the Theory of
  Sampling from Finite Populations}},\ }\href
  {https://doi.org/10.1214/aoms/1177731356} {\bibfield  {journal} {\bibinfo
  {journal} {The Annals of Mathematical Statistics}\ }\textbf {\bibinfo
  {volume} {14}},\ \bibinfo {pages} {333 } (\bibinfo {year}
  {1943})}\BibitemShut {NoStop}%
\bibitem [{\citenamefont {Li}\ \emph {et~al.}(2021)\citenamefont {Li},
  \citenamefont {Xia}, \citenamefont {Feng}, \citenamefont {Jin},\ and\
  \citenamefont {Liu}}]{PhysRevD.103.014514}%
  \BibitemOpen
  \bibfield  {author} {\bibinfo {author} {\bibfnamefont {Y.}~\bibnamefont
  {Li}}, \bibinfo {author} {\bibfnamefont {S.-C.}\ \bibnamefont {Xia}},
  \bibinfo {author} {\bibfnamefont {X.}~\bibnamefont {Feng}}, \bibinfo {author}
  {\bibfnamefont {L.-C.}\ \bibnamefont {Jin}},\ and\ \bibinfo {author}
  {\bibfnamefont {C.}~\bibnamefont {Liu}},\ }\bibfield  {title} {\bibinfo
  {title} {Field sparsening for the construction of the correlation functions
  in lattice qcd},\ }\href {https://doi.org/10.1103/PhysRevD.103.014514}
  {\bibfield  {journal} {\bibinfo  {journal} {Phys. Rev. D}\ }\textbf {\bibinfo
  {volume} {103}},\ \bibinfo {pages} {014514} (\bibinfo {year}
  {2021})}\BibitemShut {NoStop}%
\bibitem [{\citenamefont {Brice\~no}\ \emph {et~al.}(2018)\citenamefont
  {Brice\~no}, \citenamefont {Dudek},\ and\ \citenamefont
  {Young}}]{RevModPhys.90.025001}%
  \BibitemOpen
  \bibfield  {author} {\bibinfo {author} {\bibfnamefont {R.~A.}\ \bibnamefont
  {Brice\~no}}, \bibinfo {author} {\bibfnamefont {J.~J.}\ \bibnamefont
  {Dudek}},\ and\ \bibinfo {author} {\bibfnamefont {R.~D.}\ \bibnamefont
  {Young}},\ }\bibfield  {title} {\bibinfo {title} {Scattering processes and
  resonances from lattice qcd},\ }\href
  {https://doi.org/10.1103/RevModPhys.90.025001} {\bibfield  {journal}
  {\bibinfo  {journal} {Rev. Mod. Phys.}\ }\textbf {\bibinfo {volume} {90}},\
  \bibinfo {pages} {025001} (\bibinfo {year} {2018})}\BibitemShut {NoStop}%
\bibitem [{\citenamefont {Green}(2026)}]{Green:2026jgv}%
  \BibitemOpen
  \bibfield  {author} {\bibinfo {author} {\bibfnamefont {J.~R.}\ \bibnamefont
  {Green}},\ }\bibfield  {title} {\bibinfo {title} {{Hadron spectroscopy and
  interactions}},\ }\href {https://doi.org/10.22323/1.518.0005} {\bibfield
  {journal} {\bibinfo  {journal} {PoS}\ }\textbf {\bibinfo {volume}
  {LATTICE2025}},\ \bibinfo {pages} {005} (\bibinfo {year} {2026})},\ \Eprint
  {https://arxiv.org/abs/2602.23244} {arXiv:2602.23244 [hep-lat]} \BibitemShut
  {NoStop}%
\bibitem [{\citenamefont {Detmold}\ \emph {et~al.}(2019)\citenamefont
  {Detmold}, \citenamefont {Edwards}, \citenamefont {Dudek}, \citenamefont
  {Engelhardt}, \citenamefont {Lin}, \citenamefont {Meinel}, \citenamefont
  {Orginos},\ and\ \citenamefont {Shanahan}}]{Detmold:2019ghl}%
  \BibitemOpen
  \bibfield  {author} {\bibinfo {author} {\bibfnamefont {W.}~\bibnamefont
  {Detmold}}, \bibinfo {author} {\bibfnamefont {R.~G.}\ \bibnamefont
  {Edwards}}, \bibinfo {author} {\bibfnamefont {J.~J.}\ \bibnamefont {Dudek}},
  \bibinfo {author} {\bibfnamefont {M.}~\bibnamefont {Engelhardt}}, \bibinfo
  {author} {\bibfnamefont {H.-W.}\ \bibnamefont {Lin}}, \bibinfo {author}
  {\bibfnamefont {S.}~\bibnamefont {Meinel}}, \bibinfo {author} {\bibfnamefont
  {K.}~\bibnamefont {Orginos}},\ and\ \bibinfo {author} {\bibfnamefont
  {P.}~\bibnamefont {Shanahan}} (\bibinfo {collaboration} {USQCD}),\ }\bibfield
   {title} {\bibinfo {title} {{Hadrons and Nuclei}},\ }\href
  {https://doi.org/10.1140/epja/i2019-12902-4} {\bibfield  {journal} {\bibinfo
  {journal} {Eur. Phys. J. A}\ }\textbf {\bibinfo {volume} {55}},\ \bibinfo
  {pages} {193} (\bibinfo {year} {2019})},\ \Eprint
  {https://arxiv.org/abs/1904.09512} {arXiv:1904.09512 [hep-lat]} \BibitemShut
  {NoStop}%
\bibitem [{\citenamefont {Detmold}\ and\ \citenamefont
  {Orginos}(2013)}]{PhysRevD.87.114512}%
  \BibitemOpen
  \bibfield  {author} {\bibinfo {author} {\bibfnamefont {W.}~\bibnamefont
  {Detmold}}\ and\ \bibinfo {author} {\bibfnamefont {K.}~\bibnamefont
  {Orginos}},\ }\bibfield  {title} {\bibinfo {title} {Nuclear correlation
  functions in lattice qcd},\ }\href
  {https://doi.org/10.1103/PhysRevD.87.114512} {\bibfield  {journal} {\bibinfo
  {journal} {Phys. Rev. D}\ }\textbf {\bibinfo {volume} {87}},\ \bibinfo
  {pages} {114512} (\bibinfo {year} {2013})}\BibitemShut {NoStop}%
\bibitem [{\citenamefont {Chakraborty}\ \emph {et~al.}(2024)\citenamefont
  {Chakraborty}, \citenamefont {Srivastava}, \citenamefont {Kumar},\ and\
  \citenamefont {Mathur}}]{PhysRevD.110.114505}%
  \BibitemOpen
  \bibfield  {author} {\bibinfo {author} {\bibfnamefont {D.}~\bibnamefont
  {Chakraborty}}, \bibinfo {author} {\bibfnamefont {P.}~\bibnamefont
  {Srivastava}}, \bibinfo {author} {\bibfnamefont {A.}~\bibnamefont {Kumar}},\
  and\ \bibinfo {author} {\bibfnamefont {N.}~\bibnamefont {Mathur}},\
  }\bibfield  {title} {\bibinfo {title} {Nuclear correlation functions using
  first-principle calculations of lattice quantum chromodynamics},\ }\href
  {https://doi.org/10.1103/PhysRevD.110.114505} {\bibfield  {journal} {\bibinfo
   {journal} {Phys. Rev. D}\ }\textbf {\bibinfo {volume} {110}},\ \bibinfo
  {pages} {114505} (\bibinfo {year} {2024})}\BibitemShut {NoStop}%
\bibitem [{\citenamefont {Mallat}(1989{\natexlab{b}})}]{Mallat1989_1}%
  \BibitemOpen
  \bibfield  {author} {\bibinfo {author} {\bibfnamefont {S.~G.}\ \bibnamefont
  {Mallat}},\ }\bibfield  {title} {\bibinfo {title} {Multiresolution
  approximations and wavelet orthonormal bases of $l^2(\mathbb{R})$},\ }\href
  {https://doi.org/10.1090/S0002-9947-1989-1008470-5} {\bibfield  {journal}
  {\bibinfo  {journal} {Trans. Amer. Math. Soc.}\ }\textbf {\bibinfo {volume}
  {315}},\ \bibinfo {pages} {69} (\bibinfo {year}
  {1989}{\natexlab{b}})}\BibitemShut {NoStop}%
\bibitem [{\citenamefont {Daubechies}(1992)}]{Daubechies1992_1}%
  \BibitemOpen
  \bibfield  {author} {\bibinfo {author} {\bibfnamefont {I.}~\bibnamefont
  {Daubechies}},\ }\href {https://doi.org/10.1137/1.9781611970104} {\emph
  {\bibinfo {title} {Ten Lectures on Wavelets}}},\ \bibinfo {series} {CBMS-NSF
  Regional Conference Series in Applied Mathematics}, Vol.~\bibinfo {volume}
  {61}\ (\bibinfo  {publisher} {SIAM},\ \bibinfo {address} {Philadelphia},\
  \bibinfo {year} {1992})\BibitemShut {NoStop}%
\bibitem [{\citenamefont {Coifman}\ and\ \citenamefont
  {Maggioni}(2006)}]{CoifmanMaggioni2006}%
  \BibitemOpen
  \bibfield  {author} {\bibinfo {author} {\bibfnamefont {R.~R.}\ \bibnamefont
  {Coifman}}\ and\ \bibinfo {author} {\bibfnamefont {M.}~\bibnamefont
  {Maggioni}},\ }\bibfield  {title} {\bibinfo {title} {Diffusion wavelets},\
  }\href {https://doi.org/10.1016/j.acha.2006.04.004} {\bibfield  {journal}
  {\bibinfo  {journal} {Appl. Comput. Harmon. Anal.}\ }\textbf {\bibinfo
  {volume} {21}},\ \bibinfo {pages} {53} (\bibinfo {year} {2006})}\BibitemShut
  {NoStop}%
\bibitem [{\citenamefont {Hammond}\ \emph {et~al.}(2011)\citenamefont
  {Hammond}, \citenamefont {Vandergheynst},\ and\ \citenamefont
  {Gribonval}}]{Hammond2011}%
  \BibitemOpen
  \bibfield  {author} {\bibinfo {author} {\bibfnamefont {D.~K.}\ \bibnamefont
  {Hammond}}, \bibinfo {author} {\bibfnamefont {P.}~\bibnamefont
  {Vandergheynst}},\ and\ \bibinfo {author} {\bibfnamefont {R.}~\bibnamefont
  {Gribonval}},\ }\bibfield  {title} {\bibinfo {title} {Wavelets on graphs via
  spectral graph theory},\ }\href {https://doi.org/10.1016/j.acha.2010.04.005}
  {\bibfield  {journal} {\bibinfo  {journal} {Appl. Comput. Harmon. Anal.}\
  }\textbf {\bibinfo {volume} {30}},\ \bibinfo {pages} {129} (\bibinfo {year}
  {2011})},\ \Eprint {https://arxiv.org/abs/0912.3848} {arXiv:0912.3848}
  \BibitemShut {NoStop}%
\bibitem [{\citenamefont {Hasenfratz}\ and\ \citenamefont
  {Knechtli}(2001)}]{Hasenfratz:2001hp}%
  \BibitemOpen
  \bibfield  {author} {\bibinfo {author} {\bibfnamefont {A.}~\bibnamefont
  {Hasenfratz}}\ and\ \bibinfo {author} {\bibfnamefont {F.}~\bibnamefont
  {Knechtli}},\ }\bibfield  {title} {\bibinfo {title} {Flavor symmetry and the
  static potential with hypercubic blocking},\ }\href
  {https://doi.org/10.1103/PhysRevD.64.034504} {\bibfield  {journal} {\bibinfo
  {journal} {Phys. Rev. D}\ }\textbf {\bibinfo {volume} {64}},\ \bibinfo
  {pages} {034504} (\bibinfo {year} {2001})},\ \Eprint
  {https://arxiv.org/abs/hep-lat/0103029} {arXiv:hep-lat/0103029 [hep-lat]}
  \BibitemShut {NoStop}%
\bibitem [{\citenamefont {Falcioni}\ \emph {et~al.}(1985)\citenamefont
  {Falcioni}, \citenamefont {Paciello}, \citenamefont {Parisi},\ and\
  \citenamefont {Taglienti}}]{APESMEARING}%
  \BibitemOpen
  \bibfield  {author} {\bibinfo {author} {\bibfnamefont {M.}~\bibnamefont
  {Falcioni}}, \bibinfo {author} {\bibfnamefont {M.}~\bibnamefont {Paciello}},
  \bibinfo {author} {\bibfnamefont {G.}~\bibnamefont {Parisi}},\ and\ \bibinfo
  {author} {\bibfnamefont {B.}~\bibnamefont {Taglienti}},\ }\bibfield  {title}
  {\bibinfo {title} {Again on su(3) glueball mass},\ }\href
  {https://doi.org/https://doi.org/10.1016/0550-3213(85)90280-9} {\bibfield
  {journal} {\bibinfo  {journal} {Nuclear Physics B}\ }\textbf {\bibinfo
  {volume} {251}},\ \bibinfo {pages} {624} (\bibinfo {year}
  {1985})}\BibitemShut {NoStop}%
\bibitem [{\citenamefont {Bazavov}\ \emph {et~al.}(2013)\citenamefont
  {Bazavov}, \citenamefont {Bernard}, \citenamefont {Komijani}, \citenamefont
  {DeTar}, \citenamefont {Levkova}, \citenamefont {Freeman}, \citenamefont
  {Gottlieb}, \citenamefont {Zhou}, \citenamefont {Heller}, \citenamefont
  {Hetrick}, \citenamefont {Laiho}, \citenamefont {Osborn}, \citenamefont
  {Sugar}, \citenamefont {Toussaint},\ and\ \citenamefont {Van~de
  Water}}]{PhysRevD.87.054505}%
  \BibitemOpen
  \bibfield  {author} {\bibinfo {author} {\bibfnamefont {A.}~\bibnamefont
  {Bazavov}}, \bibinfo {author} {\bibfnamefont {C.}~\bibnamefont {Bernard}},
  \bibinfo {author} {\bibfnamefont {J.}~\bibnamefont {Komijani}}, \bibinfo
  {author} {\bibfnamefont {C.}~\bibnamefont {DeTar}}, \bibinfo {author}
  {\bibfnamefont {L.}~\bibnamefont {Levkova}}, \bibinfo {author} {\bibfnamefont
  {W.}~\bibnamefont {Freeman}}, \bibinfo {author} {\bibfnamefont
  {S.}~\bibnamefont {Gottlieb}}, \bibinfo {author} {\bibfnamefont
  {R.}~\bibnamefont {Zhou}}, \bibinfo {author} {\bibfnamefont {U.~M.}\
  \bibnamefont {Heller}}, \bibinfo {author} {\bibfnamefont {J.~E.}\
  \bibnamefont {Hetrick}}, \bibinfo {author} {\bibfnamefont {J.}~\bibnamefont
  {Laiho}}, \bibinfo {author} {\bibfnamefont {J.}~\bibnamefont {Osborn}},
  \bibinfo {author} {\bibfnamefont {R.~L.}\ \bibnamefont {Sugar}}, \bibinfo
  {author} {\bibfnamefont {D.}~\bibnamefont {Toussaint}},\ and\ \bibinfo
  {author} {\bibfnamefont {R.~S.}\ \bibnamefont {Van~de Water}} (\bibinfo
  {collaboration} {MILC Collaboration}),\ }\bibfield  {title} {\bibinfo {title}
  {Lattice qcd ensembles with four flavors of highly improved staggered
  quarks},\ }\href {https://doi.org/10.1103/PhysRevD.87.054505} {\bibfield
  {journal} {\bibinfo  {journal} {Phys. Rev. D}\ }\textbf {\bibinfo {volume}
  {87}},\ \bibinfo {pages} {054505} (\bibinfo {year} {2013})}\BibitemShut
  {NoStop}%
\bibitem [{\citenamefont {Follana}\ \emph {et~al.}(2007)\citenamefont
  {Follana}, \citenamefont {Mason}, \citenamefont {Davies}, \citenamefont
  {Hornbostel}, \citenamefont {Lepage}, \citenamefont {Shigemitsu},
  \citenamefont {Trottier},\ and\ \citenamefont {Wong}}]{PhysRevD.75.054502}%
  \BibitemOpen
  \bibfield  {author} {\bibinfo {author} {\bibfnamefont {E.}~\bibnamefont
  {Follana}}, \bibinfo {author} {\bibfnamefont {Q.}~\bibnamefont {Mason}},
  \bibinfo {author} {\bibfnamefont {C.}~\bibnamefont {Davies}}, \bibinfo
  {author} {\bibfnamefont {K.}~\bibnamefont {Hornbostel}}, \bibinfo {author}
  {\bibfnamefont {G.~P.}\ \bibnamefont {Lepage}}, \bibinfo {author}
  {\bibfnamefont {J.}~\bibnamefont {Shigemitsu}}, \bibinfo {author}
  {\bibfnamefont {H.}~\bibnamefont {Trottier}},\ and\ \bibinfo {author}
  {\bibfnamefont {K.}~\bibnamefont {Wong}},\ }\bibfield  {title} {\bibinfo
  {title} {Highly improved staggered quarks on the lattice with applications to
  charm physics},\ }\href {https://doi.org/10.1103/PhysRevD.75.054502}
  {\bibfield  {journal} {\bibinfo  {journal} {Phys. Rev. D}\ }\textbf {\bibinfo
  {volume} {75}},\ \bibinfo {pages} {054502} (\bibinfo {year}
  {2007})}\BibitemShut {NoStop}%
\bibitem [{\citenamefont {Higham}(2008)}]{higham2008functions}%
  \BibitemOpen
  \bibfield  {author} {\bibinfo {author} {\bibfnamefont {N.~J.}\ \bibnamefont
  {Higham}},\ }\href {https://doi.org/10.1137/1.9780898717778} {\emph {\bibinfo
  {title} {Functions of Matrices: Theory and Computation}}},\ \bibinfo {series}
  {Other Titles in Applied Mathematics}, Vol.\ \bibinfo {volume} {104}\
  (\bibinfo  {publisher} {SIAM},\ \bibinfo {address} {Philadelphia, PA},\
  \bibinfo {year} {2008})\BibitemShut {NoStop}%
\bibitem [{\citenamefont {Bali}\ \emph {et~al.}(2023)\citenamefont {Bali},
  \citenamefont {Collins}, \citenamefont {Heybrock}, \citenamefont {L\"offler},
  \citenamefont {R\"odl}, \citenamefont {S\"oldner},\ and\ \citenamefont
  {Weish\"aupl}}]{CLS}%
  \BibitemOpen
  \bibfield  {author} {\bibinfo {author} {\bibfnamefont {G.~S.}\ \bibnamefont
  {Bali}}, \bibinfo {author} {\bibfnamefont {S.}~\bibnamefont {Collins}},
  \bibinfo {author} {\bibfnamefont {S.}~\bibnamefont {Heybrock}}, \bibinfo
  {author} {\bibfnamefont {M.}~\bibnamefont {L\"offler}}, \bibinfo {author}
  {\bibfnamefont {R.}~\bibnamefont {R\"odl}}, \bibinfo {author} {\bibfnamefont
  {W.}~\bibnamefont {S\"oldner}},\ and\ \bibinfo {author} {\bibfnamefont
  {S.}~\bibnamefont {Weish\"aupl}} (\bibinfo {collaboration} {RQCD
  Collaboration}),\ }\bibfield  {title} {\bibinfo {title} {Octet baryon
  isovector charges from ${N}_{f}=2+1$ lattice qcd},\ }\href
  {https://doi.org/10.1103/PhysRevD.108.034512} {\bibfield  {journal} {\bibinfo
   {journal} {Phys. Rev. D}\ }\textbf {\bibinfo {volume} {108}},\ \bibinfo
  {pages} {034512} (\bibinfo {year} {2023})}\BibitemShut {NoStop}%
\bibitem [{\citenamefont {Hestenes}\ and\ \citenamefont
  {Stiefel}(1952)}]{HestenesStiefel1952}%
  \BibitemOpen
  \bibfield  {author} {\bibinfo {author} {\bibfnamefont {M.~R.}\ \bibnamefont
  {Hestenes}}\ and\ \bibinfo {author} {\bibfnamefont {E.}~\bibnamefont
  {Stiefel}},\ }\bibfield  {title} {\bibinfo {title} {Methods of conjugate
  gradients for solving linear systems},\ }\href
  {https://doi.org/10.6028/jres.049.044} {\bibfield  {journal} {\bibinfo
  {journal} {J. Res. Natl. Bur. Stand.}\ }\textbf {\bibinfo {volume} {49}},\
  \bibinfo {pages} {409} (\bibinfo {year} {1952})}\BibitemShut {NoStop}%
\bibitem [{\citenamefont {Saad}(2003)}]{Saad2003}%
  \BibitemOpen
  \bibfield  {author} {\bibinfo {author} {\bibfnamefont {Y.}~\bibnamefont
  {Saad}},\ }\href {https://doi.org/10.1137/1.9780898718003} {\emph {\bibinfo
  {title} {Iterative Methods for Sparse Linear Systems}}},\ \bibinfo {edition}
  {2nd}\ ed.\ (\bibinfo  {publisher} {SIAM},\ \bibinfo {address} {Philadelphia,
  PA},\ \bibinfo {year} {2003})\BibitemShut {NoStop}%
\bibitem [{\citenamefont {Stathopoulos}\ and\ \citenamefont
  {McCombs}(2010)}]{PRIMME}%
  \BibitemOpen
  \bibfield  {author} {\bibinfo {author} {\bibfnamefont {A.}~\bibnamefont
  {Stathopoulos}}\ and\ \bibinfo {author} {\bibfnamefont {J.~R.}\ \bibnamefont
  {McCombs}},\ }\bibfield  {title} {\bibinfo {title} {{PRIMME}:
  {PR}econditioned {I}terative {M}ulti{M}ethod {E}igensolver---methods and
  software description},\ }\href {https://doi.org/10.1145/1731022.1731031}
  {\bibfield  {journal} {\bibinfo  {journal} {ACM Transactions on Mathematical
  Software}\ }\textbf {\bibinfo {volume} {37}},\ \bibinfo {pages} {21:1}
  (\bibinfo {year} {2010})}\BibitemShut {NoStop}%
\bibitem [{\citenamefont {Stathopoulos}\ and\ \citenamefont
  {McCombs}(2007)}]{StathopoulosMcCombs2007}%
  \BibitemOpen
  \bibfield  {author} {\bibinfo {author} {\bibfnamefont {A.}~\bibnamefont
  {Stathopoulos}}\ and\ \bibinfo {author} {\bibfnamefont {J.~R.}\ \bibnamefont
  {McCombs}},\ }\bibfield  {title} {\bibinfo {title} {Nearly optimal
  preconditioned methods for {H}ermitian eigenproblems under limited memory.
  {P}art {II}: Seeking many eigenvalues},\ }\href
  {https://doi.org/10.1137/060661144} {\bibfield  {journal} {\bibinfo
  {journal} {SIAM Journal on Scientific Computing}\ }\textbf {\bibinfo {volume}
  {29}},\ \bibinfo {pages} {2162} (\bibinfo {year} {2007})}\BibitemShut
  {NoStop}%
\bibitem [{\citenamefont {Zhou}\ \emph {et~al.}(2006)\citenamefont {Zhou},
  \citenamefont {Saad}, \citenamefont {Tiago},\ and\ \citenamefont
  {Chelikowsky}}]{10.1016/j.jcp.2006.03.017}%
  \BibitemOpen
  \bibfield  {author} {\bibinfo {author} {\bibfnamefont {Y.}~\bibnamefont
  {Zhou}}, \bibinfo {author} {\bibfnamefont {Y.}~\bibnamefont {Saad}}, \bibinfo
  {author} {\bibfnamefont {M.~L.}\ \bibnamefont {Tiago}},\ and\ \bibinfo
  {author} {\bibfnamefont {J.~R.}\ \bibnamefont {Chelikowsky}},\ }\bibfield
  {title} {\bibinfo {title} {Self-consistent-field calculations using
  chebyshev-filtered subspace iteration},\ }\href
  {https://doi.org/10.1016/j.jcp.2006.03.017} {\bibfield  {journal} {\bibinfo
  {journal} {J. Comput. Phys.}\ }\textbf {\bibinfo {volume} {219}},\ \bibinfo
  {pages} {172–184} (\bibinfo {year} {2006})}\BibitemShut {NoStop}%
\bibitem [{\citenamefont {Zhou}\ and\ \citenamefont
  {Saad}(2007)}]{doi:10.1137/050630404}%
  \BibitemOpen
  \bibfield  {author} {\bibinfo {author} {\bibfnamefont {Y.}~\bibnamefont
  {Zhou}}\ and\ \bibinfo {author} {\bibfnamefont {Y.}~\bibnamefont {Saad}},\
  }\bibfield  {title} {\bibinfo {title} {A chebyshev–davidson algorithm for
  large symmetric eigenproblems},\ }\href {https://doi.org/10.1137/050630404}
  {\bibfield  {journal} {\bibinfo  {journal} {SIAM Journal on Matrix Analysis
  and Applications}\ }\textbf {\bibinfo {volume} {29}},\ \bibinfo {pages} {954}
  (\bibinfo {year} {2007})},\ \Eprint
  {https://arxiv.org/abs/https://doi.org/10.1137/050630404}
  {https://doi.org/10.1137/050630404} \BibitemShut {NoStop}%
\bibitem [{\citenamefont {Trefethen}(2019)}]{Trefethen2019}%
  \BibitemOpen
  \bibfield  {author} {\bibinfo {author} {\bibfnamefont {L.~N.}\ \bibnamefont
  {Trefethen}},\ }\href {https://doi.org/10.1137/1.9781611975949} {\emph
  {\bibinfo {title} {Approximation Theory and Approximation Practice, Extended
  Edition}}}\ (\bibinfo  {publisher} {Society for Industrial and Applied
  Mathematics},\ \bibinfo {address} {Philadelphia, PA},\ \bibinfo {year}
  {2019})\BibitemShut {NoStop}%
\bibitem [{\citenamefont {{MILC Collaboration}}()}]{MILC}%
  \BibitemOpen
  \bibfield  {author} {\bibinfo {author} {\bibnamefont {{MILC
  Collaboration}}},\ }\href@noop {} {\bibinfo {title} {The {MILC} code}},\
  \bibinfo {howpublished}
  {\url{https://github.com/milc-qcd/milc_qcd}}\BibitemShut {NoStop}%
\bibitem [{\citenamefont {Stathopoulos}\ \emph {et~al.}(2013)\citenamefont
  {Stathopoulos}, \citenamefont {Laeuchli},\ and\ \citenamefont
  {Orginos}}]{HProbing1}%
  \BibitemOpen
  \bibfield  {author} {\bibinfo {author} {\bibfnamefont {A.}~\bibnamefont
  {Stathopoulos}}, \bibinfo {author} {\bibfnamefont {J.}~\bibnamefont
  {Laeuchli}},\ and\ \bibinfo {author} {\bibfnamefont {K.}~\bibnamefont
  {Orginos}},\ }\bibfield  {title} {\bibinfo {title} {Hierarchical probing for
  estimating the trace of the matrix inverse on toroidal lattices},\ }\href
  {https://doi.org/10.1137/120881452} {\bibfield  {journal} {\bibinfo
  {journal} {SIAM Journal on Scientific Computing}\ }\textbf {\bibinfo {volume}
  {35}},\ \bibinfo {pages} {S299} (\bibinfo {year} {2013})},\ \Eprint
  {https://arxiv.org/abs/https://doi.org/10.1137/120881452}
  {https://doi.org/10.1137/120881452} \BibitemShut {NoStop}%
\bibitem [{\citenamefont {Laeuchli}\ and\ \citenamefont
  {Stathopoulos}(2020)}]{HProbing2}%
  \BibitemOpen
  \bibfield  {author} {\bibinfo {author} {\bibfnamefont {J.}~\bibnamefont
  {Laeuchli}}\ and\ \bibinfo {author} {\bibfnamefont {A.}~\bibnamefont
  {Stathopoulos}},\ }\bibfield  {title} {\bibinfo {title} {Extending
  hierarchical probing for computing the trace of matrix inverses},\ }\href
  {https://doi.org/10.1137/18M1176427} {\bibfield  {journal} {\bibinfo
  {journal} {SIAM Journal on Scientific Computing}\ }\textbf {\bibinfo {volume}
  {42}},\ \bibinfo {pages} {A1459} (\bibinfo {year} {2020})}\BibitemShut
  {NoStop}%
\bibitem [{\citenamefont {Papace}\ \emph {et~al.}(2026)\citenamefont {Papace},
  \citenamefont {Frommer}, \citenamefont {Jimenez-Merchan}, \citenamefont
  {Lang}, \citenamefont {Ramirez-Hidalgo},\ and\ \citenamefont
  {Schneider}}]{MultiplierBasedColoring}%
  \BibitemOpen
  \bibfield  {author} {\bibinfo {author} {\bibfnamefont {M.}~\bibnamefont
  {Papace}}, \bibinfo {author} {\bibfnamefont {A.}~\bibnamefont {Frommer}},
  \bibinfo {author} {\bibfnamefont {J.}~\bibnamefont {Jimenez-Merchan}},
  \bibinfo {author} {\bibfnamefont {B.}~\bibnamefont {Lang}}, \bibinfo {author}
  {\bibfnamefont {G.}~\bibnamefont {Ramirez-Hidalgo}},\ and\ \bibinfo {author}
  {\bibfnamefont {C.}~\bibnamefont {Schneider}},\ }\href@noop {} {\bibinfo
  {title} {{Probing and graph coloring techniques for trace estimation in
  Lattice QCD}}} (\bibinfo {year} {2026}),\ \Eprint
  {https://arxiv.org/abs/2606.00394} {arXiv:2606.00394 [hep-lat]} \BibitemShut
  {NoStop}%
\bibitem [{\citenamefont {Stathopoulos}\ and\ \citenamefont
  {Wu}(2002)}]{doi:10.1137/S1064827500370883}%
  \BibitemOpen
  \bibfield  {author} {\bibinfo {author} {\bibfnamefont {A.}~\bibnamefont
  {Stathopoulos}}\ and\ \bibinfo {author} {\bibfnamefont {K.}~\bibnamefont
  {Wu}},\ }\bibfield  {title} {\bibinfo {title} {A block orthogonalization
  procedure with constant synchronization requirements},\ }\href
  {https://doi.org/10.1137/S1064827500370883} {\bibfield  {journal} {\bibinfo
  {journal} {SIAM Journal on Scientific Computing}\ }\textbf {\bibinfo {volume}
  {23}},\ \bibinfo {pages} {2165} (\bibinfo {year} {2002})},\ \Eprint
  {https://arxiv.org/abs/https://doi.org/10.1137/S1064827500370883}
  {https://doi.org/10.1137/S1064827500370883} \BibitemShut {NoStop}%
\bibitem [{\citenamefont {Wei{\ss{}}e}\ \emph {et~al.}(2006)\citenamefont
  {Wei{\ss{}}e}, \citenamefont {Wellein}, \citenamefont {Alvermann},\ and\
  \citenamefont {Fehske}}]{RevModPhys.78.275}%
  \BibitemOpen
  \bibfield  {author} {\bibinfo {author} {\bibfnamefont {A.}~\bibnamefont
  {Wei{\ss{}}e}}, \bibinfo {author} {\bibfnamefont {G.}~\bibnamefont
  {Wellein}}, \bibinfo {author} {\bibfnamefont {A.}~\bibnamefont {Alvermann}},\
  and\ \bibinfo {author} {\bibfnamefont {H.}~\bibnamefont {Fehske}},\
  }\bibfield  {title} {\bibinfo {title} {The kernel polynomial method},\ }\href
  {https://doi.org/10.1103/RevModPhys.78.275} {\bibfield  {journal} {\bibinfo
  {journal} {Rev. Mod. Phys.}\ }\textbf {\bibinfo {volume} {78}},\ \bibinfo
  {pages} {275} (\bibinfo {year} {2006})}\BibitemShut {NoStop}%
\end{thebibliography}%

\end{document}